\documentclass[12pt]{article}
\usepackage{amsfonts,amsthm,amsmath,amssymb,upgreek,bm,mathtools}
\usepackage[bookmarks = false, linktocpage=true]{hyperref}
\usepackage[paper=letterpaper,margin=.85in]{geometry}
\usepackage{graphicx}
\usepackage{units}
\usepackage{float}
\usepackage[export]{adjustbox}
\usepackage{bbold}
\usepackage{xcolor}
\usepackage[shortlabels]{enumitem}
\usepackage{dsfont}
\usepackage[mathscr]{euscript}
\usepackage[normalem]{ulem}
\usepackage{braket}
\usepackage{slashed}
\usepackage{comment}

\newcommand{\be}{\begin{equation}}
\newcommand{\ee}{\end{equation}}

\renewcommand{\i}{\mathrm{i}}

\numberwithin{equation}{section}

\usepackage{color}
\definecolor{darkblue}{rgb}{0,0,0.6}
\definecolor{purple}{rgb}{0.4,.2,0.7}
\definecolor{darkgreen}{rgb}{0,0.5,0}

\begin{document}
\thispagestyle{empty}

\vspace*{2.5cm}
\begin{center}

{\bf {\LARGE Crosscap Contribution to Late-Time Two-Point Correlators}}

\begin{center}

\vspace{1cm}

 {\bf Cynthia Yan}\\
  \bigskip \rm

\bigskip 

Stanford Institute for Theoretical Physics,\\Stanford University, Stanford, CA 94305

\rm
  \end{center}

\vspace{2.5cm}
{\bf Abstract}
\end{center}
\begin{quotation}
\noindent

We show that in Jackiw-Teitelboim (JT) gravity, late-time two-point functions can get a leading non-decaying contribution from a spacetime with the topology of a M\"{o}bius strip (a disk with one crosscap). There is an interesting interplay between this contribution and the standard ``plateau''. The two can add together or cancel, depending on topological weighting factors. We match this behavior to Random Matrix Theory (RMT) and the N mod 8 periodicity of Sachdev-Kitaev-Ye (SYK) results.

\end{quotation}

 \newcommand{\psltwor}{\text{PSL}(2,\mathbb{R})}
  \newcommand{\psltwoc}{\text{PSL}(2,\mathbb{C})}
 \newcommand{\volhtwo}{\text{vol}(H^2)}
 \newcommand{\osp}{\text{OSp}(1|2)/\mathbb{Z}_2}

\setcounter{page}{0}
\setcounter{tocdepth}{3}
\setcounter{footnote}{0}
\newpage

\setcounter{page}{2}
\tableofcontents
\pagebreak

\section{Introduction}

Maldacena \cite{Maldacenaeternalbh} formulated a version of the black hole information paradox using simple correlation functions in AdS black holes. He observed that the bulk theory's prediction of exponential decay \cite{Horowitz:1999jd, Goheer:2002vf, Dyson:2002pf, Barbon:2003aq} of a two-point correlation function cannot be consistent with unitarity on the boundary CFT. He proposed that this paradox can be resolved if one sums over all geometries with prescribed boundary conditions on the bulk side. Saad \cite{Saadsingleauthor} \footnote{continuing the idea of \cite{Blommaert:2019hjr}} carried out this proposal by considering geometries in the context of Jackiw-Teitelboim (JT) gravity \cite{Teitelboim, Jackiw, AlmheiriPolchinski} coupled with matter. More specifically, he computed bosonic two-point correlation functions using the techniques developed by Yang \cite{Yangsingleauthor} \footnote{for other approaches see \cite{Lam:2018pvp, Mertens:2017mtv, Blommaert:2018oro, Blommaert:2019hjr, Iliesiu:2019xuh}} on the bulk side  and compared with Random Matrix Theory (RMT) predictions for operators satisfying Eigenstate Thermalization Hypothesis (ETH) \cite{ethDeutsch, ethSrednicki} on the boundary side.  

\begin{figure}[h]
\centering
\includegraphics[width=0.7\textwidth]{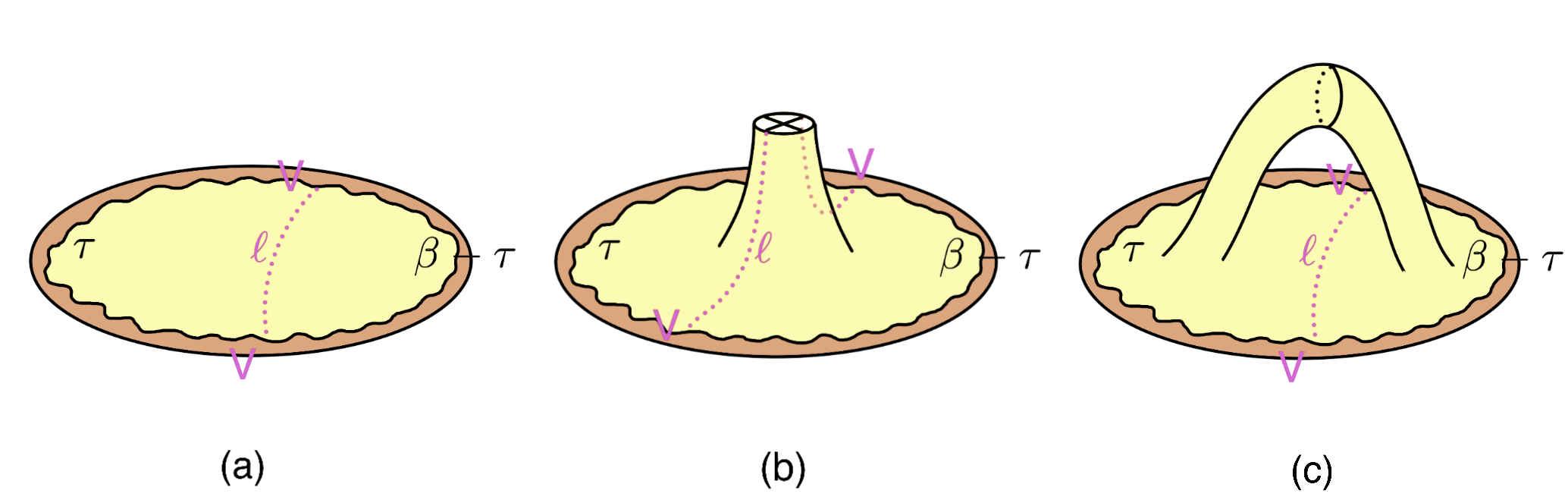}
\caption{Lowest order contributions to two-point correlation functions. (a) The contribution of the disk topology decays at late time \cite{AltlandBagrets, MertensVerlinde, KitaevSuh, Yangsingleauthor}. (b) The crosscap contribution gives a leading non-decaying contribution proportional to $e^{-S_0}$. (c) The contribution of a genus one surface with one boundary, i.e. a handle-disk, is non-decaying and proportional to $te^{-2S_0}$ \cite{Saadsingleauthor}.}
\label{intro}
\end{figure}

Saad \cite{Saadsingleauthor} showed on the boundary side using RMT that with respect to time $t$ bosonic two-point correlation function should first decay exponentially (called a slope), then climb up linearly (called a ramp), and finally stay constant (called a plateau). To explain the late-time non-decaying feature on the bulk side, Saad studied a genus one surface with one boundary (a handle-disk), which gives a contribution to the two-point correlation function proportional to $te^{-2S_0}$, and so the handle-disk is responsible for the ramp on the graph of two-point correlator v.s. time.

In this paper, we extend Saad's result to non-orientable geometries and fermionic two-point correlation functions. A key geometry we study is a disk with a crosscap, which is topologically equivalent to a M\"{o}bius band. We find that the crosscap gives a non-decaying contribution to the bosonic two-point correlation function proportional to $e^{-S_0}$, which is similar to the original plateau, and can either enhance it or cancel it depending on the weighting factors we attach in front of this contribution. From there we crosscheck with predictions from RMT with appropriate symmetry classes on the boundary side. We then go to fermionic two-point correlation functions. There we consider a disk with one crosscap and a disk with two crosscaps. On each particular geometry, we sum over all Spin/Pin$^-$ structures with appropriate weightings related to their corresponding topological invariants. This leads to an N mod 8 periodicity. Our result confirms known numerics results \cite{9authors} of two-point correlation functions in Sachdev-Kitaev-Ye (SYK) model \cite{SachdevYe, KitaevTalks, KitaevSuh} \footnote{for more works on SYK model see \cite{Polchinski:2016xgd, Maldacena:2016hyu, Jensen:2016pah, 9authors, SaadShenkerStanford18}}.

\begin{figure}[H]
\centering
\includegraphics[width=0.5\textwidth]{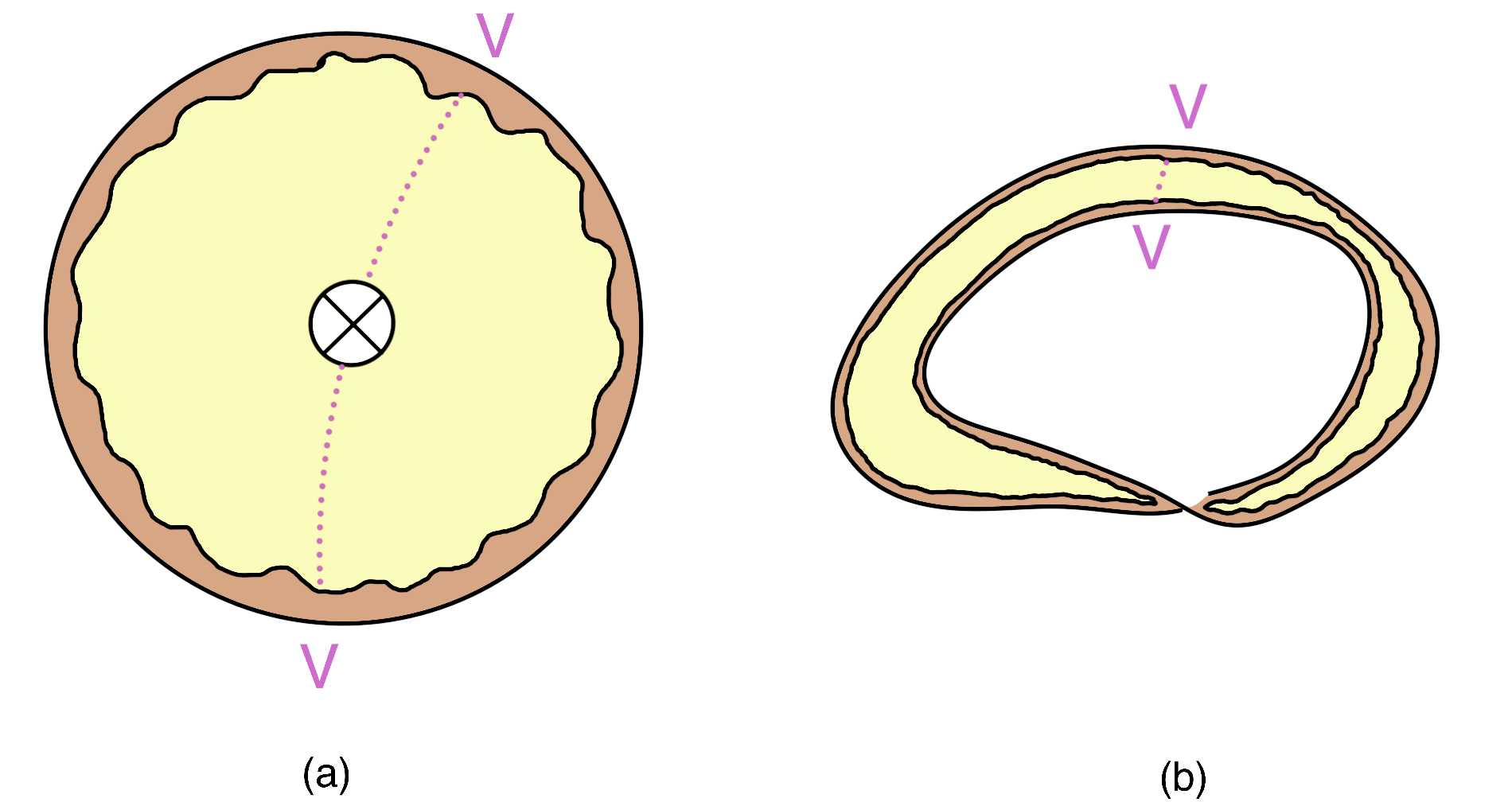}
\caption{a disk with a crosscap is topologically equivalent to a M\"{o}bius band}
\label{ccintro}
\end{figure}

In section 2, we review some important tools involved in calculating two-point correlation functions in JT, calculate crosscap contributions to bosonic two-point functions, and compare the results with RMT computations. In section 3, we review Spin structure on orientable geometries, introduce Pin$^-$ structure on non-orientable geometries and from there examine the contributions of a disk with one crosscap or two crosscaps to the fermionic two-point functions, and compare with RMT computations as well as SYK numerics. 

\section{Bosonic Two-Point Correlation Functions}
\subsection{Review}

In this paper we are working in a simple model of holography, where the bulk theory is JT gravity with probe matter fields, and the boundary theory is formally an ordinary quantum system, or an ensemble of such systems. From the boundary perspective, we are interested in computing the thermal two-point function at late times $t$:
\begin{equation}
\braket{V(t)V(0)}=\mathrm{Tr}[e^{-\beta H}V(t)V(0)]=\sum_{n,m}e^{-\beta E_m}e^{-it(E_n-E_m)}|\braket{E_n|V|E_m}|^2
\end{equation}
We are hoping to match expected features of this function to bulk computations of correlation functions in JT gravity. To compute the bulk correlation function, we are suppose to hold fixed the boundary conditions (including the operators $V(0)$ and $V(t)$ that we are inserting) and sum over two-dimensional bulk topologies. We are particularly interested in the contribution of the topology of a disk with a crosscap inserted, but we will review the computations of the disk topology and the disk with a handle inserted. 

The 2d gravity theory we will study consists of the Einstein-Hilbert action + JT gravity action + action from matter. JT gravity on a 2d manifold $M$ has Euclidean action
\begin{equation}
I_{JT}=-\frac{1}{2}\left(\int_M\phi(R+2)+2\int_{\partial M}\phi_b (K-1)\right)
\end{equation}
Classically, the equation of motion fixes the bulk geometry to be AdS$_2$ with $R=-2$ and the action reduces to a Schwarzian action on the boundary \cite{MaldacenaStanfordYang}. In 2d, the Einstein-Hilbert action is purely topological and can be written as 
\begin{equation}
I_{EH}=-\chi S_0
\end{equation}
where $\chi=2-2g-n$ is the Euler character for manifold $M$ with $g$ the genus and $n$ the number of boundaries, and $S_0$ is the zero-temperature bulk entropy which is a constant. The Einstein-Hilbert action then contributes an overall factor $e^{\chi S_0}$ to the partition function. In all of our figures the orange disks represent infinite hyperbolic space (or its quotient) and yellow geometries inside represent the physical Euclidean spacetimes, with wiggly regularized boundaries described by the Schwarzian theory \cite{MaldacenaStanfordYang}.

\begin{figure}[H]
\centering
\includegraphics[width=0.5\textwidth]{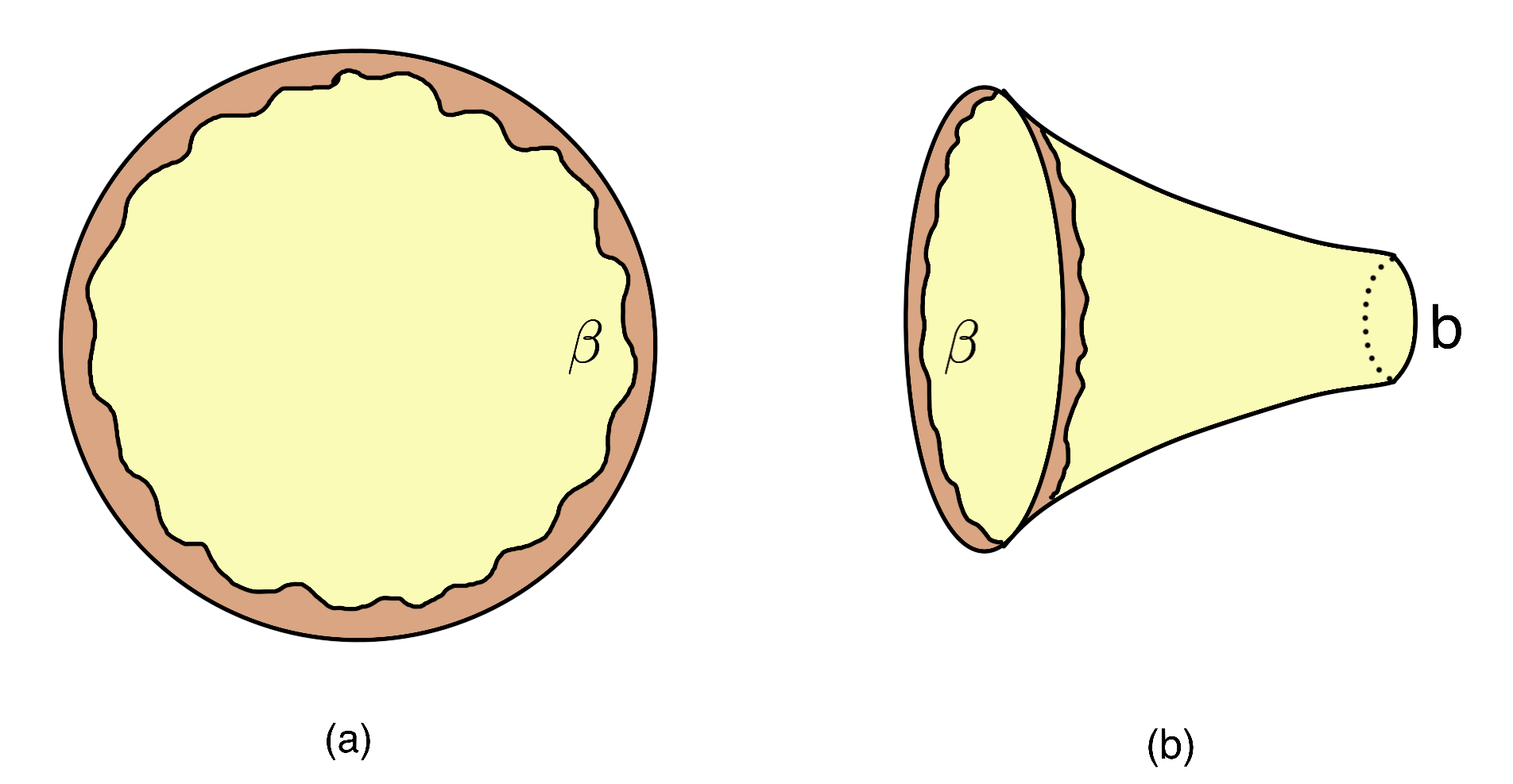}
\caption{(a) disk partition function (b) trumpet partition function}
\label{partitionfunction}
\end{figure}

The two main shapes of Euclidean AdS we consider in this review are a hyperbolic disk which has one asymptotically boundary with renormalized length $\beta$, and a hyperbolic trumpet which has one asymptotic boundary with renormalized length $\beta$ and one geodesic boundary with length $b$ (see figure~\ref{partitionfunction}). That is because a disk is the simplest hyperbolic geometry with one asymptotic boundary and a trumpet can be thought of as a building block of more complicated geometries via attaching a Riemann surface with one geodesic boundary to the geodesic boundary of the trumpet. 

JT path integrals without operator insertions can be computed directly by doing the path integral over the wiggly boundary of the disk and the trumpet explicitly. Disk \cite{BagretsAltlandKamenev16, 9authors, BagretsAltlandKamenev17, StanfordWitten17, SchwarzianBelokurovShavgulidze, SchwarzianMertensTuriaciVerlinde, KitaevSuhBH, Yangsingleauthor, IliesiuPufuVerlindeWang} and trumpet partition functions \cite{StanfordWitten17, Saadsingleauthor} are given respectively by
\begin{equation}
Z_{\text{Disk}}(\beta)=e^{S_0}\frac{e^{\frac{2\pi^2}{\beta}}}{\sqrt{2\pi}\beta^{3/2}}=e^{S_0}\int_0^\infty dE\,\underbrace{\frac{\sinh(2\pi\sqrt{2E})}{2\pi^2}}_{\rho_0(E)}e^{-\beta E}\label{zdisk}
\end{equation}
and 
\begin{equation}
Z_{\text{Trumpet}}(\beta,b)=\frac{e^{-\frac{b^2}{2\beta}}}{\sqrt{2\pi\beta}}=\int_0^\infty dE\,\frac{\cos(b\sqrt{2E})}{\pi\sqrt{2E}}e^{-\beta E}\label{ztrumpet}
\end{equation}
where $\rho_0(E)$ denotes the density of state.

To compute path integrals with operator insertions we need more tools. Before we do that, we should note that there is another way of computing the disk partition function. A disk can be decomposed into two Hartle-Hawking wavefunctions by the following procedure

\begin{align}
Z_{\text{Disk}}(\beta)&=\includegraphics[valign=c,width=0.2\textwidth]{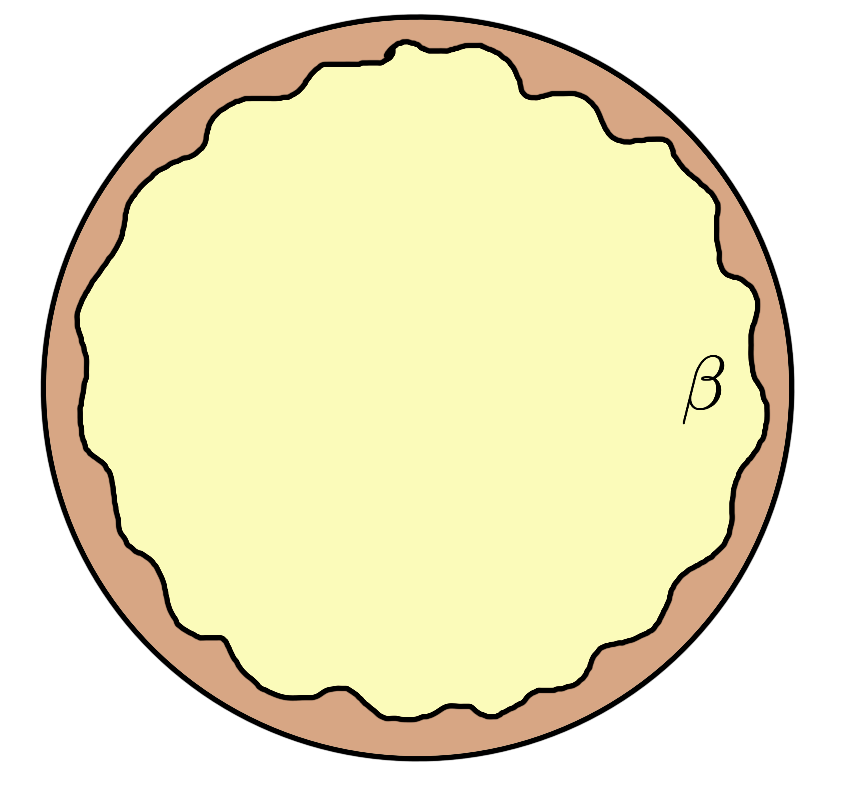}\\
&=\int\,e^\ell d\ell\includegraphics[valign=c,width=0.2\textwidth]{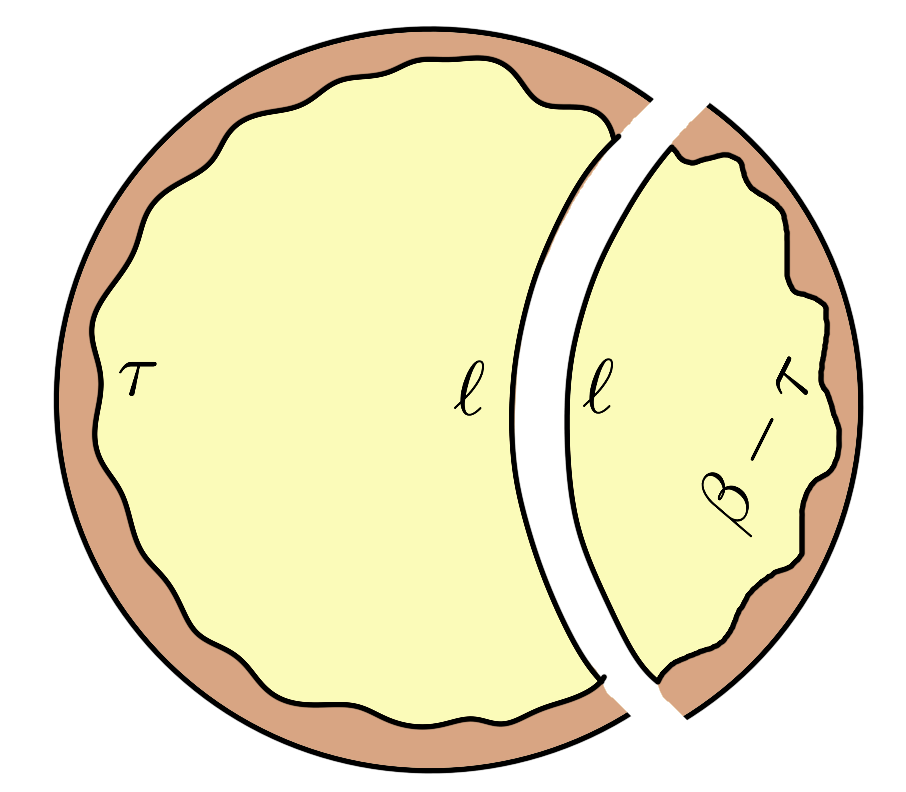}\\
&=e^{S_0}\int\,e^\ell d\ell\,\varphi_{\text{Disk},\tau}(\ell)\varphi_{\text{Disk},\beta-\tau}(\ell)
\end{align}

This decomposition may seem redundant since we already know how to calculate $Z_{\mathrm{disk}}$ but this procedure teaches us how to calculate two-point correlation functions. To do that, we just need another factor of $e^{-\Delta\ell}$ in the integral, which is the QFT two-point correlation function of two boundary operators $V$ of conformal weight $\Delta$ with renormalized geodesic distance $\ell$ apart. Disk contribution to two-point correlation functions at time $t=-i\tau$ is then given by
\begin{align}
\braket{V(t=-i\tau)V(0)}_{\chi=1}&=\includegraphics[valign=c,width=0.2\textwidth]{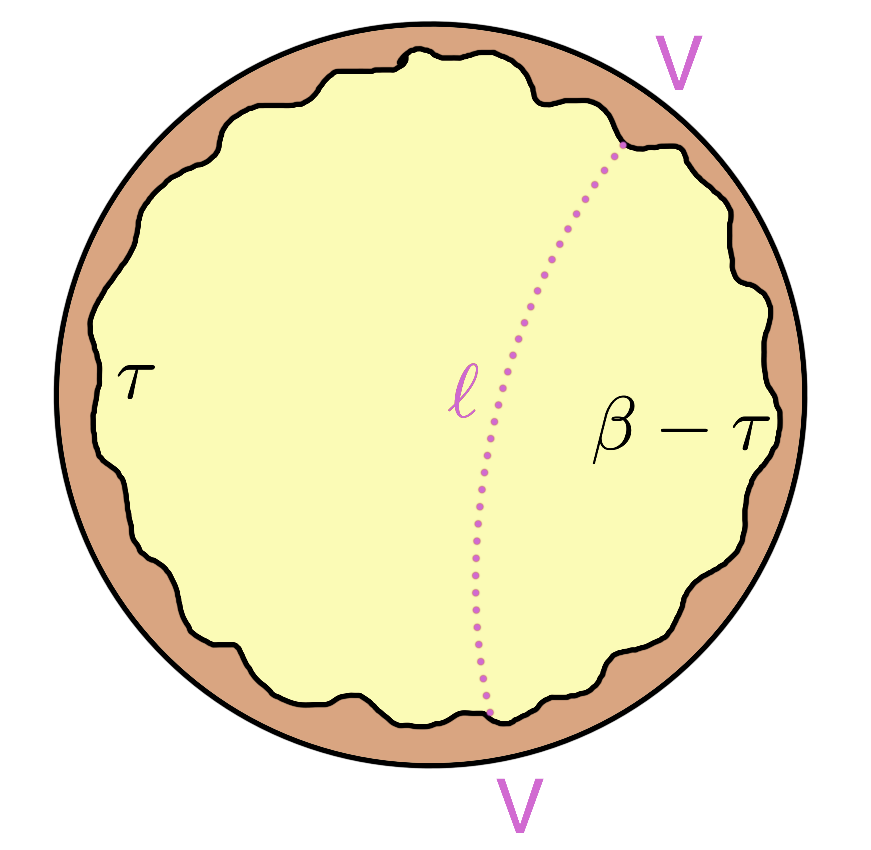}\\
&=\int\,e^\ell d\ell\includegraphics[valign=c,width=0.2\textwidth]{figures/splitcopy.png}e^{-\Delta\ell}\\
&=e^{S_0}\int\,e^\ell d\ell\,\varphi_{\text{Disk},\tau}(\ell)\varphi_{\text{Disk},\beta-\tau}(\ell)e^{-\Delta\ell}
\end{align}
Note that this two-point correlator is not normalized by dividing out the disk partition function which is of order $e^{S_0}$, in this paper the notation $\braket{VV}$ in JT is not normalized. Hartle-Hawing wavefunctions can be written in a simple closed form by first writing the wavefunctions with fixed energy boundary conditions given by
\begin{equation}
\varphi_E(\ell)=\braket{\ell|E}=4e^{-\ell/2}K_{i\sqrt{8E}}(4e^{-\ell/2})
\end{equation}
where $K$ is a Bessel-K function. Hartle-Hawking wavefunctions i.e. wavefunctions with fixed length boundary condition are given by \cite{Yangsingleauthor,Saadsingleauthor}
\begin{align}
\varphi_{\text{Disk},\tau}(\ell)&=\int_0^\infty dE\,\rho_0(E)e^{-\tau E}\varphi_E(\ell)\label{hhdisk}\\
\varphi_{\text{Trumpet},\tau}(\ell,b)&=\int_0^\infty dE\,\frac{\cos(b\sqrt{2E})}{\pi\sqrt{2E}}e^{-\tau E}\varphi_E(\ell)\label{hhtrumpet}
\end{align}

\begin{figure}[h]
\centering
\includegraphics[width=0.5\textwidth]{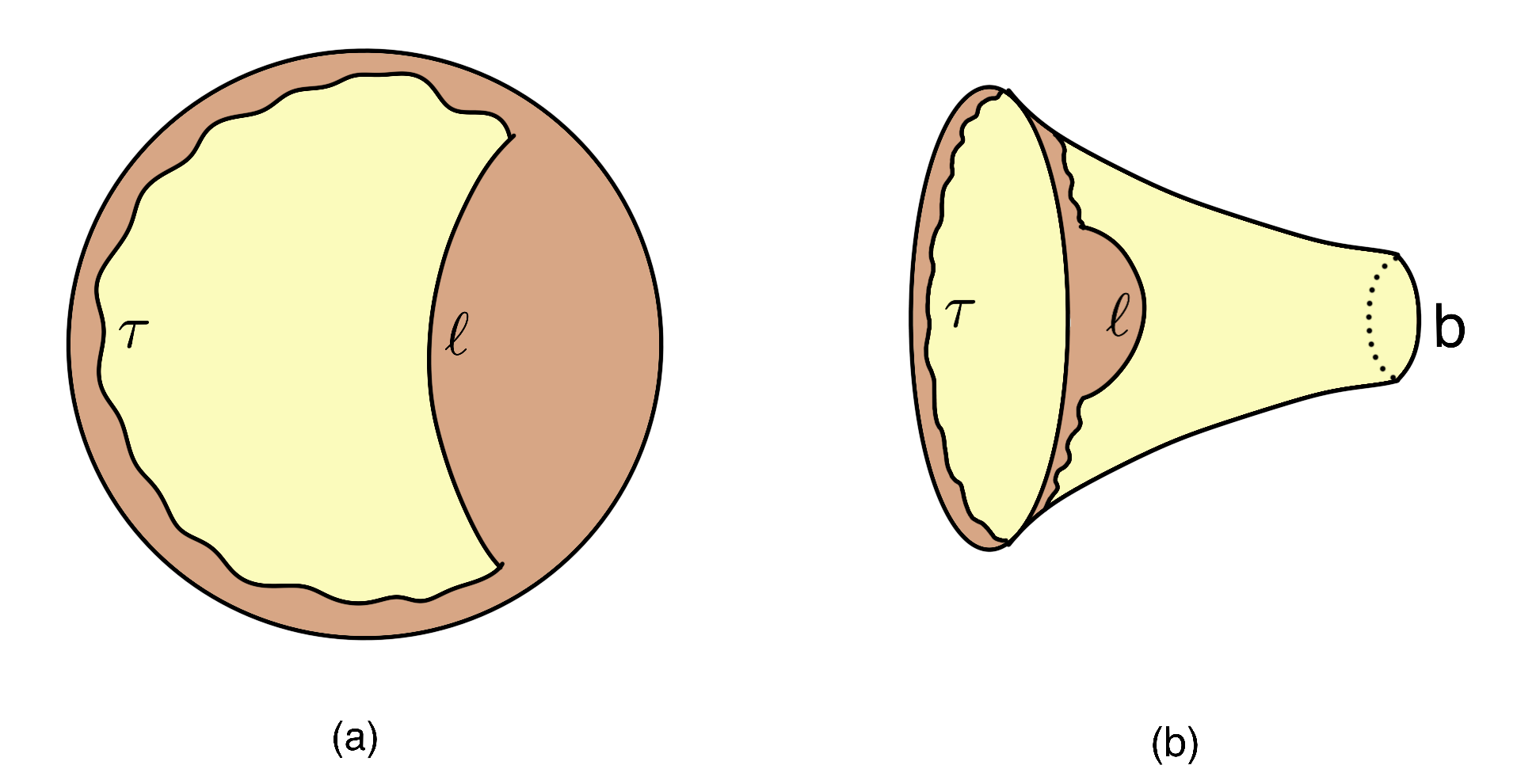}
\caption{(a) disk Hartle-Hawing state $\varphi_{\text{Disk},\tau}(\ell)$ (b) trumpet Hartle-Hawing state $\varphi_{\text{Trumpet},\tau}(\ell,b)$}
\end{figure}

Now we review two important relations that the Hartle-Hawking wavefunctions satisfy:
\begin{align}
\int_{-\infty}^\infty e^\ell d\ell\,\varphi_E(\ell)\varphi_{E'}(\ell)&=\frac{\delta(E-E')}{\rho_0(E)}\label{hhrelation1}\\
\int_{-\infty}^\infty e^\ell d\ell\,\varphi_E(\ell)\varphi_{E'}(\ell)e^{-\Delta\ell}&=|V_{E,E'}|^2=\frac{\left|\Gamma\left(\Delta+i(\sqrt{2E}+\sqrt{2E'})\right)\Gamma\left(\Delta+i(\sqrt{2E}-\sqrt{2E'})\right)\right|^2}{2^{2\Delta+1}\Gamma(2\Delta)}
\end{align}
In particular using (\ref{hhrelation1}) we can verify 
\begin{align}
Z_{\text{Disk}}(\beta)&=e^{S_0}\int e^\ell d\ell\,\varphi_{\text{Disk},\tau}(\ell)\varphi_{\text{Disk},\beta-\tau}(\ell)\\
Z_{\text{Trumpet}}(\beta,b)&=\int e^\ell d\ell\,\varphi_{\text{Disk},\tau}(\ell)\varphi_{\text{Trumpet},\beta-\tau}(\ell,b)
\end{align}
by plugging in (\ref{zdisk}, \ref{ztrumpet}, \ref{hhdisk}, \ref{hhtrumpet}). In addition to partition functions and Hartle-Hawking states, we review a final and important tool we use: propagators i.e. time evolution operators of Hartle-Hawking wavefunctions such that
\begin{align}
\varphi_{\text{Disk},\beta+\beta_1+\beta_2}(\ell)&=\int e^{\ell'}d\ell'\,P_{\text{Disk}}(\beta_1,\beta_2,\ell,\ell')\varphi_{\text{Disk},\beta}(\ell')\\
\varphi_{\text{Trumpet},\beta+\beta_1+\beta_2}(\ell)&=\int e^{\ell'}d\ell'\,P_{\text{Trumpet}}(\beta_1,\beta_2,b,\ell,\ell')\varphi_{\text{Disk},\beta}(\ell')
\end{align}
we can check \cite{Saadsingleauthor} that the above relations are solved by
\begin{align}
P_{\text{Disk}}(\beta_1,\beta_2,\ell,\ell')&=\int dE\,\rho_0(E)e^{-(\beta_1+\beta_2)E}\varphi_E(\ell)\varphi_E(\ell')\label{diskpropagator}\\
P_{\text{Trumpet}}(\beta_1,\beta_2,b,\ell,\ell')&=\int_0^\infty dE\frac{\cos(b\sqrt{2E})}{\pi\sqrt{2E}}e^{-(\beta_1+\beta_2)E}\varphi_E(\ell)\varphi_E(\ell')
\end{align}

\begin{figure}[h]
\centering
\includegraphics[width=0.5\textwidth]{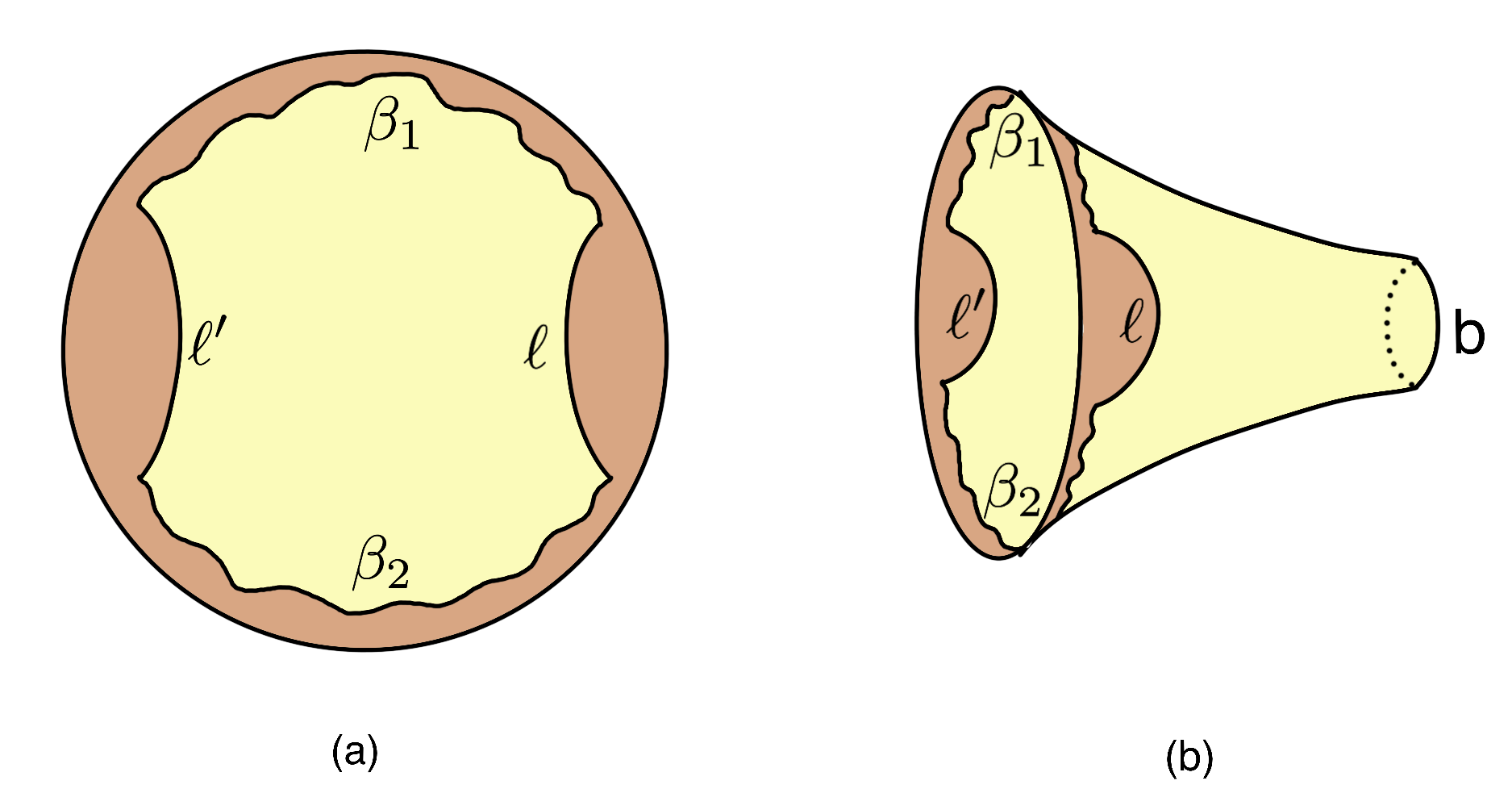}
\caption{(a) disk propagator (b) trumpet propagator}
\end{figure}

Recall that the disk two-point correlators is given by
\begin{align}
\braket{V(t=-i\tau)V(0)}_{\chi=1}&=e^{S_0}\int e^\ell d\ell\,\varphi_{\text{Disk},\tau}(\ell)\varphi_{\text{Disk},\beta-\tau}(\ell)e^{-\Delta\ell}\\
&=e^{S_0}\int dE_1dE_2\,\rho_0(E_1)\rho_0(E_2)e^{-\tau E_1}e^{-(\beta-\tau)E_2}|V_{E_1,E_2}|^2\\
&\sim \frac{e^{S_0}}{t^3}|V_{0,0}|^2 \quad\quad t\rightarrow\infty
\end{align}
This is dominated by $E_1,E_2$ close to zero as time goes to infinity, so we get a decay proportional to $t^{-3}$, showing that the disk does not contribute to the late-time behavior of two-point correlation functions. 

Saad \cite{Saadsingleauthor} showed that for a handle-disk geometry a procedure similar to the disk can be done by separating the geometry into two trumpet Hartle-Hawking wavefunctions. Then the contribution of a single example of handle-disk to two-point correlator is given by 
\begin{equation}
\braket{V(t=-i\tau)V(0)}_{\chi=-1}\supset e^{-S_0}\varphi_{\text{Trumpet},\tau}(\ell,b)\varphi_{\text{Trumpet},\beta-\tau}(\ell,b)e^{-\Delta\ell}\label{hdgeodesic}
\end{equation}
we need to integrate over all geodesics and also integrate over $b$ according to the Mapping Class Group. We will explain how to do that in section3. 

\begin{figure}[h]
\centering
\includegraphics[width=0.5\textwidth]{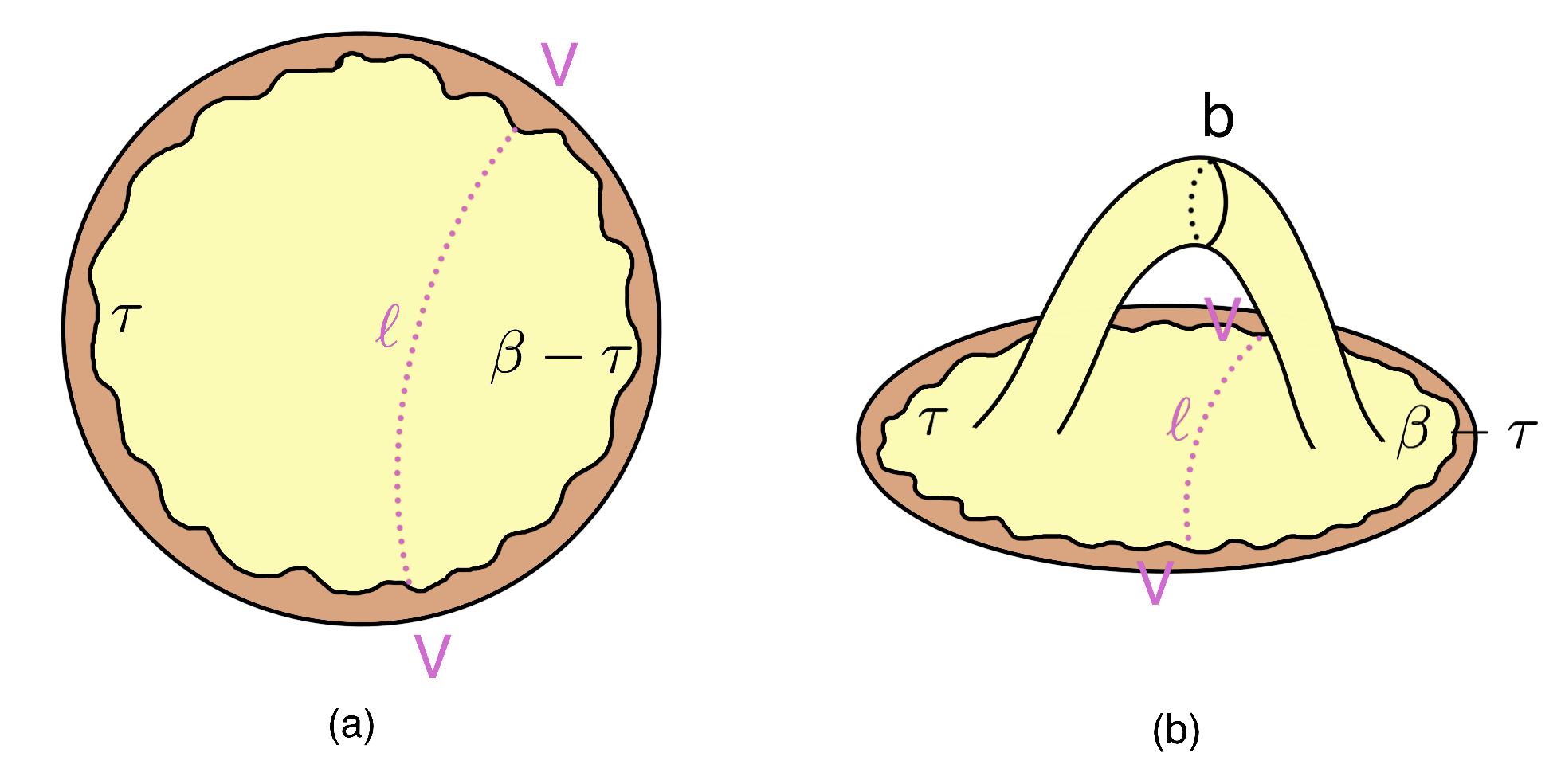}
\caption{(a) one example of genus-0 contribution to 2pt correlator (b) one example of genus-1 contribution to 2pt correlator }
\end{figure}

\subsection{Crosscap}

\begin{figure}[h]
\centering
\includegraphics[width=0.8\textwidth]{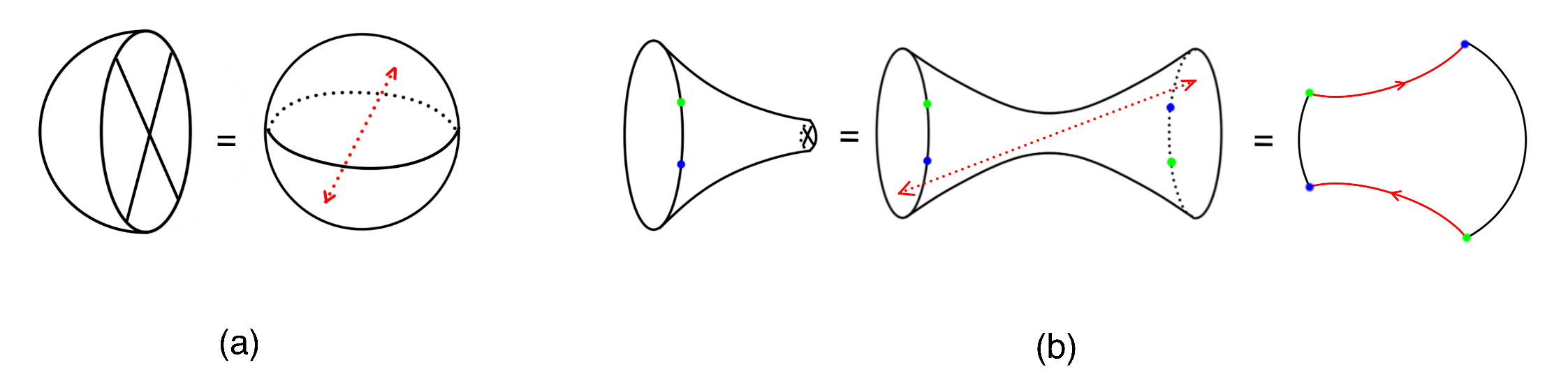}
\caption{(a) $\mathbb{RP}^2$ i.e. a sphere with a crosscap is a sphere with all pairs of antipodal points identified (b) a disk with a crosscap is a double-trumpet with all pairs of antipodal points identified which is topologically equivalent to a M\"{o}bius band}
\label{ccdrawing}
\end{figure}

If we allow non-orientable geometries, there is another contribution to the two-point correlation function given by a disk with a crosscap. A crosscap is a hole ($S^1$) with antipodal points identified. A useful fact is that each crosscap adds genus $1/2$ to the topology, so a disk with a crosscap has Euler characteristic $\chi=0$.  A disk with a crosscap is topologically equivalent to a M\"{o}bius band. To see this, we look at a more familiar topology which is $\mathbb{RP}^2$, i.e. a sphere with a crosscap. This is topologically equivalent to a sphere with all pairs of antipodal points identified. A disk with a crosscap is similarly topologically equivalent to a double-trumpet with all pairs of antipodal points identified, as shown in figure~\ref{ccdrawing}. A double-trumpet then contains two copies of disk+crosscap. If we take one of those copies by cutting the cylinder twice longitudinally and identifying antipodal points on the two cuts, we get a M\"{o}bius band. Geometrically this M\"{o}bius band can be embedded on a hyperbolic disk, so its shape reminds us of the disk propagator we defined in equation (\ref{diskpropagator}).

\begin{figure}[h]
\centering
\includegraphics[width=0.65\textwidth]{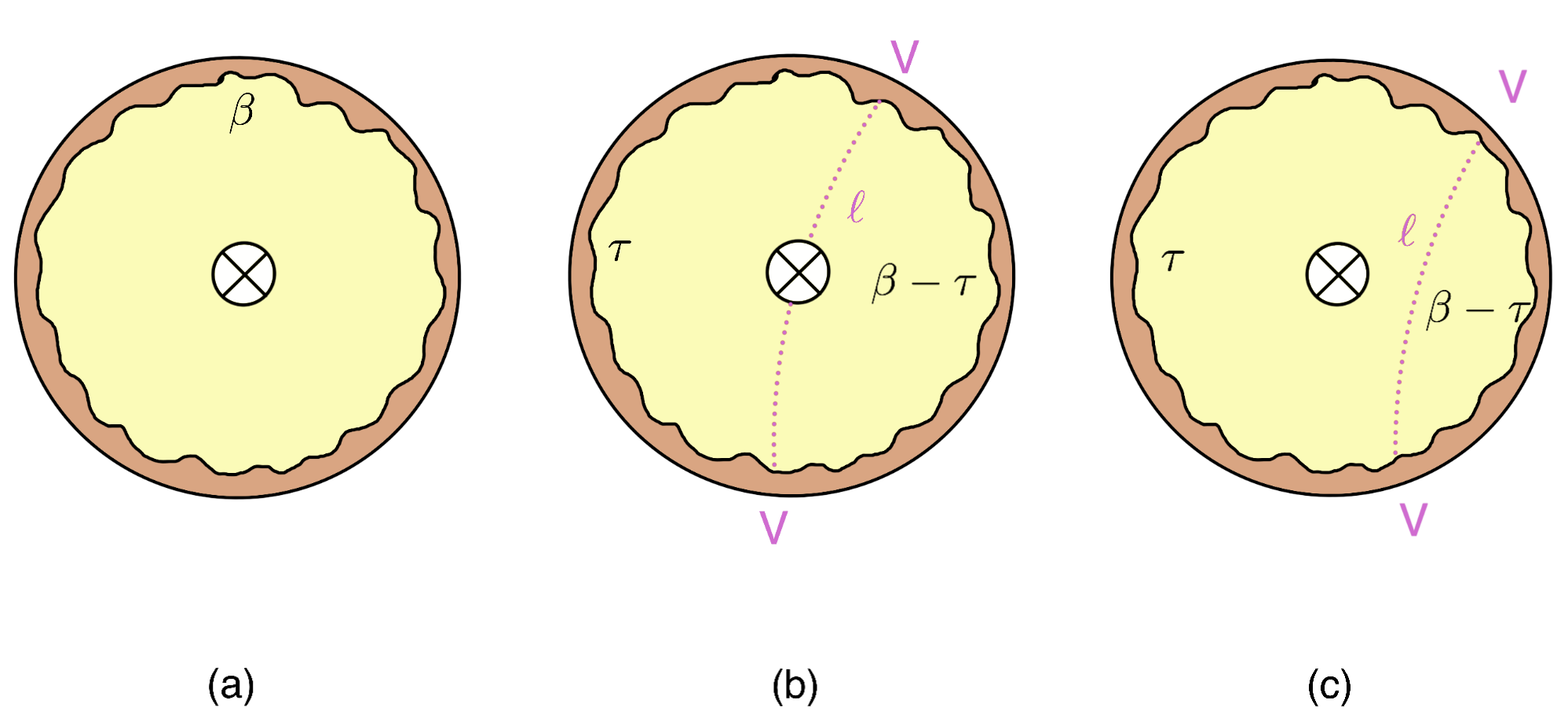}
\caption{(a) a hyperbolic disk with a crosscap (b) a geodesic going through the crosscap (c) a geodesic not going through the crosscap}
\label{diskcc}
\end{figure}

There can be two types of geodesics on a disk with a crosscap: a geodesic going through the crosscap, and a geodesic not going through the crosscap as shown in figure~\ref{diskcc}.

 In the following two subsections we consider two-point function contributions arising from these two types of geodeiscs separately.

\begin{figure}[h]
\centering
\includegraphics[width=0.25\textwidth]{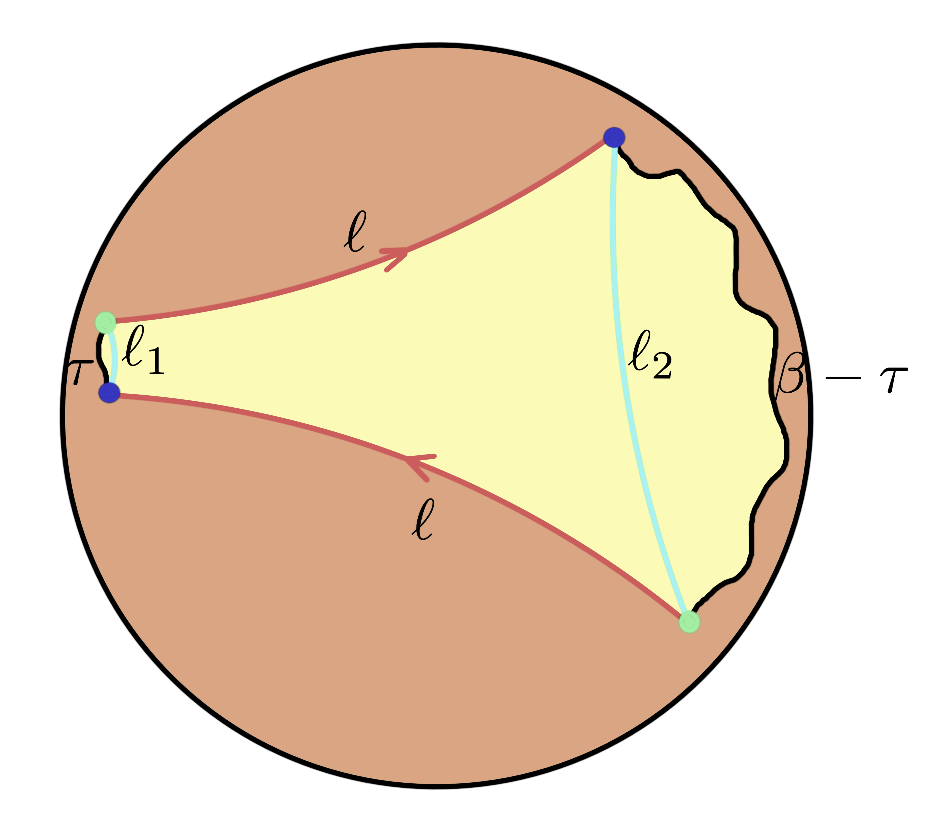}
\caption{On the M\"{o}bius band representation of disk+crosscap, if the two operators are located at the blue and the green dots respectively, a geodesic going through the crosscap is given by $\ell$ and two geodesics not going through the crosscap are given by $\ell_1$ and $\ell_2$.}
\label{ccgeodesics}
\end{figure}

\subsubsection{Non-decaying part}

In this subsection, we consider geodesics going through the crosscap. A disk+crosscap again can be represented as a M\"{o}bius strip. If we cut along the geodesic $\ell$ which goes through the crosscap, we get a disk propagator.

\begin{figure}[h]
\centering
\includegraphics[width=0.54\textwidth]{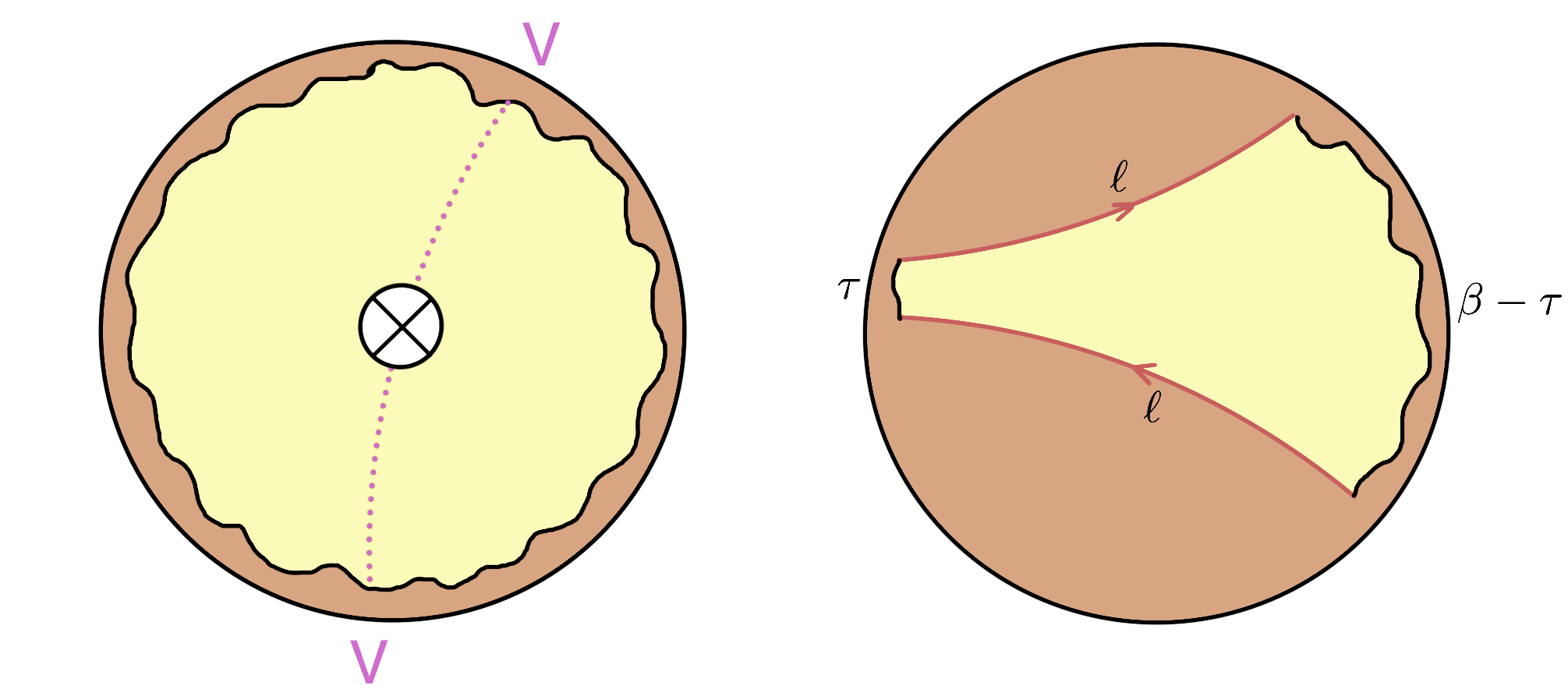}
\caption{A geodesic going through the crosscap on disk+crosscap representation and M\"{o}bius band representation respectively with the geodesic given by $\ell$.}
\label{ccgeodesics1}
\end{figure}

Integrating over the length of the geodesic weighted by the free propagator $e^{-\Delta\ell}$ we get a contribution to 2-point correlator. 

\begin{align}
\braket{V(t=-i\tau)V(0)}_{\text{cc},0}&= \int e^\ell d\ell\,P_{\text{Disk}}(\tau,\beta-\tau,\ell,\ell)e^{-\Delta\ell}\\
&=\int e^\ell d\ell\,\int dE\,\rho_0(E)e^{-\beta E}\varphi_E(\ell)^2e^{-\Delta\ell}\\
&=\int dE\,\rho_0(E)e^{-\beta E}|V_{E,E}|^2
\end{align}
This is one of the main results of this paper. Note that the result is manifestly independent of time $t$. Thus disk+crosscap gives a non-decaying contribution to two-point correlation function of order $e^{-S_0}$ after normalization (by dividing by the disk partition function).

\subsubsection{Decaying part}

\begin{figure}[H]
\centering
\includegraphics[width=0.65\textwidth]{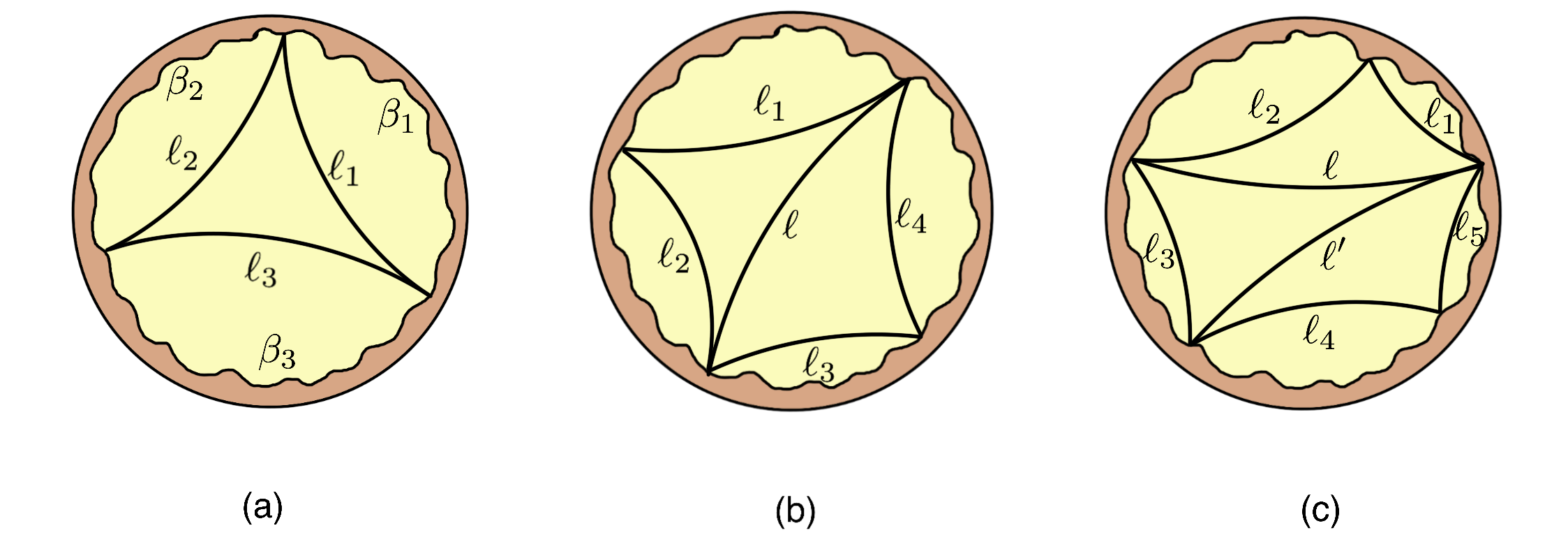}
\caption{(a) $\ell_1$, $\ell_2$, $\ell_3$ enclose a hyperbolic triangle $I(\ell_1,\ell_2,\ell_3)$ (b) $\ell_1$, $\ell_2$, $\ell_3$, $\ell_4$ enclose a hyperbolic quadrilateral $I(\ell_1,\ell_2,\ell_3,\ell_4)$ (c) $\ell_1$, $\ell_2$, $\ell_3$, $\ell_4$, $\ell_5$ enclose a hyperbolic pentagon $I(\ell_1,\ell_2,\ell_3,\ell_4,\ell_5)$}
\label{I3I4I5}
\end{figure}

In this subsection, we consider geodesics not going through the crosscap. Before calculating their contributions to two-point correlation functions, we introduce two quantities. The first is $I(\ell_1,\ell_2,\ell_3)$: the path integral of the hyperbolic triangle as shown in figure~\ref{I3I4I5}(a) (see \cite{Yangsingleauthor} for more information)
\begin{align}
I(\ell_1,\ell_2,\ell_3)&=e^{\ell_1/2}e^{\ell_2/2}e^{\ell_3/2}\int_0^\infty ds\,s\frac{\sinh(2\pi s)}{2\pi^2}\varphi_{s^2/2}(\ell_1)\varphi_{s^2/2}(\ell_2)\varphi_{s^2/2}(\ell_3)\\
&=e^{\ell_1/2}e^{\ell_2/2}e^{\ell_3/2}\int_0^\infty dE\rho_0(E)\varphi_E(\ell_1)\varphi_E(\ell_2)\varphi_E(\ell_3)
\end{align}
We can check that integrating over the product of a hyperbolic triangle and three Hartle-Hawking states we get back the disk partition function
\begin{multline}
e^{S_0}\int e^{\ell_1/2+\ell_2/2+\ell_3/2}d\ell_1d\ell_2d\ell_3\,\varphi_{\text{Disk},\beta_1}(\ell_1)\varphi_{\text{Disk},\beta_2}(\ell_2)\varphi_{\text{Disk},\beta_3}(\ell_3)I(\ell_1,\ell_2,\ell_3)\\
=e^{S_0}\int e^{\ell_1}d\ell\,\varphi_{\text{Disk},\beta_1}(\ell_1)\varphi_{\text{Disk},\beta_2+\beta_3}(\ell_1)=Z_{\text{Disk}}
\end{multline}

We can generalize the hyperbolic triangle to any hyperbolic polygons. To give a simplest example, gluing two hyperbolic triangles together gives a hyperbolic quadrilateral
\begin{align}
I(\ell_1,\ell_2,\ell_3,\ell_4)&=\int d\ell\,I(\ell_1,\ell_2,\ell)I(\ell,\ell_3,\ell_4)\\
&=\int d\ell dEdE'\,e^{\ell_1/2+\ell_2/2+\ell_3/2+\ell_4/2}e^{\ell}\rho_0(E)\rho_0(E')\varphi_E(\ell_1)\varphi_E(\ell_2)\varphi_E(\ell)\varphi_{E'}(\ell)\varphi_{E'}(\ell_3)\varphi_{E'}(\ell_4)\\
&=\int dE\,e^{\ell_1/2+\ell_2/2+\ell_3/2+\ell_4/2}\rho_0(E)\varphi_E(\ell_1)\varphi_E(\ell_2)\varphi_E(\ell_3)\varphi_E(\ell_4)
\end{align}
In general
\begin{equation}
I(\ell_1,\ldots,\ell_n)=\int dE\,\rho_0(E)\prod e^{\ell_i/2}\varphi_E(\ell_i)
\end{equation}

A second important quantity we introduce is the crosscap correction to the disk partition function $Z_{cc}$. And we can define the correction $\rho_{1/2}(E)$ to density of states $\rho_0(E)$ accordingly
\begin{equation}
Z_{\mathrm{disk}+\mathrm{cc}}(\beta)=Z_{\mathrm{disk}}(\beta)+Z_{\mathrm{cc}}(\beta)=\int dE\, \left(e^{S_0}\rho_0(E)+\rho_{1/2}(E)\right)e^{-\beta E}\label{rho1}
\end{equation}
Now let us calculate the crosscap partition function using the $I$ function we just defined. Analyzing figure~\ref{ccgeodesics} we get
\begin{align}
Z_{\mathrm{cc}}&=\int d\ell d\ell_1d\ell_2\,e^{\ell_1/2+\ell_2/2}I(\ell,\ell_1,\ell,\ell_2)\varphi_{\text{Disk},\tau}(\ell_1)\varphi_{\text{Disk},\beta-\tau}(\ell_2)\\
&=\int dE\,\rho_0(E)\int e^\ell d\ell\,\varphi_E(\ell)\varphi_E(\ell)e^{-\beta E}
\end{align}
Together with (\ref{rho1}), this gives a relation analogous to (\ref{hhrelation1}) but with $E=E'$
\begin{equation}
\int_{-\infty}^\infty e^\ell d\ell\,\varphi_E(\ell)\varphi_E(\ell)=\frac{\rho_{1/2}(E)}{\rho_0(E)}
\end{equation}
This is a divergent integral, so (\ref{hhrelation1}) only makes sense when $E\neq E'$. To deal with the problem of divergence in the crosscap partition function, we compute it another way by integrating over trumpet partition functions with geodesic boundary replaced by a crosscap with different perimeter lengths.

\begin{figure}[h]
\centering
\includegraphics[width=0.2\textwidth]{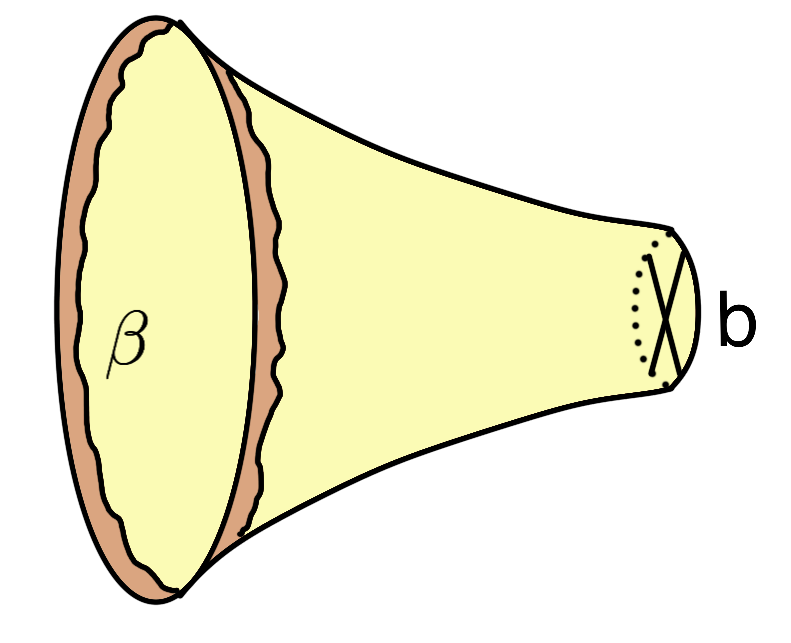}
\caption{a trumpet with its geodesic boundary replaced by a crosscap}
\end{figure}

The geodesic boundary length is $b$. In Appendix~\ref{appendixmeasure}, we show that the measure to integrate over $b$ for crosscaps is $db/2\tanh\frac{b}{4}$. Using this we get
\begin{equation}
Z_{\text{cc}}=\int \frac{db}{2\tanh\frac{b}{4}}Z_{\mathrm{Trumpet}}(\beta,b)=\int_0^\infty dE\,\int_0^\infty \frac{db}{2\tanh\frac{b}{4}}\frac{\cos(b\sqrt{2E})}{\pi\sqrt{2E}}e^{-\beta E}
\end{equation}
We then get an explicit expression of $\rho_{1/2}(E)$ i.e.
\begin{equation}
\rho_{1/2}(E)=\int_0^\infty \frac{db}{2\tanh\frac{b}{4}}\frac{\cos(b\sqrt{2E})}{\pi\sqrt{2E}}
\end{equation}
Note that the lower limit of this integral $b\rightarrow0$ gives a divergence. That is when the crosscap end of the trumpet becomes very long and thin. The divergence could be regulated by studying a different theory of gravity for example $(2,p)$ Liouville gravity \cite{Seiberg:2004at} (see Appendix F of \cite{StanfordWitten19}). But the way it is regulated won't be important because after regularization it will still be a small correction to the disk partition function since it is $e^{S_0}$ smaller. 

We now show that after regulation the contribution of geodesics not going through the crosscap to the two-point correlation function is decaying with time. The contribution of geodesic $\ell_1$ is 
\begin{align}
\braket{V(t=-i\tau)V(0)}_{\text{cc},1}&=\int dL d\ell_1d\ell_2\,e^{\ell_1/2+\ell_2/2}e^{-\Delta\ell_1}I(L,\ell_1,L,\ell_2)\varphi_{\text{Disk},\tau}(\ell_1)\varphi_{\text{Disk},\beta-\tau}(\ell_2)\\
&=\int dEdE_1\rho_0(E_1)e^{-\tau E_1}e^{-(\beta-\tau)E}\rho_{1/2}(E)|V_{E,E_1}|^2
\end{align}
and similarly the contribution of geodesic $\ell_2$ is 
\begin{align}
\braket{V(t=-i\tau)V(0)}_{\text{cc},2}&=\int dL d\ell_1d\ell_2\,e^{\ell_1/2+\ell_2/2}e^{-\Delta\ell_2}I(L,\ell_1,L,\ell_2)\varphi_{\text{Disk},\tau}(\ell_1)\varphi_{\text{Disk},\beta-\tau}(\ell_2)\\
&=\int dEdE_2\rho_0(E_2)e^{-\tau E}e^{-(\beta-\tau)E_2}\rho_{1/2}(E)|V_{E,E_2}|^2
\end{align}

When $t$ is large, $\braket{V(t=-i\tau)V(0)}_{\text{cc},1}$ is dominated by $E,E_1$ close to zero. In that case $|V_{E,E_1}|^2$ approaches a constant and $\rho_0(E)\approx\sqrt{E}$ so we can do the integral approximately
\begin{equation}
\braket{V(t=-i\tau)V(0)}_{\text{cc},1}\approx\int dEdE_1\sqrt{E_1}e^{-\tau E_1}e^{-(\beta-\tau)E}\rho_{1/2}(E)|V_{0,0}|^2\sim\frac{1}{t^{5/2}}\rho_{1/2}(0)|V_{0,0}|^2
\end{equation}
and similar for $\braket{V(t=-i\tau)V(0)}_{\text{cc},2}$. They decay with time as $t^{-5/2}$, so these geodesics not going through the crosscap do not contribute to the late-time two-point correlation function. 

Therefore, geodesics going through the crosscap gives a non-decaying contribution to the two-point correlation function $\braket{V(t=-i\tau)V(0)}_{\text{cc},0}$. On the disk+crosscap, there are also other geodesics that wind around the crosscap multiple times so that they self-intersect. At large time, they do not contribute. The classical solution to the disk+crosscap configuration with two operator insertions is given by the quotient of the configuration calculated in Appendix B of \cite{Stanfordwormhole} by identifying antipodal points. At late time the classical solution of $b$ scales with time $t$, so the main contribution to the quantum wavefunction calculation comes from large $b$. This is saying that the geodesic tends to be long and its contribution decays over time.

But before we conclude, we should note that there are two versions of sums over geometries with or without a weighting factor $(-1)^{n_c}$ where $n_c$ is the number of crosscaps \cite{StanfordWitten19}. Thus in the case of disk+one crosscap, this factor is simply $-1$, which gives the two-point correlator $\pm\braket{V(t=-i\tau)V(0)}_{\text{cc},0}$. We will see that these correspond to GOE and GSE-like matrix integrals on the boundary.

\subsection{RMT}
\label{bosonicRMT}
\cite{SaadShenkerStanford19} showed that there is a correspondence between JT gravity path integrals and Hermitian matrix integrals
\begin{equation}
Z_{\mathrm{JT}}(\beta_1,\ldots,\beta_n)\leftrightarrow \frac{1}{\mathcal{Z}}\int dH\,e^{-L\mathrm{Tr} V(H)}\mathrm{Tr}e^{-\beta_1 H}\cdots\mathrm{Tr}e^{-\beta_n H}
\end{equation}
where $H$ are Hermitian $L\times L$ matrices \footnote{More precisely, these matrices should be double-scaled.}, $V(H)$ is a potential function and $\mathcal{Z}=\int dH\,e^{-L\mathrm{Tr} V(H)}$ is the matrix integral partition function. The left-hand side is the JT gravity partition function with $n$ asymptotic boundaries of regularized lengths $\beta_1,\ldots,\beta_n$, and the right-hand side is an average of products of thermal partition functions over an appropriate random matrix ensemble. 

There are three Dyson $\beta$-ensembles of random matrices: GUE, GOE, GSE \cite{Dyson}. GUE is an ensemble of random Hermitian matrices and the ensemble is invariant under unitary transformations, i.e. $U(L)$ satisfying $U^\dagger U=1$. GOE and GSE can be derived by adding time-reversal symmetry (i.e. the time-reversal operator $T$ commuting with $U$) to GUE with different anomaly conditions $T^2=\pm1$. Recall that Lorentzian time-reversal $T$ is antilinear and antiunitary. Since we know that complex conjugate $K$ is antilinear and antiunitary \footnote{Complex conjugate operator is an antilinear and antiunitary operator $K:\mathcal{H}\rightarrow\mathcal{H}$ because $K\sum\alpha_i\ket{i}=\sum\alpha_i^*\ket{i}$ and if we write $\ket{\psi}=\sum\alpha_i\ket{i}, \quad \ket{\chi}=\sum\beta_i\ket{i}$ then we have the inner product $\braket{K\chi| K\psi}=\sum_{i,j}(\bra{j}\beta_j)(\alpha^*_i\ket{i})=\sum_i\alpha_i^*\beta_i\braket{i|i}=\sum_{i,j}(\bra{j}\alpha^*_j)(\beta_i\ket{i})=\braket{\psi|\chi}$.}, it is natural to model $T$ with a factor of $K$ in it. For $T^2=1$, we can just take $T=K$ and the condition $TUT^{-1}=U$ reduces $U^\dagger U=1$ to $U^TU=1$, which is equivalent to saying $U\in O(L)$, and the ensemble becomes GOE. For $T^2=-1$, we instead take $T=K\omega$ where $\omega=\begin{pmatrix}0&1\\-1&0\end{pmatrix}$. The condition $TUT^{-1}=U$ reduces $U^\dagger U=1$ to $U^T\omega U=\omega$, which is equivalent to saying $U\in Sp(L)$, and the ensemble becomes GSE. Now we examine the three ensembles one by one.

\subsubsection*{GUE}
Now that we have considered two-point correlation functions in JT, we want to calculate $\braket{V(t)V(0)}=\braket{U(t)^\dagger V(0) U(t)V(0)}=\frac{1}{L}\mathrm{tr}[U(t)^\dagger V(0) U(t)V(0)]$ in RMT. A simple model \cite{Blommaert, Weingartens} of the time evolution operator $U(t)$ is to approximate it by
\begin{equation}
U(t)= e^{-i(u^\dagger hu)t}=u^\dagger e^{-iht} u
\end{equation}
where $h$ is diagonal and $u$ is Haar random. Note that this model only works at late times when the time evolution is random enough. Time evolution of an operator $W$ is then given by
\begin{equation}
W(t)=U^\dagger W(0) U=u^\dagger e^{iht} u W(0) u^\dagger e^{-iht} u
\end{equation}
On the RMT side, the two-point correlation function can be calculated via an integral over $u$
\begin{equation}
\int du\,\braket{V(t)V(0)}=\int du\,\braket{e^{iHt}V(0)e^{-iHt}V(0)}=\int du\,\braket{u^\dagger e^{iht}uV(0)u^\dagger e^{-iht}uV(0)}
\end{equation}
This can be calculated using the Weingarten formula for unitary matrices $u$ 
\begin{multline}
\int du\,u_{i_1}^{\phantom{i_1}j_1}u_{i_2}^{\phantom{i_2}j_2}(u^\dagger)_{k_1}^{\phantom{k_1}l_1}(u^\dagger)_{k_2}^{\phantom{k_2}l_2}=\frac{1}{L^2-1}\Big(\delta_{i_1,l_1}\delta_{i_2,l_2}\delta_{k_1,j_1}\delta_{k_2,j_2}+\delta_{i_1,l_2}\delta_{i_2,l_1}\delta_{k_1,j_2}\delta_{k_2,j_1}\\
-\frac{1}{L}\delta_{i_1,l_1}\delta_{i_2,l_2}\delta_{k_1,j_2}\delta_{k_2,j_1}-\frac{1}{L}\delta_{i_1,l_2}\delta_{i_2,l_1}\delta_{k_1,j_1}\delta_{k_2,j_2}\Big)\label{weingartenu}
\end{multline}
To leading order
\begin{equation}
\int du\,u_{i_1}^{\phantom{i_1}j_1}u_{i_2}^{\phantom{i_2}j_2}(u^\dagger)_{k_1}^{\phantom{k_1}l_1}(u^\dagger)_{k_2}^{\phantom{k_2}l_2}\approx\frac{1}{L^2}\big(\delta_{i_1,l_1}\delta_{i_2,l_2}\delta_{k_1,j_1}\delta_{k_2,j_2}+\delta_{i_1,l_2}\delta_{i_2,l_1}\delta_{k_1,j_2}\delta_{k_2,j_1}\big)
\end{equation}
then two-point correlator is given by
\begin{multline}
\int du\,\braket{V(t)V(0)}=\frac{1}{L^2-1}\Big(L^2\braket{e^{iht}}\braket{e^{-iht}}\braket{VV}+L^2\braket{V}\braket{V}\Big)\\
-\frac{1}{L(L^2-1)}\Big(L\braket{VV}+L^3\braket{e^{iht}}\braket{e^{-iht}}\braket{V}\braket{V}\Big)
\end{multline}
Then in the limit $L\rightarrow\infty$ and assuming $\braket{V}=0$
\begin{equation}
\int du\,\braket{V(t)V(0)}\approx \braket{e^{iht}}\braket{e^{-iht}}\braket{VV}\label{VVu}
\end{equation}
We know from RMT computation \cite{9authors} that
\begin{equation}
\braket{e^{iht}}\braket{e^{-iht}}\sim\min\{\frac{t}{2\pi L^2},\frac{1}{\pi L}\}\sim\begin{cases}
t/(2\pi L^2)& t<2L\\1/(\pi L)& t\geq 2L\end{cases}\label{ramprelation}
\end{equation}
Thus (\ref{VVu}) becomes approximately 
\begin{equation}
\min\{\frac{t}{L^2},\frac{1}{L}\}\braket{VV}
\end{equation}
It exhibits a ramp connected to a plateau in time confirming the analysis in \cite{Saadsingleauthor}. Also note that \cite{Saadsingleauthor} showed that the ramp part is originated from contribution from handle-disk in JT gravity with two-point correlation function given by
\begin{equation}
\braket{VV}_{\mathrm{handle-disk}}=e^{-S_0}\int dEdE'\,\rho_2(E,E')e^{-\tau E}e^{-(\beta-\tau)E'}|V_{E,E'}|^2
\end{equation}
where
\begin{equation}
\rho_2(E,E')=\int_0^\infty bdb\,\frac{\cos(b\sqrt{2E})\cos(b\sqrt{2E'})}{\pi^2\sqrt{2E}\sqrt{2E'}}
\end{equation}
In particular, if we identify $\rho_0(E)|V_{E,E}|^2=\braket{E|VV|E}$, we can compute the normalized contribution to the two-point correlation function from handle-disk 
\begin{equation}
\frac{\braket{VV}_{\mathrm{handle-disk}}}{\braket{1}_{\mathrm{disk}}}=\frac{\braket{VV}_{\mathrm{handle-disk}}}{\int dE\,e^{S_0}\rho_0(E)e^{-\beta E}}\sim \frac{t}{(\rho_0(E)e^{S_0})^2}\braket{E|VV|E}\sim \frac{t}{L^2}\braket{E|VV|E}
\end{equation}
where we have identified $L$ with $\rho_0(E)e^{S_0}$. This confirms the claim that the handle-disk gives the ramp \cite{Saadsingleauthor}.

\subsubsection*{GOE}
In our case, the geometries of JT are non-orientable. For the case where we do not include the weighting factor $(-1)^{n_c}$, on the RMT side we replace the complex unitary matrix $u$ by a real orthogonal matrix $o$ so that there is time-reversal symmetry. The two-point correlation function is then given by
\begin{equation}
\int do\,\braket{V(t)V(0)}=\int do\,\braket{e^{iHt}V(0)e^{-iHt}V(0)}=\int do\,\braket{o^\dagger e^{iht}oV(0)o^\dagger e^{-iht}oV(0)}
\end{equation}
We can derive a formula analogous to Weingarten formula for orthogonal matrices
\begin{multline}
\int do\,o_{i_1}^{\phantom{i_1}j_1}o_{i_2}^{\phantom{i_2}j_2}(o^\dagger)_{k_1}^{\phantom{k_1}l_1}(o^\dagger)_{k_2}^{\phantom{k_2}l_2}\\
=\frac{(L+1)}{L(L-1)(L+2)}\Big(\delta_{i_1,l_1}\delta_{i_2,l_2}\delta_{k_1,j_1}\delta_{k_2,j_2}+\delta_{i_1,l_2}\delta_{i_2,l_1}\delta_{k_1,j_2}\delta_{k_2,j_1}+\delta_{i_1,i_2}\delta_{l_1,l_2}\delta_{k_1,k_2}\delta_{j_1,j_2}\\
-\frac{1}{L+1}\delta_{i_1,l_1}\delta_{i_2,l_2}\delta_{k_1,j_2}\delta_{k_2,j_1}-\frac{1}{L+1}\delta_{i_1,l_1}\delta_{i_2,l_2}\delta_{k_1,k_2}\delta_{j_1,j_2}-\frac{1}{L+1}\delta_{i_1,l_2}\delta_{i_2,l_1}\delta_{k_1,j_1}\delta_{k_2,j_2}\\
-\frac{1}{L+1}\delta_{i_1,l_2}\delta_{i_2,l_1}\delta_{k_1,k_2}\delta_{j_1,j_2}-\frac{1}{L+1}\delta_{i_1,i_2}\delta_{l_1,l_2}\delta_{k_1,j_1}\delta_{k_2,j_2}-\frac{1}{L+1}\delta_{i_1,i_2}\delta_{l_1,l_2}\delta_{k_1,j_2}\delta_{k_2,j_1}\Big)\label{weingarteno}
\end{multline}
To leading order
\begin{equation}
\int do\,o_{i_1}^{\phantom{i_1}j_1}o_{i_2}^{\phantom{i_2}j_2}(o^\dagger)_{k_1}^{\phantom{k_1}l_1}(o^\dagger)_{k_2}^{\phantom{k_2}l_2}\approx\frac{1}{L^2}\big(\delta_{i_1,l_1}\delta_{i_2,l_2}\delta_{k_1,j_1}\delta_{k_2,j_2}+\delta_{i_1,l_2}\delta_{i_2,l_1}\delta_{k_1,j_2}\delta_{k_2,j_1}+\delta_{i_1,i_2}\delta_{l_1,l_2}\delta_{k_1,k_2}\delta_{j_1,j_2}\big)
\end{equation}
then using the above formula the two-point correlator is given by
\begin{multline}
\int do\,\braket{V(t)V(0)}=\frac{(L+1)}{L(L-1)(L+2)}\Big(L^2\braket{e^{iht}}\braket{e^{-iht}}\braket{VV}+L^2\braket{V}\braket{V}+L\braket{VV^T}\Big)\\
-\frac{1}{L(L-1)(L+2)}\Big(L\braket{VV}+L\braket{VV}+L\braket{VV^T}\\
+L^2\braket{e^{iht}}\braket{e^{-iht}}\braket{VV^T}+L^2\braket{V}\braket{V}+L^3\braket{e^{iht}}\braket{e^{-iht}}\braket{V}\braket{V}\Big)
\end{multline}
Then in the limit $L\rightarrow\infty$ and assuming $\braket{V}=0$
\begin{equation}
\int do\,\braket{V(t)V(0)}\approx\braket{e^{iht}}\braket{e^{-iht}}\braket{VV}+\frac{1}{L}\braket{VV^T}\label{VVo}
\end{equation}
The first term is the same as (\ref{VVu}) and again gives a ramp at early time and a plateau at late time and in early time corresponds to a handle-disk in JT. Since we are dealing with non-orientable geometries, we write $\rho_0(E)|V_{E,E}|^2=\braket{E|VV|E}$ if we identify the two sides of the geodesic without twists and write $\rho_0(E)|V_{E,E}|^2=\braket{E|VV^T|E}$ if we identify the two sides of the geodesic with a one-half twist which translates to a reflection across the horizontal axis or Euclidean time-reversal. Thus the normalized contribution to the two-point correlation function from crosscap is given by
\begin{equation}
\frac{\braket{VV}_{\mathrm{cc}}}{\braket{1}_{\mathrm{disk}}}=\frac{\int dE\,e^{-\beta E}\braket{E|VV^T|E}}{\int dE\,e^{S_0}\rho_0(E)e^{-\beta E}}\sim \frac{1}{\rho_0(E)e^{S_0}}\braket{E|VV^T|E}\sim \frac{1}{L}\braket{E|VV^T|E}
\end{equation}
which identifes with the third term of equation~(\refeq{VVo}). This doubles the original plateau.

\subsubsection*{GSE}
For the case where we do include the weighting factor $(-1)^{n_c}$, on the RMT side we replace the orthogonal matrix $o$ by a symplectic matrix $s$. The two-point correlation function is then given by
\begin{equation}
\int ds\,\braket{V(t)V(0)}=\int ds\,\braket{e^{iHt}V(0)e^{-iHt}V(0)}=\int do\,\braket{s^\dagger e^{iht}sV(0)s^\dagger e^{-iht}sV(0)}
\end{equation}
We can derive a formula for symplectic matrices $s$ analogous to (\ref{weingarteno})
\begin{multline}
\int ds\,s_{i_1}^{\phantom{i_1}j_1}s_{i_2}^{\phantom{i_2}j_2}(s^\dagger)_{k_1}^{\phantom{k_1}l_1}(s^\dagger)_{k_2}^{\phantom{k_2}l_2}\\
=\frac{(L+1)}{L(L-1)(L+2)}\Big(\delta_{i_1,l_1}\delta_{i_2,l_2}\delta_{k_1,j_1}\delta_{k_2,j_2}+\delta_{i_1,l_2}\delta_{i_2,l_1}\delta_{k_1,j_2}\delta_{k_2,j_1}+\omega_{i_1,i_2}\omega_{l_1,l_2}\omega_{k_1,k_2}\omega_{j_1,j_2}\\
-\frac{1}{L+1}\delta_{i_1,l_1}\delta_{i_2,l_2}\delta_{k_1,j_2}\delta_{k_2,j_1}+\frac{1}{L+1}\delta_{i_1,l_1}\delta_{i_2,l_2}\omega_{k_1,k_2}\omega_{j_1,j_2}-\frac{1}{L+1}\delta_{i_1,l_2}\delta_{i_2,l_1}\delta_{k_1,j_1}\delta_{k_2,j_2}\\
+\frac{1}{L+1}\delta_{i_1,l_2}\delta_{i_2,l_1}\omega_{k_1,k_2}\omega_{j_1,j_2}+\frac{1}{L+1}\omega_{i_1,i_2}\omega_{l_1,l_2}\delta_{k_1,j_1}\delta_{k_2,j_2}+\frac{1}{L+1}\omega_{i_1,i_2}\omega_{l_1,l_2}\delta_{k_1,j_2}\delta_{k_2,j_1}\Big)\label{weingartens}
\end{multline}
to leading order
\begin{equation}
\int ds\,s_{i_1}^{\phantom{i_1}j_1}s_{i_2}^{\phantom{i_2}j_2}(s^\dagger)_{k_1}^{\phantom{k_1}l_1}(s^\dagger)_{k_2}^{\phantom{k_2}l_2}
\approx\frac{1}{L^2}\big(\delta_{i_1,l_1}\delta_{i_2,l_2}\delta_{k_1,j_1}\delta_{k_2,j_2}+\delta_{i_1,l_2}\delta_{i_2,l_1}\delta_{k_1,j_2}\delta_{k_2,j_1}+\omega_{i_1,i_2}\omega_{l_1,l_2}\omega_{k_1,k_2}\omega_{j_1,j_2}\big)
\end{equation}
where $\omega=\begin{pmatrix}0&1\\-1&0\end{pmatrix}$ and a symplectic matrix $s$ satisfies the relations $s\omega s^T=\omega$ and $ss^\dagger=1$. The two-point correlation function is then given by
\begin{multline}
\int ds\,\braket{V(t)V(0)}=\frac{(L+1)}{L(L-1)(L+2)}\Big(L^2\braket{e^{iht}}\braket{e^{-iht}}\braket{VV}+L^2\braket{V}\braket{V}+L\braket{\omega ^Te^{iht}\omega e^{-iht}}\braket{V \omega V^T \omega}\Big)\\
-\frac{1}{L(L-1)(L+2)}\Big(L\braket{VV}-L\braket{\omega ^Te^{iht}\omega e^{-iht}}\braket{VV}-L\braket{V \omega V^T \omega}\\
-L^2\braket{e^{iht}}\braket{e^{-iht}}\braket{V \omega V^T \omega}-L^2\braket{\omega ^Te^{iht}\omega e^{-iht}}\braket{V}\braket{V}+L^3\braket{e^{iht}}\braket{e^{-iht}}\braket{V}\braket{V}\Big)
\end{multline}
Then in the limit $L\rightarrow\infty$ and assuming $\braket{V}=0$
\begin{equation}
\int ds\,\braket{V(t)V(0)}\approx\braket{e^{iht}}\braket{e^{-iht}}\braket{VV}-\frac{1}{L}\braket{V \omega V^T \omega^{-1}}\label{VVs}
\end{equation}
where we have used the relation $\omega^T e^{iht}\omega=e^{iht}$ since matrix $h$ has no symplectic structure.

Again, the first term is the same as (\ref{VVu}) and in early time corresponds to a handle-disk in JT. In a topological field theory with weighting factor $(-1)^{n_c}$, if we identify the two sides of the geodesic with a one-half twist we should write $\rho_0(E)|V_{E,E}|^2=\braket{E|V \omega V^T \omega^{-1}|E}$. Thus the normalized contribution to the two-point correlation function from crosscap is given by
\begin{equation}
\frac{\braket{VV}_{\mathrm{cc}}}{\braket{1}_{\mathrm{disk}}}=\frac{-\int dE\,e^{-\beta E}\braket{E|V\omega V^T\omega^{-1}|E}}{\int dE\,e^{S_0}\rho_0(E)e^{-\beta E}}\sim -\frac{1}{\rho_0(E)e^{S_0}}\braket{E|V\omega V^T\omega^{-1}|E}\sim -\frac{1}{L}\braket{E|V\omega V^T\omega^{-1}|E}
\end{equation}
which identifes with the third term of equation~(\refeq{VVs}). This cancels the original plateau. 
\section{Fermionic Two-Point Correlation Functions}

We now shift gears from bosonic 2-point functions to fermionic 2-point functions. On the bulk side, fermions introduce Spin structures for orientable manifolds and Pin structures for non-orientable manifolds. On the boundary side, anomalies of two discrete symmetries of the boundary theory classify it into different RMT classes. In this section we show that JT gravity and RMT calculations of fermionic two-point correlation functions match and they together confirm numerical computations in SYK \cite{9authors}. There, SYK Hamiltonian for $N$ Majorana fermions $\psi_a$ is given by
\begin{equation}
H=\frac{1}{4!}\sum_{a,b,c,d}J_{abcd}\psi_a\psi_b\psi_c\psi_d=\sum_{a<b<c<d}J_{abcd}\psi_a\psi_b\psi_c\psi_d\label{syk}
\end{equation}
where $J_{abcd}$ are totally anti-symmetric independent parameters drawn from Gaussian distribution. In SYK, the two-point correlation function averaged over couplings $J$ is given by
\begin{equation}
G(t)=\frac{1}{N}\sum_{i=1}^N\frac{\braket{\mathrm{Tr}[e^{-\beta H}\psi_i(t)\psi_i(0)]}_J}{\braket{Z(\beta)}_J}
\end{equation}
The graph of $G(t)$ v.s.~time $t$ has combinations of ramp and plateau features depending on $N\mod 8$ summarized in table~\ref{SYKresult}.

\begin{table}[h]
\centering
\begin{tabular}{ c c c c c c c }
\\
N & ramp & plateau\\
\hline\\
$0\mod8$& $\times$ & $\times$ \\
\\
$2\mod8$& $\checkmark$ & $\checkmark$\\
\\
$4\mod8$&  $\times$ & $\times$ \\
\\
$6\mod8$& $\checkmark$ & $\times$\\
\\
\end{tabular}
\caption{SYK numerics result \cite{9authors}}
\label{SYKresult}
\end{table}

Note that this table only contains even $N$. We will match JT and RMT calculations for both even and odd $N$, and for even $N$ we are able to confirm the results in the table.

\subsection{Review}
Before we delve into Pin structures on non-orientable geometries and compute fermionic 2-point correlation functions, let us first review Spin structures on oriented geometries by computing a simpler example: the product of two fermionic one-point functions. This subsection is based on \cite{StanfordWitten19} and \cite{Witten16}. 

To compute a particular quantity in JT involving fermions, we need to sum over all Spin structures on a particular geometry with an appropriate weighting factor characterized by topological invariant $\zeta$. More specifically, if we consider a manifold $Y$ with boundary $X$ we do that sum by fixing a Spin structure on $X$ and sum over compatible Spin structures on $Y$. For SYK model with $N$ Majorana fermions, the weighting factor is given by $(-1)^{N\zeta}$. The topological invariant $\zeta$ is defined as follows: If $Y$ has no boundary, for a bulk fermion field $\Psi$, we can consider its Dirac equation on $Y$, $\slashed{D}\Psi=0$. The number of zero modes of this equation mod2 is $\zeta$.

\begin{figure}[H]
\centering
\includegraphics[width=0.5\textwidth]{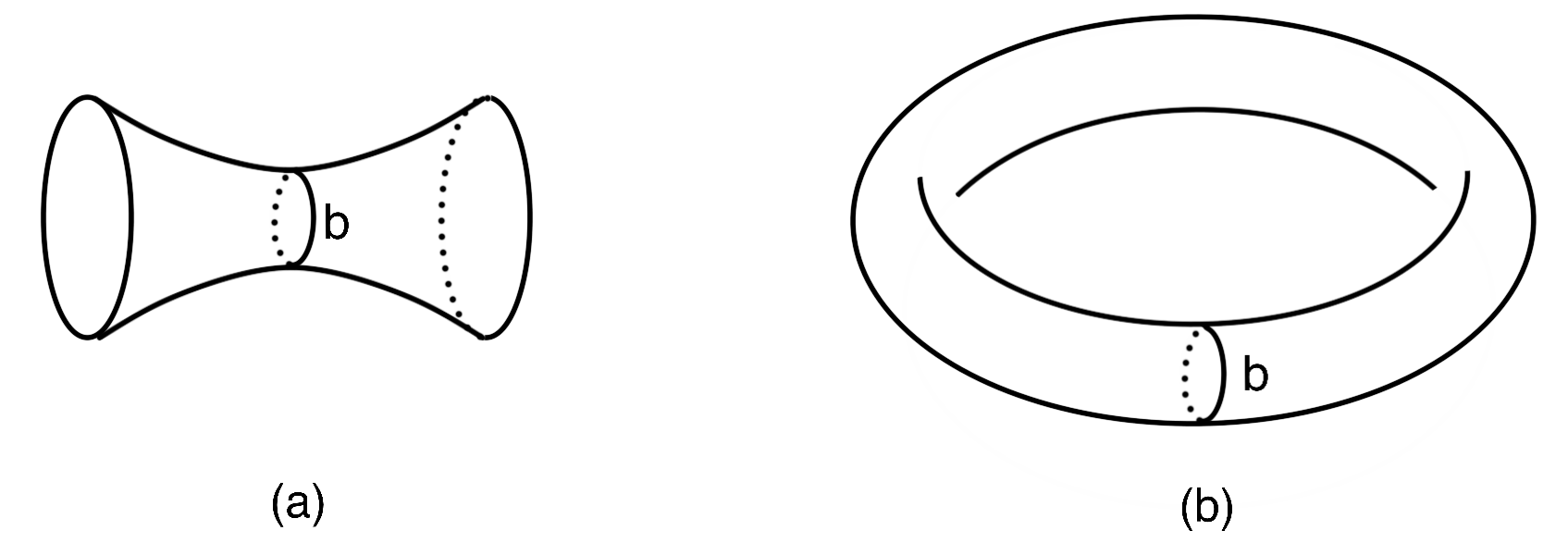}
\caption{(a) double trumpet (b) torus}
\label{doubletrumpet}
\end{figure}

Now we consider an example illustrating the Spin structure: the product of two fermionic one-point functions. Before looking at fermions, we first recall Saad \cite{Saadsingleauthor} showed that for bosons, the product of two one-point functions have a non-decaying contribution from a connected geometry called a double-trumpet (see figure~\ref{doubletrumpet}a), and the result is given by 
\begin{align}
\braket{V}_\beta\braket{V}_{\beta'}&= \int e^\ell d\ell\,P_{\text{Disk}}(\beta,\beta',\ell,\ell)e^{-\Delta\ell}\\
&=\int e^\ell d\ell\,\int dE\,\rho_0(E)e^{-(\beta+\beta')E}\varphi_E(\ell)^2e^{-\Delta\ell}\\
&=\int dE\,\rho_0(E)e^{-(\beta+\beta')E}|V_{E,E}|^2\label{1pointprod}
\end{align}
In this case, the manifold $Y$ is a double trumpet and $X$ are two asymptotic circles. On each circle, there are two Spin structures: antiperiodic and periodic fermions on a circle are called Neveu-Schwarz (NS) and Ramond (R) respectively. Note that the Spin structures on the two boundaries of $Y$ should be the same since they can both be slided to the center of the double trumpet. Thus we can bring the two boundary circles together and identify them, so that the Spin structures on a double trumpet are the same as those on a torus.

Topologically we can draw a torus as a square with $x\simeq x+1$ and $y\simeq y+1$. Then a spin field $\Psi$ on $Y$ satisfies
\begin{equation}
\Psi(x+1,y)=\color{green}\pm\color{black}\Psi(x,y)\quad\quad\Psi(x,y+1)=\color{red}\pm\color{black} \Psi(x,y)
\end{equation}
+ for R and - for NS. 

\begin{figure}[H]
\centering
\includegraphics[width=0.5\textwidth]{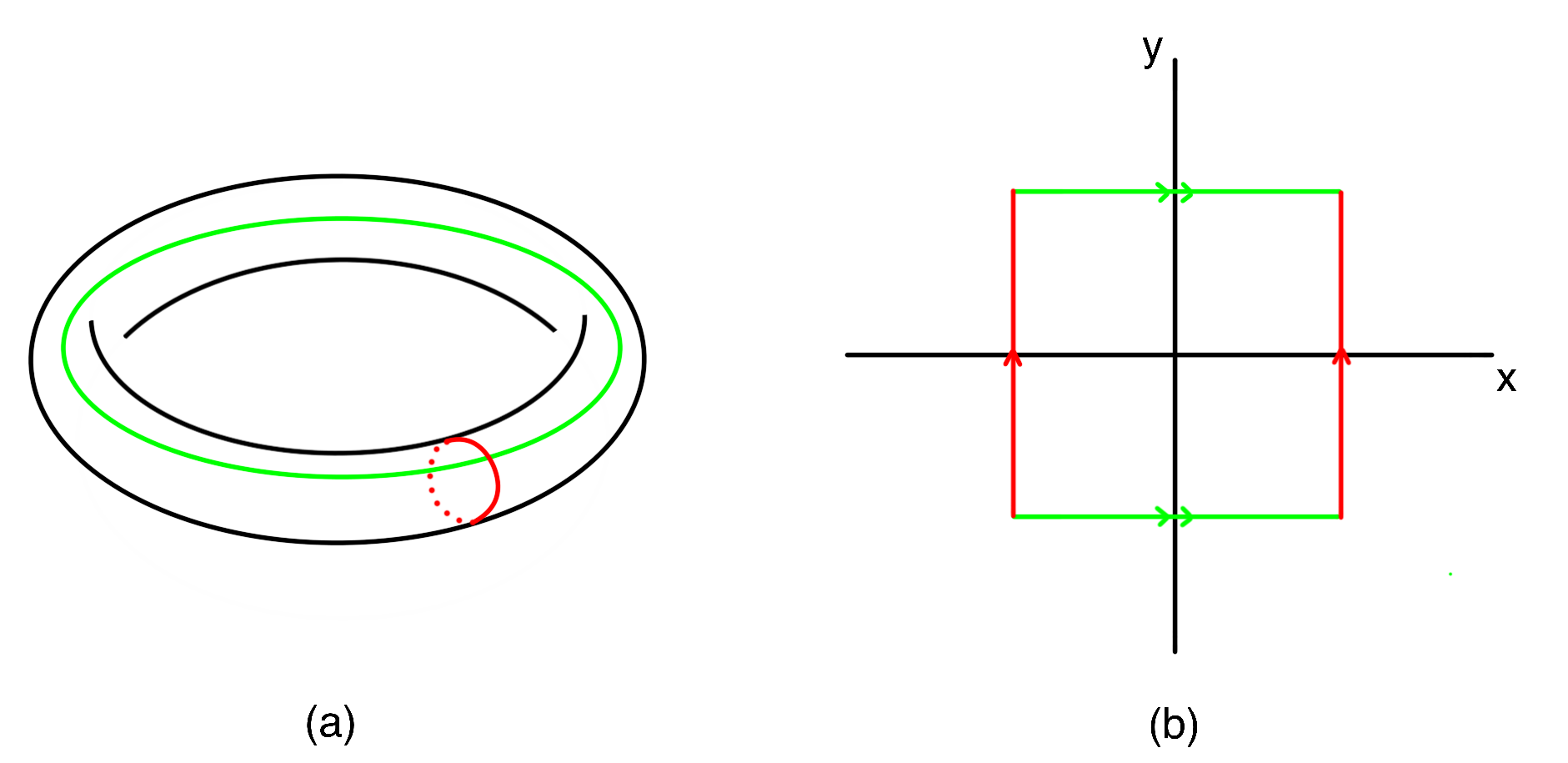}
\caption{torus spin structure}
\end{figure}

Let $z=x+iy$, then the Dirac equation of $\Psi$ can be written as $\partial\Psi/\partial\overline{z}=0$ and the topological invariant $\zeta$, i.e. the number of solutions to the Dirac equation mod2, are given in table~\ref{table:torusspin}. We can see that a torus has 4 Spin structures.

\begin{table}[h]    
\centering                                                         
\begin{tabular}{ c | c c }
\\
$\color{blue}\zeta$ & \color{red}R & \color{red}NS\\
\hline\\
\color{green}R & \color{blue}1 & \color{blue}0 \\
\\
\color{green}NS & \color{blue}0 & \color{blue}0 \\
\\
\end{tabular}
\caption{$\zeta$ of a torus}
\label{table:torusspin}
\end{table}

Let $F$ denote the sum over Spin structures on $Y$ keeping the Spin structure on $X$ fixed. We can then just sum $(-1)^{N\zeta}$ over rows for each column of the above table and get
\begin{equation}
F_{\mathrm{Torus}}^{\mathrm{\color{red}R}}=(-1)^N+1\quad\quad F_{\mathrm{Torus}}^{\mathrm{\color{red}NS}}=2
\end{equation}

Now we go back to double-trumpet contribution to the product of two one-point functions.
\begin{figure}[H]
\centering
\includegraphics[width=0.2\textwidth]{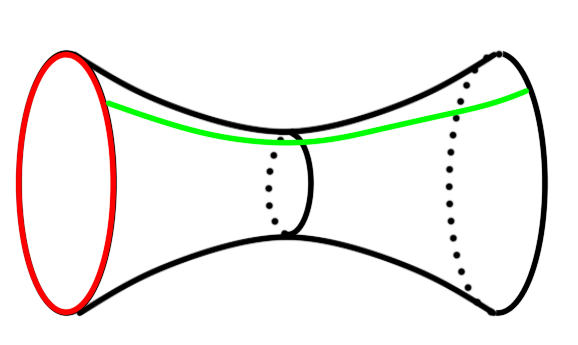}
\caption{double-trumpet with Spin structure}
\label{dtspin}
\end{figure}

We showed at the beginning of this subsection that without Spin structures the product of two bosonic one-point functions is given by
\begin{equation}
\braket{V}\braket{V}=\braket{VV}_{\mathrm{double-trumpet}}=\int dE\,\rho_0(E)e^{-(\beta+\beta')E}|V_{E,E}|^2
\end{equation}
With Spin structures, we just multiply the above expression by the appropriate weighting factor $(-1)^{N\zeta}$ and get the following table.

\begin{table}[H]
\centering
\begin{tabular}{ c | c c }
\\
$\braket{V}\braket{V}$ & $\braket{V}_\mathrm{\color{red}R}\braket{V}_\mathrm{\color{red}R}$ & $\braket{V}_\mathrm{\color{red}NS}\braket{V}_\mathrm{\color{red}NS}$\\
\hline\\
\color{green}R & $(-1)^{N\color{blue}1}$$\braket{VV}_{\mathrm{double-trumpet}}$ & $(-1)^{N\color{blue}0}$$\braket{VV}_{\mathrm{double-trumpet}}$ \\
\\
\color{green}NS & $(-1)^{N\color{blue}0}$$\braket{VV}_{\mathrm{double-trumpet}}$ & $(-1)^{N\color{blue}0}$$\braket{VV}_{\mathrm{double-trumpet}}$ \\
\\
\end{tabular}
\caption{product of two bosonic one-point functions}
\end{table}

Summing over Spin structures with boundary Spin structure fixed, i.e. suming over the \color{green}green \color{black}Spin structures, we get
\begin{equation}
\braket{V}_{\color{red}\mathrm{R}}\braket{V}_{\color{red}\mathrm{R}}=((-1)^N+1)\braket{VV}_{\mathrm{double-trumpet}}\quad\quad\braket{V}_{\color{red}\mathrm{NS}}\braket{V}_{\color{red}\mathrm{NS}}=2\braket{VV}_{\mathrm{double-trumpet}}
\end{equation}

We now look at the product of two one-point functions of boundary fermions $\psi$, again if we ignore Spin structures, the product of two fermionic one-point functions is given by
\begin{equation}
\braket{\psi\psi}_{\mathrm{double-trumpet}}=\int dE\,\rho_0(E)e^{-(\beta+\beta')E}|\psi_{E,E}|^2\label{1pointprodf}
\end{equation}
With Spin structures, in addition to multiplying the above expression by the appropriate weighting factor $(-1)^{N\zeta}$, there is another factor we need to take into account for fermions. According to the \color{green}green \color{black}Spin structure in figure~\ref{dtspin} the two fermions on left/right boundaries are the same if the Spin structure is \color{green}R \color{black} and would differ by a minus sign if the Spin structure is \color{green}NS\color{black}.  We summarize the result in the following table.

\begin{table}[h]
\centering
\begin{tabular}{ c | c c }
\\
$\braket{\psi}\braket{\psi}$ & $\braket{\psi}_\mathrm{\color{red}R}\braket{\psi}_\mathrm{\color{red}R}$ & $\braket{\psi}_\mathrm{\color{red}NS}\braket{\psi}_\mathrm{\color{red}NS}$\\
\hline\\
\color{green}R & $(-1)^{N\color{blue}1}$$\braket{\psi\psi}_{\mathrm{double-trumpet}}$ & $(-1)^{N\color{blue}0}$$\braket{\psi\psi}_{\mathrm{double-trumpet}}$ \\
\\
\color{green}NS & $(-1)^{N\color{blue}0}$$\color{green}(-1)$$\braket{\psi\psi}_{\mathrm{double-trumpet}}$ & $(-1)^{N\color{blue}0}\color{green}(-1)$$\braket{\psi\psi}_{\mathrm{double-trumpet}}$ \\
\\
\end{tabular}\caption{product of two fermionic one-point functions}
\end{table}

Summing over Spin structures with boundary Spin structure fixed, i.e. suming over the \color{green}green \color{black}Spin structures, we get
\begin{equation}
\braket{\psi}_{\color{red}\mathrm{R}}\braket{\psi}_{\color{red}\mathrm{R}}=((-1)^N-1)\braket{\psi\psi}_{\mathrm{double-trumpet}}\quad\quad\braket{\psi}_{\color{red}\mathrm{NS}}\braket{\psi}_{\color{red}\mathrm{NS}}=0
\end{equation}

\subsection{Pin$^-$ Structure and Crosscaps}

We now consider non-orientable geometries, where there is a structure called Pin structures analogous to the Spin structure for orientable geometries. There are two ways of defining these Pin structures called Pin$^+$ and Pin$^-$, respectively. We know that bulk fermions transform under Lorentzian time-reversal and spacial reflection respectively as (see \cite{Witten16} section 5 and Appendix A)
\begin{equation}
T:\Psi(t,x)\mapsto \gamma^0\Psi(-t,x)\quad\quad R:\Psi(t,x)\mapsto \gamma^1\Psi(t,-x)
\end{equation}
with one of $T$ and $R$ squares to 1 and the other squares to $(-1)^F$. Pin$^-$ or Pin$^+$ depends on which one squares to 1 as summarized in table~\ref{pin+-}.

\begin{table}[H]
\centering
\begin{tabular}{ c c c c c c }
\\
Pin structures & $T^2$ & $R^2$\\
\hline\\
Pin$^-$ &  1 & $(-1)^F$\\
\\
Pin$^+$ & $(-1)^F$ & 1\\
\\
\end{tabular}
\caption{Pin$^-$ vs Pin$^+$}
\label{pin+-}
\end{table}

\noindent To reproduce the behavior of the standard SYK model, we are interested in the Pin$^-$ case, as explained in \cite{StanfordWitten19}. In this case, based on table~\ref{pin+-} we would like $(\gamma^0)^2=1$ and $(\gamma^1)^2=-1$, which can be satisfied with the following choice of real matrices \footnote{Note that our $\gamma$ matrices satisfy $\{\gamma^\mu,\gamma^\nu\}=-2\eta^{\mu\nu}$, which leads to a minus sign in the spin matrix as explained in Appendix~\ref{appendixfermion}.}. 
\begin{equation}
\gamma^0=\begin{pmatrix}0&1\\1&0\end{pmatrix}\quad\quad\gamma^1=\begin{pmatrix}0&-1\\1&0\end{pmatrix}
\end{equation}
and we define the quantity
\begin{equation}
\overline{\gamma}=\gamma^0\gamma^1=\begin{pmatrix}1&0\\0&-1\end{pmatrix}\label{gammabar}
\end{equation}
to be used later. In Euclidean signature, the corresponding gamma matrices are given by
\begin{equation}
\gamma^1=\begin{pmatrix}0&-1\\1&0\end{pmatrix}\quad\quad\gamma^2=\begin{pmatrix}0&i\\i&0\end{pmatrix}
\end{equation}

In Euclidean signature, we can obtain our non-orientable geometry from a quotient of the hyperbolic disk, so we start from looking at fermions on a hyperbolic disk. There are two sets of commonly used coordinates for the hyperbolic disk as shown in figure~\ref{coordinate} (see Appendix~\ref{appendixmeasure} for more details): $xy$-coordinates given by 
\begin{equation}
ds^2=dx^2+\cosh^2x\,dy^2\label{diskmetricxy}
\end{equation}
and $\rho\theta$-coordinates given by
\begin{equation}
ds^2=d\rho^2+\sinh^2\rho\,d\theta^2\label{diskmetricrho}
\end{equation}

\begin{figure}[h]
\centering
\includegraphics[width=0.5\textwidth]{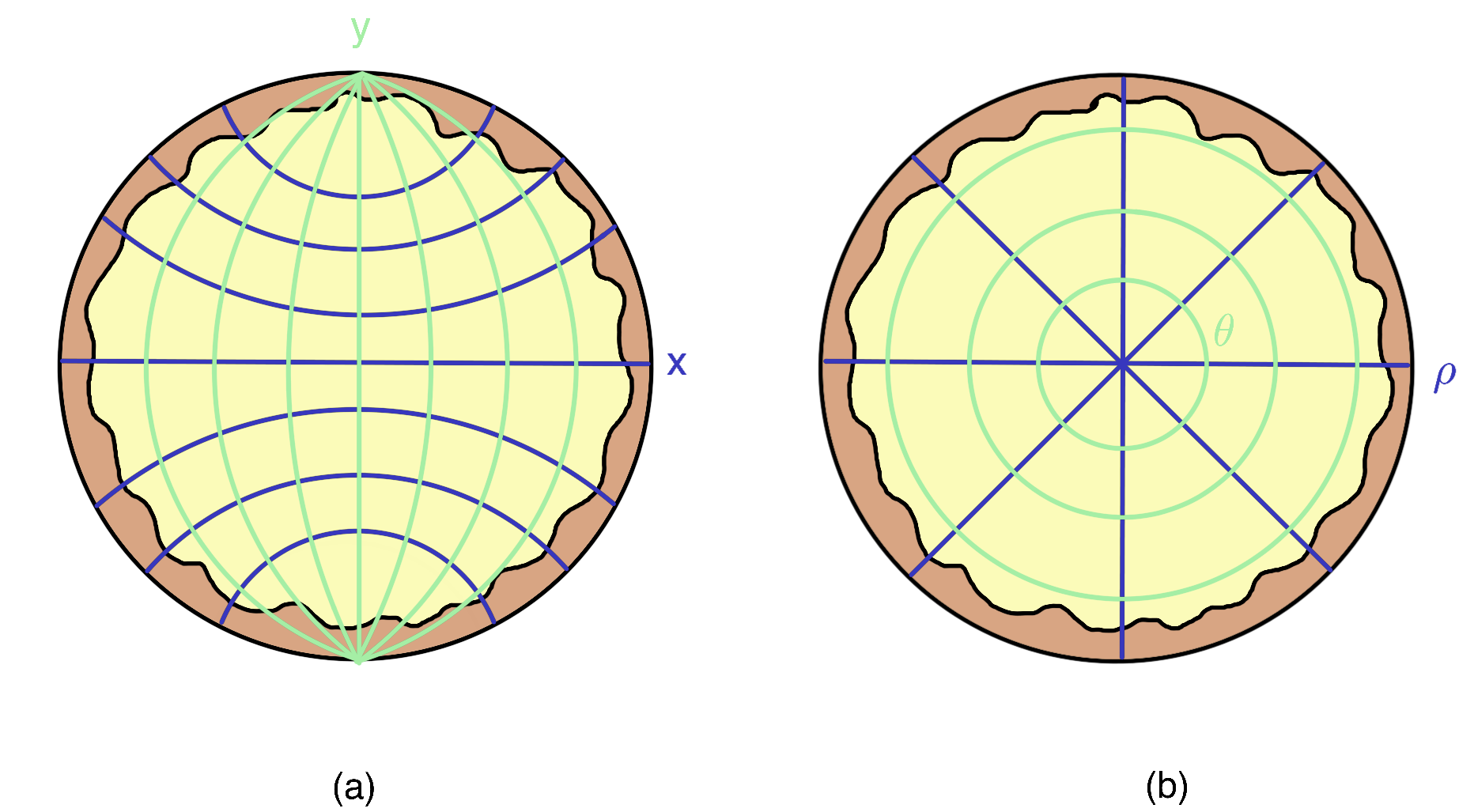}
\caption{(a) $xy$-coordinates (b) $\rho\theta$-coordinates}
\label{coordinate}
\end{figure}

For bulk fermions $\Psi$, we work with the $xy$-frame. Euclidean time-reversal in $y$ and reflection in $x$ are given by
\begin{equation}
T':\Psi(y,x)\mapsto \gamma^2\Psi(-y,x)\quad\quad R:\Psi(y,x)\mapsto \gamma^1\Psi(y,-x)
\end{equation}
Near $y=0$, boundary fermions $\psi$ on the $xy$-frame and $\rho\theta$-frame are the same on one side of $x\rightarrow\pm\infty$ and related by $\mathrm{Rot}(\pi)$, i.e. a rotation by $180^\circ$, on the other side. The rotation can be written as a product of two reflections $\mathrm{Rot}(\pi)=T'R$.

In the AdS/CFT correspondence, the behavior of the bulk fermion field near the boundary is important. Near the ``right'' boundary $x\rightarrow\infty$, the behavior is:
\begin{equation}
\Psi(y,x\rightarrow\infty)=e^{(-\frac{1}{2}+M)x}\psi_+(y)\eta_++e^{(-\frac{1}{2}-M)x}\psi_-(y)\eta_-\label{psiexpansion1}
\end{equation}
where $\eta_+=\begin{pmatrix}1\\1\end{pmatrix}$ and $\eta_-=\begin{pmatrix}1\\-1\end{pmatrix}$. We will take $M\geq 0$. One can impose boundary conditions that set to zero either $\psi_+$ (Dirichlet-like) or $\psi_-$ (Neumann-like). In each case, the operator that is not set to zero becomes the boundary fermion operator. The conformal dimensions of the boundary operator is $\frac{1}{2}+M$ or $\frac{1}{2}-M$ in the two cases. Although it doesn't actually affect what follows, for concreteness we will use the Neumann-like boundary conditions that are necessary to get a boundary fermion with dimension $\frac{1}{4}$ as in the SYK model.

Similarly, near the ``left'' boundary $x\rightarrow-\infty$, the asymptotic behavior is
\begin{equation}
\Psi(y,x\rightarrow-\infty)=e^{(\frac{1}{2}+M)x}\psi_+(y)\eta_++e^{(\frac{1}{2}-M)x}\psi_-(y)\eta_-\label{psiexpansion2}
\end{equation}
(see Appendix~\ref{appendixfermion} for details on how we get these asymptotic behaviors). 

The actions of the discrete symmetry operators $T'$ and $R$ on the bulk fermion induce corresponding actions on the boundary operators, which we can work out as follows. $T'$ and $R$ act on bulk fermions by conjugation as
\begin{equation}
T'\Psi(y,x)T'^{-1}=\gamma^2\Psi(-y,x)\quad\quad R\Psi(y,x)R^{-1}=\gamma^1\Psi(y,-x)
\end{equation}
We substitute (\ref{psiexpansion1}) and (\ref{psiexpansion2}) into the above equations
\begin{align}
R\Psi(y,x\rightarrow\pm\infty)R^{-1}&=e^{(\mp\frac{1}{2}+M)x}R\psi_+(y)R^{-1}\eta_++e^{(\mp\frac{1}{2}-M)x}R\psi_-(y)R^{-1}\eta_-\\
&=e^{(\pm\frac{1}{2}-M)(-x)}R\psi_+(y)R^{-1}\gamma^1\eta_-+e^{(\pm\frac{1}{2}+M)(-x)}R\psi_-(y)R^{-1}(-\gamma^1\eta_+)\\
&=\gamma^1\Psi(y,-x\rightarrow\mp\infty)
\end{align}
gives
\begin{equation}
R\psi_+(y) R^{-1}=\psi_-(y)\quad\quad R\psi_-(y) R^{-1}=-\psi_+(y)
\end{equation}
And similarly, 
\begin{align}
T'\Psi(y,x\rightarrow\pm\infty)T'^{-1}&=e^{(\mp\frac{1}{2}+M)x}T'\psi_+(y)T'^{-1}\eta_++e^{(\mp\frac{1}{2}-M)x}T'\psi_-(y)T'^{-1}\eta_-\\
&=e^{(\mp\frac{1}{2}+M)x}T'\psi_+(y)T'^{-1}(-i\gamma^2\eta_+)+e^{(\mp\frac{1}{2}-M)x}T'\psi_-(y)T'^{-1}(i\gamma^2\eta_-)\\
&=\gamma^2\Psi(-y,x\rightarrow\pm\infty)
\end{align}
gives
\begin{equation}
T'\psi_+(y)T'^{-1}=i\psi_+(-y)\quad\quad T'\psi_-(y) T'^{-1}=-i\psi_-(-y)
\end{equation}
On other thing we should keep in mind is that according to AdS/CFT the bulk field should decay as they approach the boundary, so we choose the mode $\psi_+$ as $x\rightarrow\infty$ and the mode $\psi_-$ as $x\rightarrow-\infty$ on the boundary of the hyperbolic disk (see Appendix~\ref{appendixfermion} for more details).

There is a topological invariant $\eta$ for non-orientable manifolds analogous to $\zeta$ for orientable manifolds. For Pin$^-$, the Dirac equation is given by
\begin{equation}
(i\overline{\gamma}\slashed{D}+iM)\Psi=0
\end{equation}
Let $\{\lambda_k\}$ be eigenvalues of $i\overline{\gamma}\slashed{D}$. When $M>0$ is large, the regularized partition function is given by
\begin{equation}
Z_{\mathrm{reg}}=\det{i\overline{\gamma}\slashed{D}}=\prod_k\frac{\lambda_k}{\lambda_k+iM}=|Z|\exp(-\frac{i\pi}{2}\sum_k\mathrm{sign}(\lambda_k))=|Z|\exp(-\frac{i\pi}{2}\eta)
\end{equation}
(for more information see \cite{Witten16}). We can then define APS invariant $\eta$ as
\begin{equation}
\eta=\lim_{s\rightarrow0}\sum_k\mathrm{sign}(\lambda_k)|\lambda_k|^{-s}=\lim_{\epsilon\rightarrow0}\sum_k\mathrm{sign}(\lambda_k)e^{-\epsilon\lambda_k}
\end{equation}
More specifically, to match anomaly class of SYK model with $N$ Majorana fermions on the boundary, we can generalize the weighting factor from $(-1)^{N\zeta}=e^{-i\pi N\zeta}$ for orientable geometries to $e^{-i\pi N\eta/2}$ for non-orientable geometries in the bulk (they agree in the orientable case). One vague motivation for the form of the weighting factors is that if we have $N$ bulk Majorana fermions each with the same weighting factor overall they would give that to the $N$th power. But this is not to be taken literally. SYK is not dual to JT with $N$ bulk fermions, only that they have the same anomalies. 

\subsubsection{Genus one-half: constant shift}
\begin{figure}[h]
\centering
\includegraphics[width=0.5\textwidth]{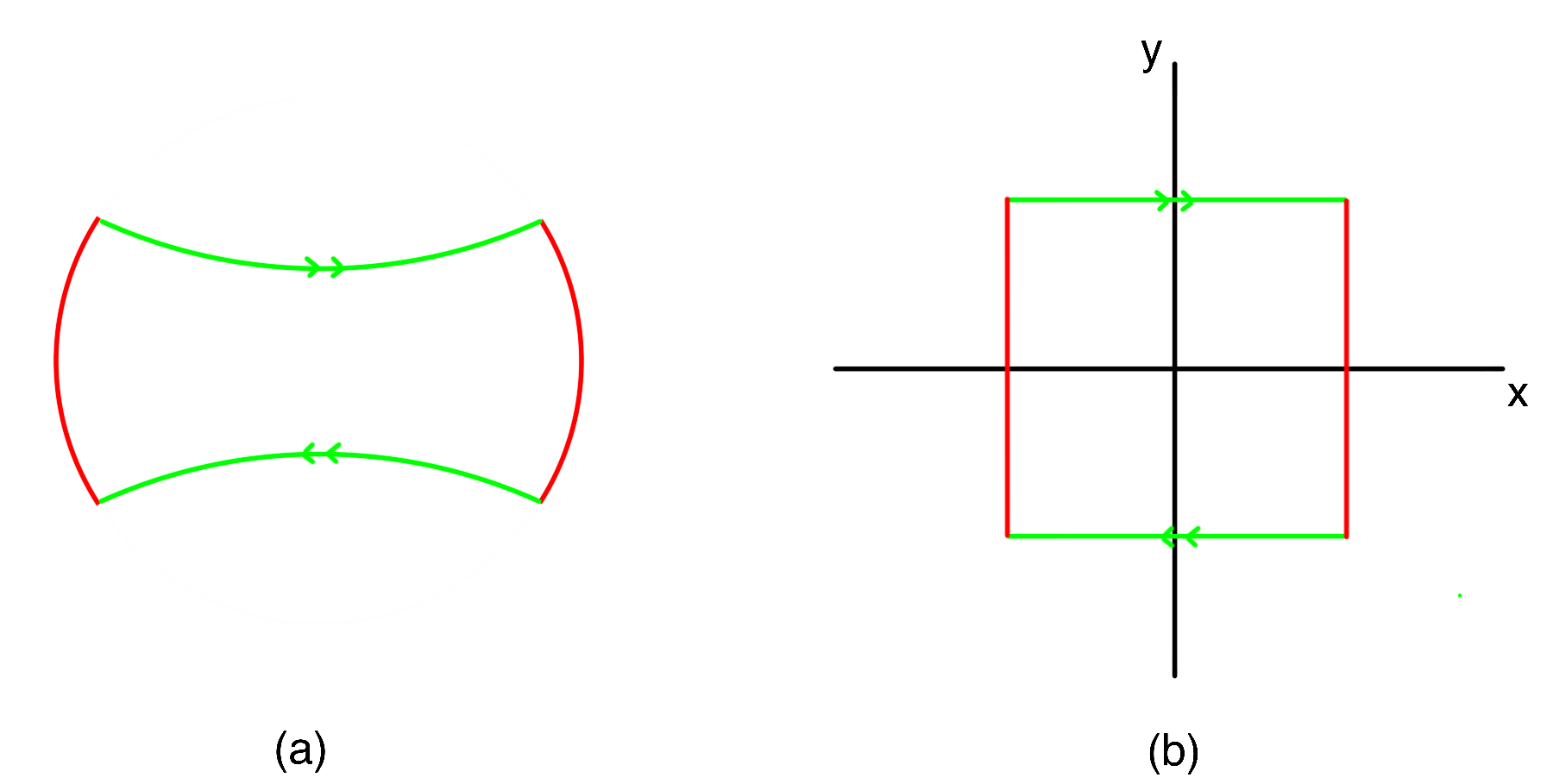}
\caption{crosscap as a square diagram}
\label{ccsquare}
\end{figure}

It can be shown that a crosscap has $\eta=\pm\frac{1}{2}$ (see Appendix~\ref{appendixeta} for more information). In particular, going around the boundary $S^1$ of a crosscap looks like reflection squared
\begin{equation}
R^2: \Psi\mapsto(\pm\gamma^1)^2\Psi
\end{equation}
we have $\pm\gamma^1$ corresponding to the two possible $\eta$'s. We can then define the Pin$^-$ sum of weighting factors over $\eta=\pm1/2$ for a crosscap 
\begin{equation}
F_{cc}(N)=\sum_{\text{pin$^-$ structures}}e^{-\frac{i\pi}{2}N\eta}=\sum_{\eta=\pm\frac{1}{2}}e^{-\frac{i\pi}{2}N\eta}=2\cos(2\pi N/8)
\end{equation}
We can draw a crosscap topologically as a square with two opposite edges identified with a reflection as in figure~\ref{ccsquare}(b). 

A bulk fermion field $\Psi$ on flat space with the same kind of square diagram would satisfy
\begin{equation}
\Psi(-x,y+1)=(-1)^{\color{green}\alpha}\gamma^1\Psi(x,y)\label{ccid}
\end{equation}
with $\alpha=0,1$ specifying two Pin$^-$ structures $\eta=\pm1/2$. If we embed the crosscap in the hyperbolic disk with $xy$-coordinates then instead of identifying $y=\pm1/2$, we are indentifying $y=\pm b/4$ as in figure~\ref{ccsquare}(a), so the equation (\ref{ccid}) should instead look like
\begin{equation}
\Psi(-x,y+\frac{b}{2})=(-1)^{\color{green}\alpha}\gamma^1\Psi(x,y)
\end{equation}
Before calculating the disk+crosscap contribution to fermionic two-point correlation functions, we first need to describe what geodesics look like on a disk+crosscap connecting two boundary fermions. Here we only focus on the geodesic that gives non-decaying contribution to the two-point correlator.

\begin{figure}[H]
\centering
\includegraphics[width=0.5\textwidth]{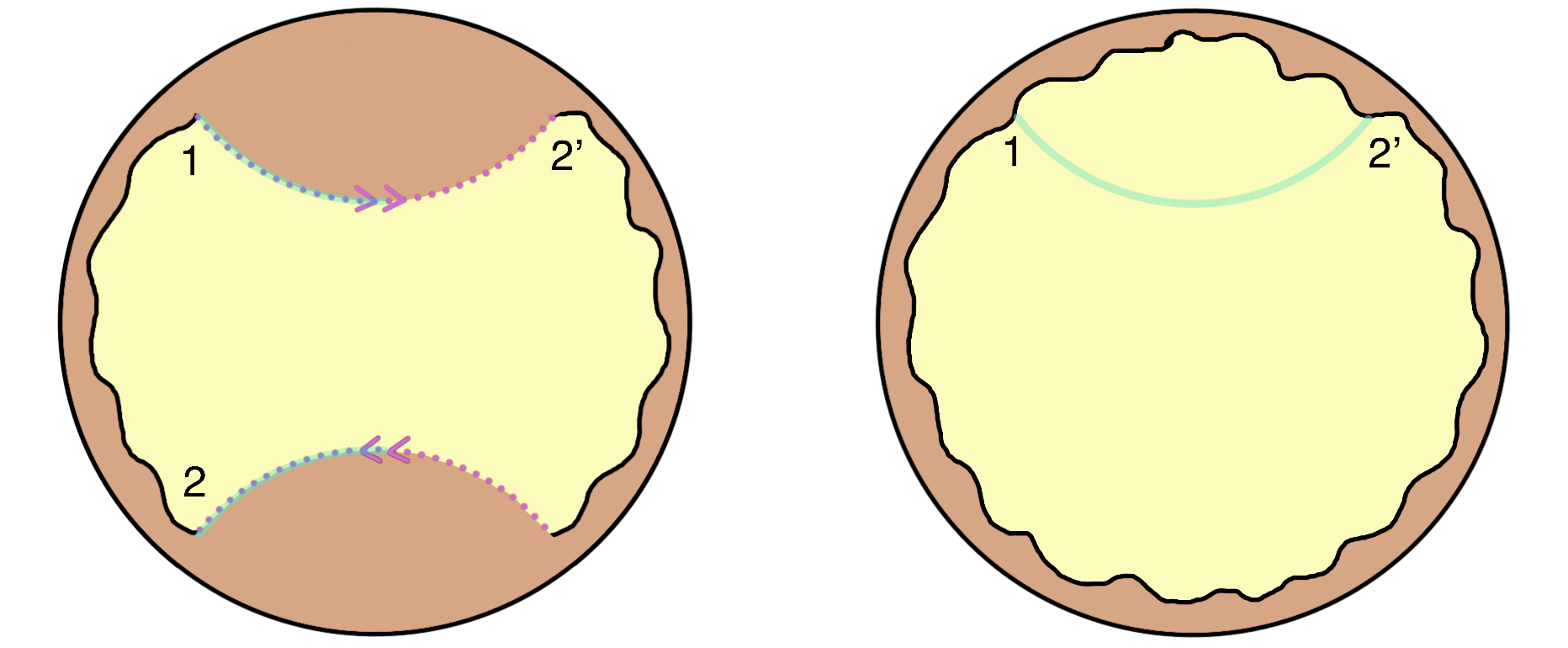}
\caption{Geodesic (light blue curve) on a crosscap v.s. on a disk}
\label{reflection}
\end{figure}

Recall that disk+crosscap is topologically a M\"{o}bius band and can be drawn as a quotient of the hyperbolic disk. A geodesics going through the crosscap consists of two pieces connected together across the crosscap (see figure~\ref{reflection}a). The two pieces of the geodesics can be put together into a horizontal geodesics (see figure~\ref{reflection}b) but with a reflection on the fermion. Thus a two-point correlation function on the crosscap can be related to that on the disk via
\begin{align}
\braket{\psi_{1-}\psi_{2-}}_{cc}&=\braket{\psi_{1-}R\psi_{2'+}R^{-1}}_{\text{Disk},xy}\\
&=\braket{\psi_{1-}\mathrm{Rot}(\pi)R\psi_{2'+}R^{-1}\mathrm{Rot}(\pi)^{-1}}_{\text{Disk},\rho\theta}\\
&=-\braket{\psi_{1-}T'\psi_{2'+}T'^{-1}}_{\text{Disk},\rho\theta}\\
&=-i\braket{\psi_{1-}\psi_{2'+}}_{\text{Disk},\rho\theta}
\end{align}
where $1$, $2$, and $2'$ are shown as in figure~\ref{reflection}. 

Here we ignore the decaying contributions to the two-point correlation function. Let $\braket{\psi\psi}_{\chi=0,0}$ be the 2-point function without considering the Spin/Pin$^-$ structure
\begin{equation}
\braket{\psi\psi}_{\chi=0,0}=\int dE\,\rho_0e^{-\beta E}|\psi_{E,E}|^2
\end{equation}
then with Pin$^-$ structures, the 2-point function is given by
\begin{equation}
\braket{\psi\psi}_{cc}=\sum_{\eta=\pm\frac{1}{2}}e^{-\frac{i\pi}{2}N\eta}(\pm1)(-i)\braket{\psi \psi}_{\chi=0,0}=2\sin\left(\frac{2\pi N}{8}\right)\braket{\psi\psi}_{\chi=0,0}
\end{equation}

\subsubsection{Genus one: ramp}

Any non-orientable two-manifold is an oriented manifold with one or two crosscaps glued in, so after considering disk+one crosscap, it is natural to consider disk+two crosscaps. But before doing that, let us review the Pin$^-$ structures of a reflected-double-trumpet. 

We now consider a double trumpet glued from two trumpets but with a reflection on one of the interfaces being glued (see figure~\ref{doubletrumpet}). This has the same Pin$^-$ structure as a Klein bottle. 

\begin{figure}[h]
\centering
\includegraphics[width=0.5\textwidth]{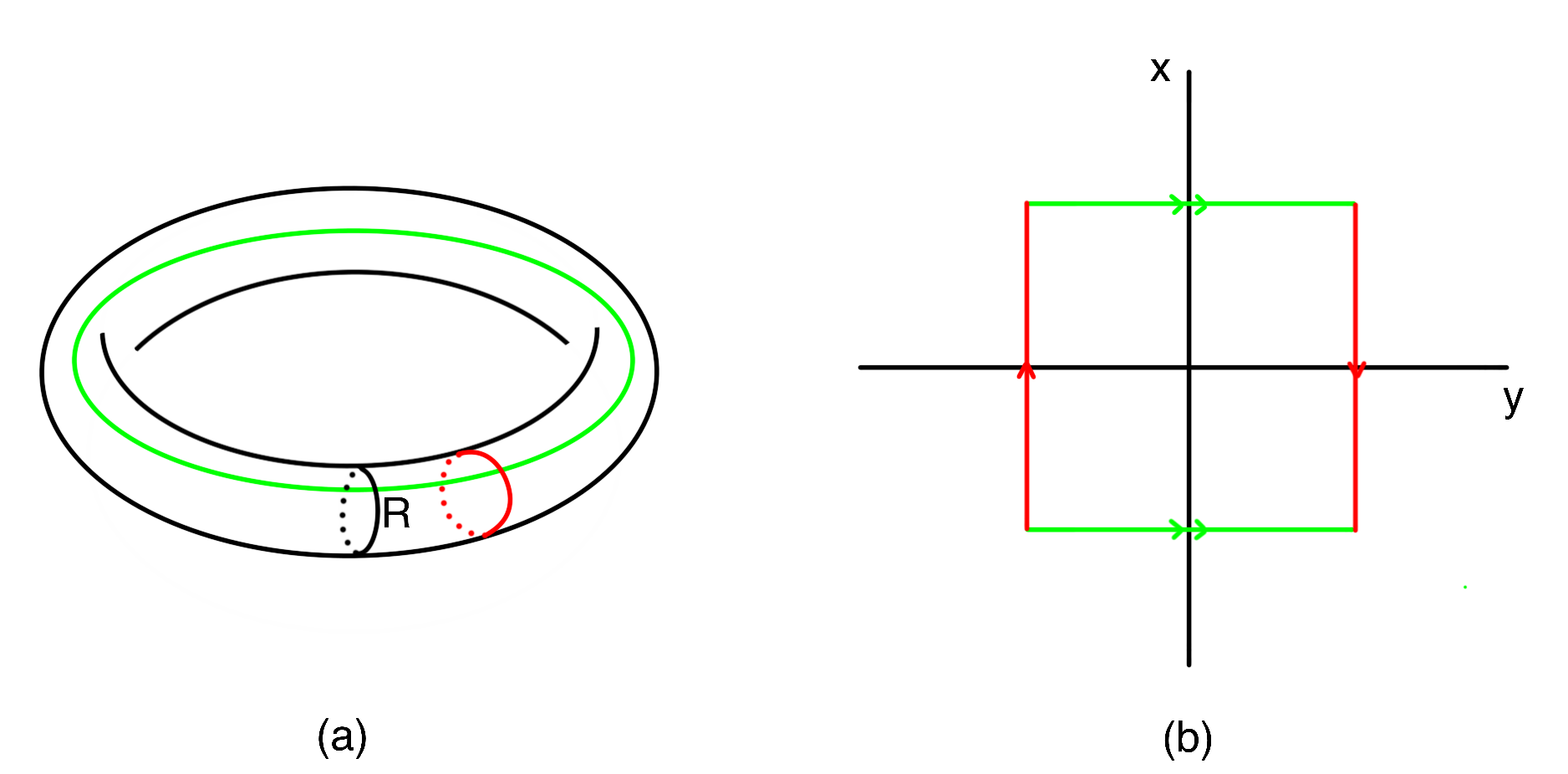}
\caption{(a) Klein bottle (here R means reflecion) (b) square diagram of a Klein bottle, note that here the vertical axis is the $x$-axis while the horizontal axis is the $y$-axis}
\end{figure}

\begin{figure}[h]
\centering
\includegraphics[width=0.5\textwidth]{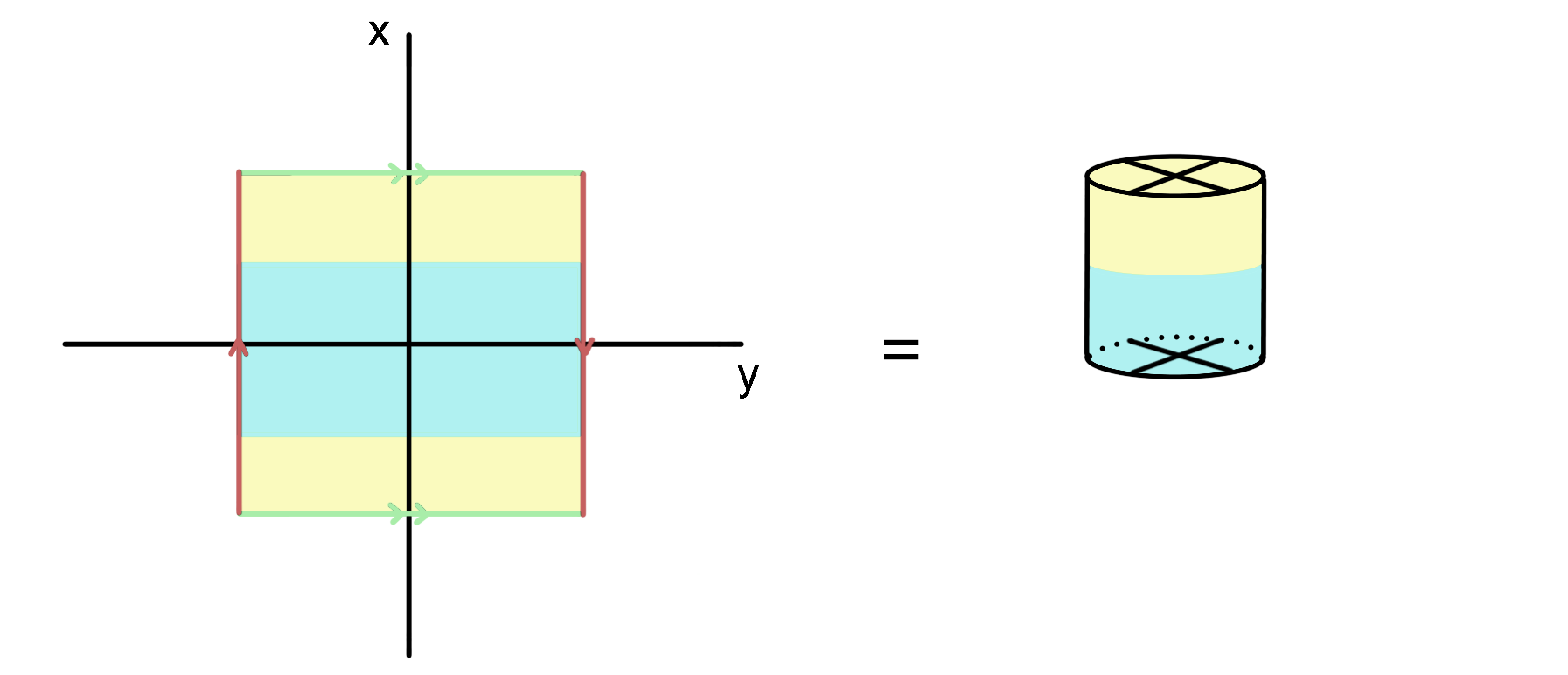}
\caption{Klein bottle is topologically equivalent to a cylinder with two crosscaps}
\label{kb2cc}
\end{figure}

Topologically we can draw a Klein bottle as a square with the sides identified according to
\begin{equation}
(x,y)\simeq (x+1,y)\quad\quad (x,y)\simeq(-x,y+1)
\end{equation}
Fermion field $\Psi$ on Klein bottle satisfies
\begin{align}
\Psi(x+1,y)&=(-1)^{\color{red}\alpha}\Psi(x,y)\label{kbf1}\\
\Psi(-x,y+1)&=(-1)^{\color{green}\beta}\gamma^1\Psi(x,y)\label{kbf2}
\end{align}
As shown in figure~\ref{kb2cc}, a Klein bottle is topologically equivalent to two crosscaps (one at $x=0$ and the other at $x=1/2$) on a cylinder. 

Therefore $F_{\mathrm{KB}}(N)=F_{\mathbb{RP}^2}(N)^2$. Equations (\refeq{kbf1}) and (\refeq{kbf2}) tell us that
\begin{align}
\Psi(0,y+1)&=(-1)^{\color{green}\beta}\gamma^1\Psi(0,y)\\
\Psi(\frac{1}{2},y+1)&=(-1)^{\color{red}\alpha+\color{green}\beta}\gamma^1\Psi(\frac{1}{2},y)
\end{align}
This is telling us if $\alpha=0$, i.e. if the boundaries are of R type, the two crosscaps have the same Pin$^-$ structure while if $\alpha=1$, i.e. if the boundaries are of NS type, the two crosscaps have opposite Pin$^-$ structures. Then
\begin{align}
F_{\mathrm{KB}}^{\mathrm{R}}&=\sum_{\eta=\pm\frac{1}{2}}\left(e^{-\frac{i\pi}{2}N\eta}\right)^2=2\cos(2\pi N/4)\\
F_{\mathrm{KB}}^{\mathrm{NS}}&=\sum_{\eta=\pm\frac{1}{2}}e^{-\frac{i\pi}{2}N\eta}e^{-\frac{i\pi}{2}N(-\eta)}=2
\end{align}

The Pin$^-$ structures of torus and Klein bottle are summarized in table~\ref{tkbtab}.

\begin{table}[h]
\centering
\begin{tabular}{ c | c c }
\\
\color{blue}pin$^-$ structure & \color{red}R & \color{red}NS\\
\hline\\
$F_{\mathrm{KB}}$ & $2\cos(2\pi N/4)$ & $2$ \\
\\
$F_{\mathrm{T}}$ & $1+(-1)^N$ & $2$ \\
\\
$F_{\mathrm{KB}}+F_{\mathrm{T}}$ & $4\delta_{N\mod4,0}$ & $4$ \\
\\
\end{tabular}
\caption{Spin/Pin$^-$ structure of torus and Klein bottle}
\label{tkbtab}
\end{table}

\begin{figure}[h]
\centering
\includegraphics[width=0.5\textwidth]{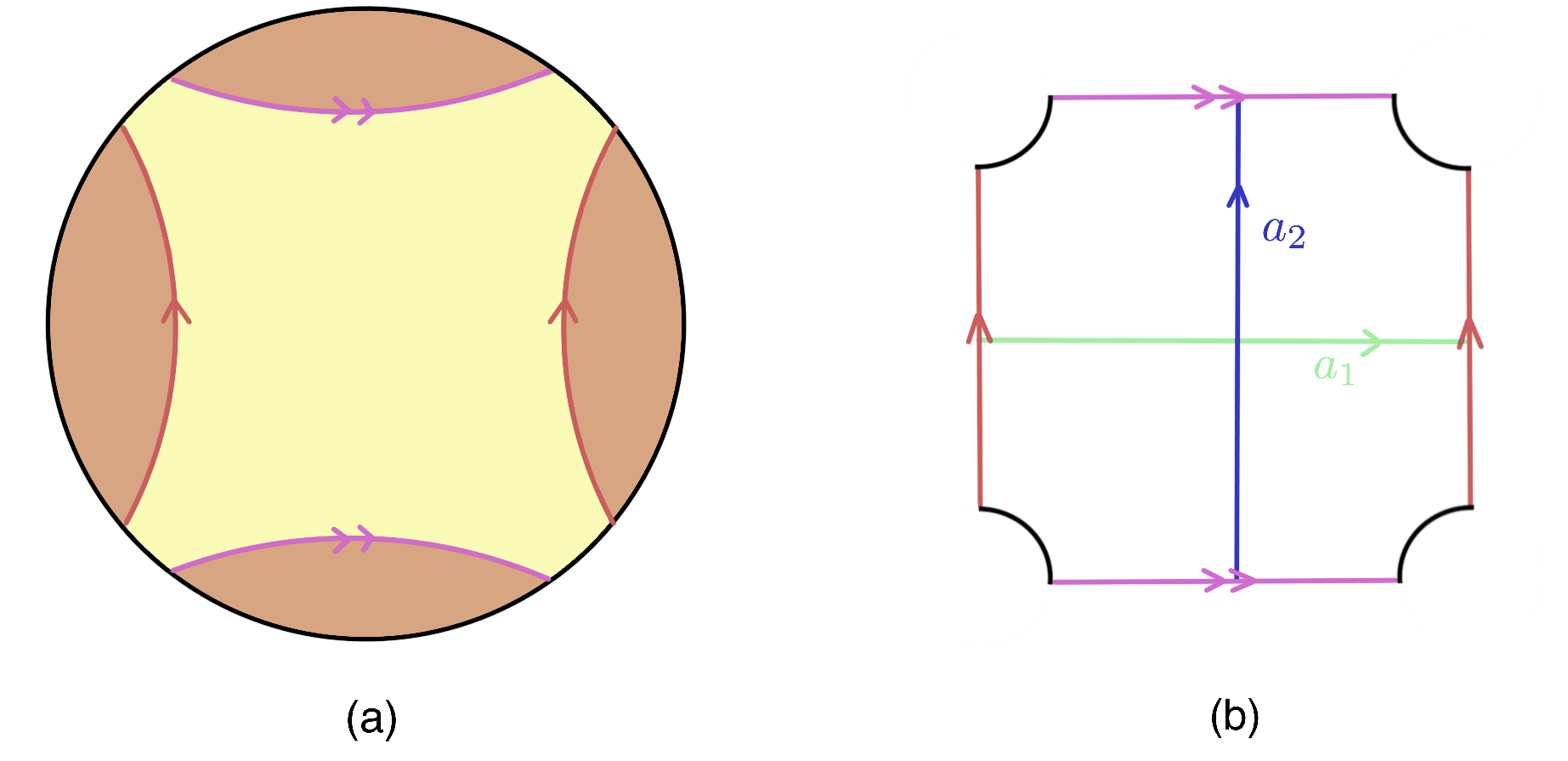}
\caption{(a) a handle-disk represented on a hyperbolic disk (b) square diagram of a handle-disk with $a_1$ and $a_2$ generators of $\pi_1$.}
\label{torusMCG}
\end{figure}

For order $\mathcal{O}(e^{-2S_0})$ contribution to the 2-point function, let's first consider the handle-disk given in \cite{Saadsingleauthor}. There are two types of geodesics, ones that go through the handle and ones that do not go through the handle. Saad showed that the former gives non-decaying contribution to the two-point correlation functions. Without loss of generality, we can represent the handle-disk as part of the hyperbolic disk as shown in figure~\ref{torusMCG}, with the pink line our geodesic. We already know that each single geodesic contributes (\ref{hdgeodesic}) to bosonic 2-pt correlator. To get the full contribution we need to integrate over all possible geodesics on all possible configurations that look like a handle-disk. 

The mapping class group (MCG) is a group of transformations that preserve a certain geometry. The mapping class group of a handle-disk is generated by two Dehn twist $t_1:\{a_1\mapsto a_1,a_2\mapsto a_1a_2\}$ and $t_2:\{a_1\mapsto a_1a_2,a_2\mapsto a_2\}$. Fixing the geodesic (thus also fixing the length of $b$), the only elements in MCG that preserves the shape of the handle-disk as well as the geodesic are generated by one of the Dehn twist $t_2$. Thus we need to integrate over twist $\tau$ but only from $0$ to $b$. To integrate over all geodesics we need to also integrate over the length of the geodesic from $0$ to infinity and to integrate $b$ from $0$ to infinity. Then the total non-decaying contribution from handle-disk to bosonic 2-point correlation function is
\begin{align}
\braket{V(-i\tau)V(0)}_{\chi=-1}&=e^{-S_0}\int_0^\infty bdb\int_{-\infty}^\infty e^\ell d\ell\,\varphi_{\text{Trumpet},\tau}(\ell,b)\varphi_{\text{Trumpet},\beta-\tau}(\ell,b)e^{-\Delta\ell}\\
&=e^{-S_0}\int_0^\infty bdb\int dE\,\int dE'\frac{\cos(b\sqrt{2E})\cos(b\sqrt{2E'})}{\pi^2\sqrt{2E}\sqrt{2E'}}\,e^{-\tau E}e^{-(\beta-\tau)E'}|V_{E,E'}|^2
\end{align}
But we should note that fermion 2-pt correlator need to be multiplied by the Spin structure $F_T$. Define $\braket{\psi\psi}_{\chi=-1,0}$ to be the handle-disk contribution to 2-pt fermion correlator ignoring Spin/Pin$^-$ structure, i.e.
\begin{equation}
\braket{\psi\psi}_{\chi=-1,0}=e^{-S_0}\int_0^\infty bdb\int dE\,\int dE'\frac{\cos(b\sqrt{2E})\cos(b\sqrt{2E'})}{\pi^2\sqrt{2E}\sqrt{2E'}}\,e^{-\tau E}e^{-(\beta-\tau)E'}|\psi_{E,E'}|^2
\end{equation}
Again in addition to multiplying $\braket{\psi\psi}_{\chi=-1,0}$ by the spin structure $F_T$, fermions connected by NS differ by a sign. We can summarize the result in table~\ref{tabletoruspsipsi}.

\begin{table}[H]
\centering
\begin{tabular}{ c | c c }
\\
 & $\braket{\psi\psi}_\mathrm{\color{red}R}$ & $\braket{\psi\psi}_\mathrm{\color{red}NS}$\\
\hline\\
\color{green}R & $(-1)^{N\color{blue}1}$$\braket{\psi\psi}_{\chi=-1,0}$ & $(-1)^{N\color{blue}0}$$\braket{\psi\psi}_{\chi=-1,0}$ \\
\\
\color{green}NS & $(-1)^{N\color{blue}0}$$\color{green}(-1)$$\braket{\psi\psi}_{\chi=-1,0}$ & $(-1)^{N\color{blue}0}\color{green}(-1)$$\braket{\psi\psi}_{\chi=-1,0}$ \\
\\
\end{tabular}\\
\caption{handle-disk two-point function}
\label{tabletoruspsipsi}
\end{table}

\noindent Summing over spin structures, we get
\begin{equation}
\braket{\psi\psi}_{\mathrm{\color{red}R},\text{handle-disk}}=((-1)^N-1)\braket{\psi\psi}_{\chi=-1,0}\quad\quad \braket{\psi\psi}_{\mathrm{\color{green}NS},\text{handle-disk}}=0
\end{equation}
so there is no contribution from handle-disk for $N$ even.

\begin{figure}[H]
\centering
\includegraphics[width=0.5\textwidth]{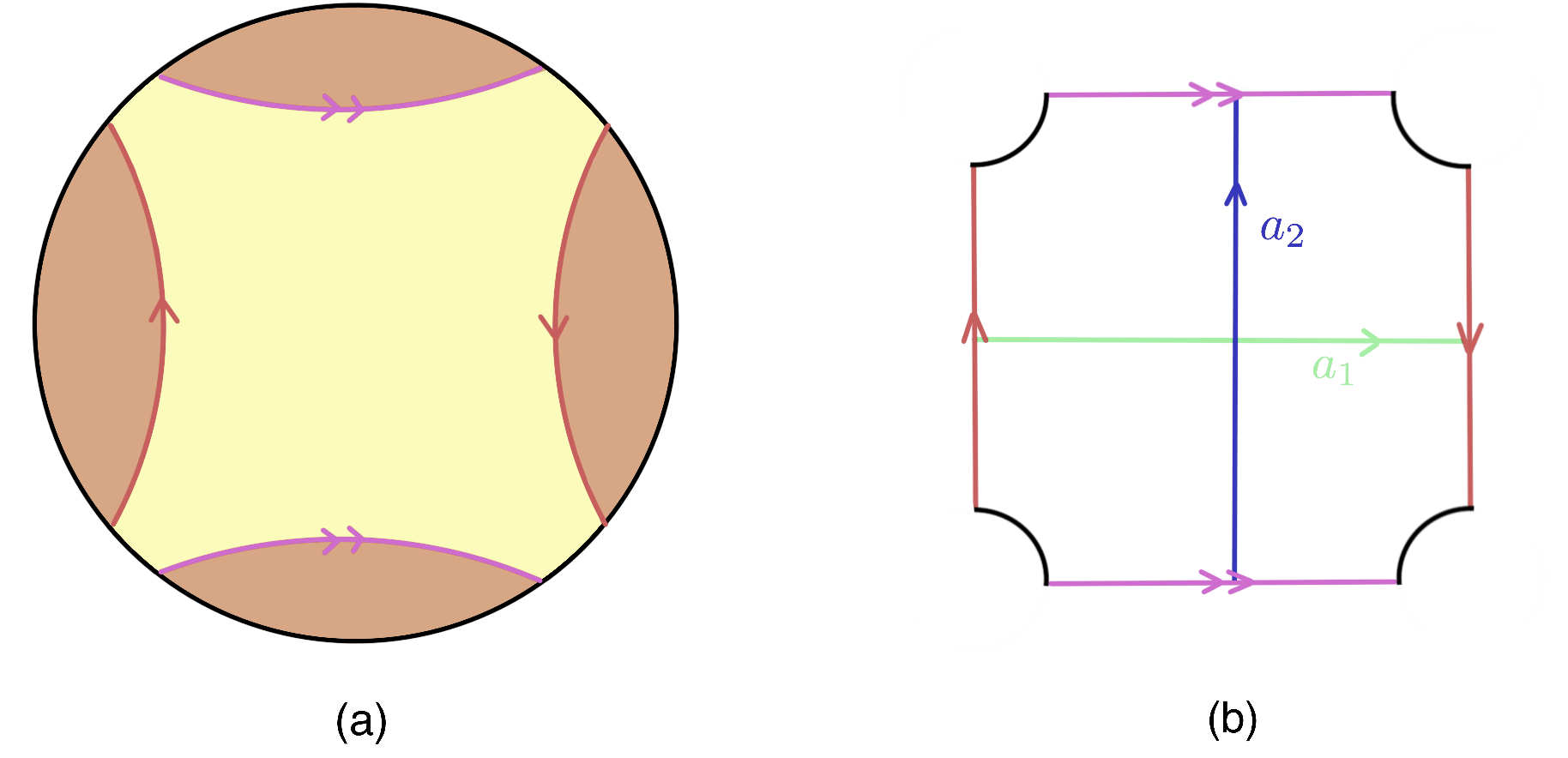}
\caption{(a) a reflected-handle-disk (Klein-bottle with a hole) represented on a hyperbolic disk (b) square diagram of a reflected-handle-disk with $a_1$ and $a_2$ generators of $\pi_1$.}
\label{kbMCG}
\end{figure}

In the non-orientable case, there is not only a handle-disk but also a reflected-handle-disk (rhd) as drawn in figure~\ref{kbMCG}. The mapping class group of such a geometry is generated by a Dehn twist $t_2:\{a_1\mapsto a_1a_2,a_2\mapsto a_2\}$, and two reflections $y:\{a_1\mapsto a_1^{-1},a_2\mapsto a_2\}$ and $\omega_1:\{a_1\mapsto a_1,a_2\mapsto a_2^{-1}\}$. (For more information on MCG see \cite{gomez2017classification}.)

\noindent There are four possible types of geodesics on a reflected-handle-disk, we analyze them one by one. Note that we will show that only the first type contributes to the ramp.

\textbf{(1)} A geodesic that goes through two crosscaps

\begin{figure}[H]
\centering
\includegraphics[width=0.6\textwidth]{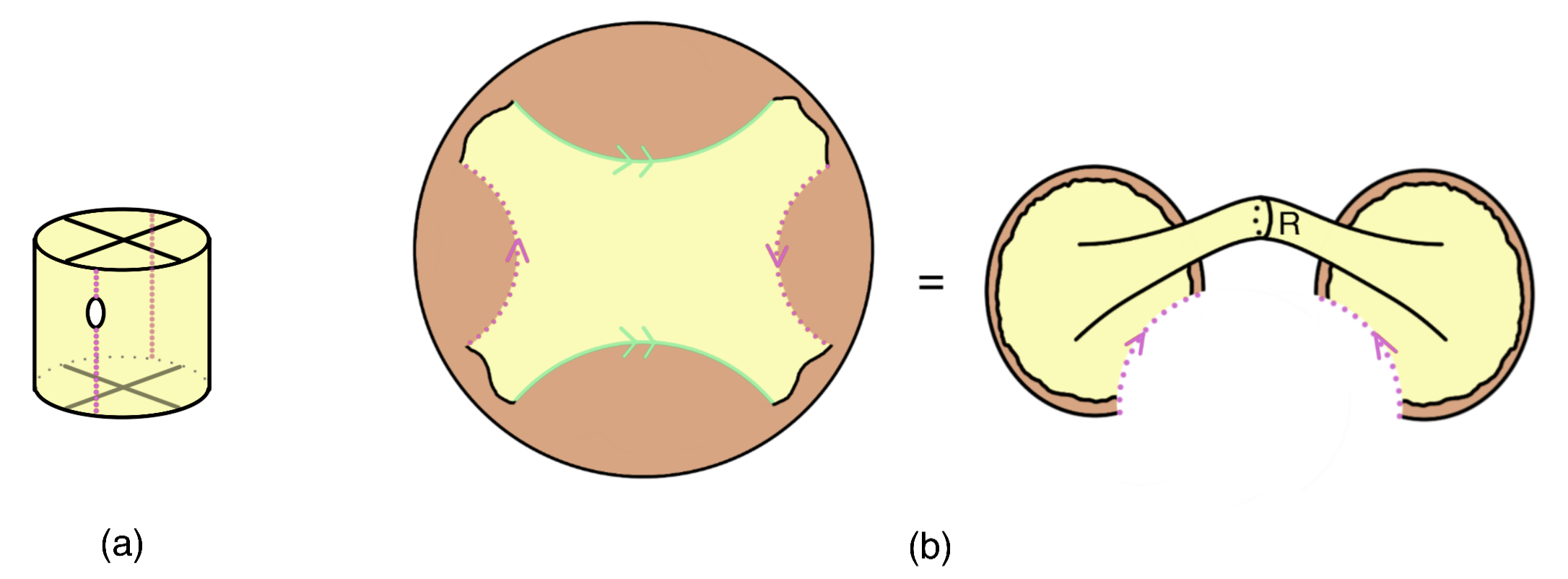}
\caption{(a) On a reflected-handle-disk, type (1) geodesic goes through two crosscaps (b) draw a reflected-handle-disk as a quotient of a hyperbolic disk}
\end{figure}

Elements of MCG that preserves this configuration are $t_2$ and $y$. Since $y$ generates a finite group, we need to focus on $t_2$. The resulting integral is the same as the case for handle-disk, so we get the same two-point function as handle disk $\braket{\psi\psi}_{\chi=-1,0}$ if we do not consider the Pin$^-$ structure. With the Pin$^-$ structure, from section 3.2.1, we know that the geodesic going through one crosscap would change the two-point correlation function from $\braket{\psi\psi}$ to $2\sin\left(\frac{2\pi N}{8}\right)\braket{\psi \psi}$. Thus naturally, geodesic going through two crosscaps would repeat this procedure twice on a fermionic two-point function, i.e.
\begin{equation}
\braket{\psi\psi}_{\text{rhd}(1)}=4\sin^2\left(\frac{2\pi N}{8}\right)\braket{\psi\psi}_{\chi=-1,0}
\end{equation}

\pagebreak

\textbf{(2)} A geodesic that goes through one crosscap

\begin{figure}[H]
\centering
\includegraphics[width=0.8\textwidth]{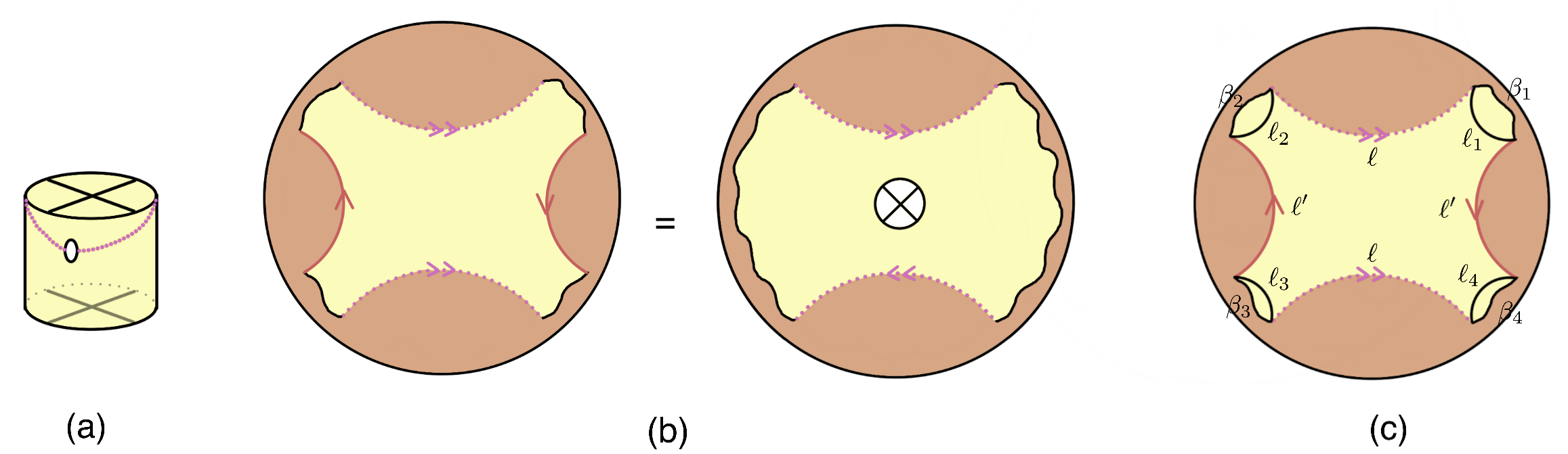}
\caption{(a) On a reflected-handle-disk, type (2) geodesic goes through one crosscap (b) draw a reflected-handle-disk as a quotient of a hyperbolic disk (c) dividing the geometry into pieces we know the function forms of}
\label{kb2calc}
\end{figure}

This geometry can be thought of as a crosscap correction to the disk+one crosscap contribution to two-point correlators. We can divide this geometry into five pieces as in figure~\ref{kb2calc}(c). Then its contribution to the two-point correlation function without considering the Pin$^-$ structure is 

\begin{align}
\braket{\psi\psi}_{\text{rhd}(2),0}&=e^{-S_0}\int d\ell d\ell'e^{-\Delta \ell}\int d\ell_1d\ell_2d\ell_3d\ell_4\,e^{\ell_1/2+\ell_2/2+\ell_3/2+\ell_4/2}I(\ell_1,\ell,\ell_2,\ell',\ell_3,\ell,\ell_4,\ell')\nonumber\\
&\quad\times \varphi_{\text{Disk},\beta_1}(\ell_1)\varphi_{\text{Disk},\beta_2}(\ell_2)\varphi_{\text{Disk},\beta_3}(\ell_3)\varphi_{\text{Disk},\beta_4}(\ell_4)\\
&=e^{-S_0}\int d\ell d\ell'e^{-\Delta \ell}dE\,e^{\ell}e^{\ell'}\rho_0(E)\varphi_E(\ell)\varphi_E(\ell)\varphi_E(\ell')\varphi_E(\ell')e^{-(\beta_1+\beta_2+\beta_3+\beta_4)E}\\
&=e^{-S_0}\int dE\,e^{-(\beta_1+\beta_2+\beta_3+\beta_4)E}|\psi_{E,E}|^2\rho_{1/2}(E)
\end{align}

The only MGC element that preserves this configuration is $\omega_1$ but this only generates a discrete group so no further integration is needed.

Since the geodesic goes through one crosscap, again this has the effect of changing the two-point correlation function from $\braket{\psi\psi}$ to $2\sin\left(\frac{2\pi N}{8}\right)\braket{\psi \psi }$ but at the same time we need to sum over Pin$^-$ structure of the other crosscap, which gives a multiplicative factor $F_{\text{cc}}(N)=2\cos\left(\frac{2\pi N}{8}\right)$. Thus type (2) geodesic contributes to fermionic two-point correlation function by
\begin{equation}
\braket{\psi\psi}_{\text{rhd}(2)}=4\sin\left(\frac{2\pi N}{8}\right)\cos\left(\frac{2\pi N}{8}\right)\braket{\psi \psi }_{\text{rhd}(2),0}
\end{equation}
In particular, this equals to zero for $N$ even. For odd $N$, we should note that this contribution is of order $e^{-2S_0}$ after normalization and is independent of time, so it will still be subleading.

\pagebreak

\textbf{(3)} A geodesic that goes through no crosscap and divide the two crosscaps into two disconnected pieces

\begin{figure}[H]
\includegraphics[width=\textwidth]{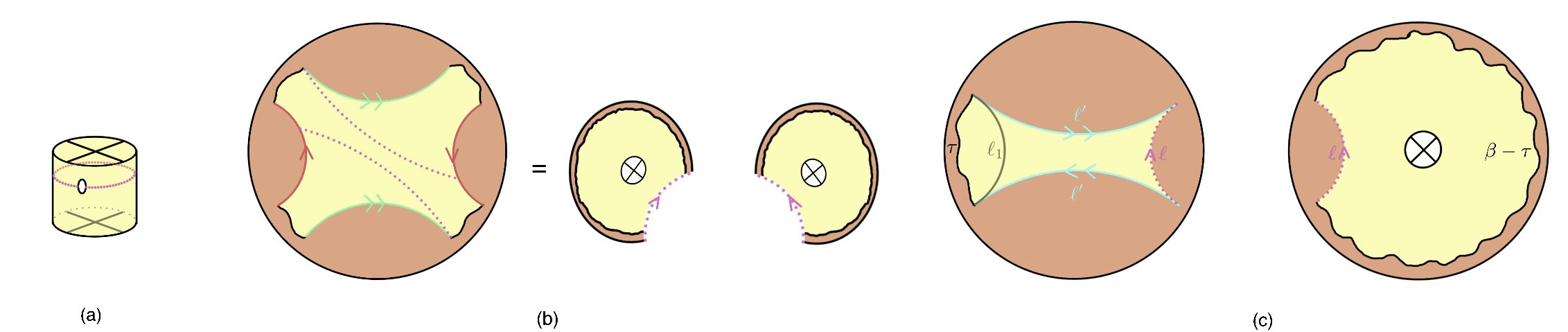}
\caption{(a) On a reflected-handle-disk, type (3) geodesic goes through no crosscap and divide the two crosscaps into two disconnected pieces (b) draw a reflected-handle-disk as a quotient of a hyperbolic disk (c) dividing the geometry into pieces we know the function forms of}
\label{kb3calc}
\end{figure}

To compute the 2-point function we can divide the configuration as in figure~\ref{kb3calc}(c), i.e. We divide this config into left part and right part. We can first write out the left part as
\begin{align}
\mathrm{LeftPart}&=\int d\ell'd\ell_1\,e^{\ell_1/2}I(\ell,\ell',\ell_1,\ell')\psi^{HH}_\tau(\ell_1)\\
&=\int d\ell'd\ell_1\,e^{\ell_1+\ell'+\ell/2}dEdE'\,\rho_0(E)\rho_0(E')\varphi_E(\ell)\varphi_E(\ell')\varphi_E(\ell_1)\varphi_E(\ell')\varphi_{E'}(\ell_1)e^{-\tau E'}\\
&=e^{\ell/2}\int dE\,e^{-\tau E}\varphi_E(\ell)\rho_{1/2}(E)
\end{align}
Right part is similar with $\tau$ changed to $\beta-\tau$. Then ignoring Pin$^-$ structure, we get
\begin{align}
\braket{\psi\psi}_{\text{rhd}(3),0}&=e^{-S_0}\int d\ell e^{-\Delta\ell}\mathrm{LeftPart}\cdot\mathrm{RightPart}\\
&=e^{-S_0}\int dE_1dE_2\int d\ell\,e^{\ell}e^{-\Delta\ell}e^{-\tau E_1}\psi_{E_1}(\ell_1)\delta(0)e^{-(\beta-\tau) E_2}\psi_{E_2}(\ell)\delta(0)\\
&=e^{-S_0}\int dE_1dE_2\,|\psi_{E_1,E_2}|^2e^{-\tau E_1}e^{-(\beta-\tau) E_2}\rho_{1/2}(E_1)\rho_{1/2}(E_2)
\end{align}
If $\tau$ is large and imaginary, this is dominated by $E_1,E_2$ close to zero. In that case $|\psi_{E_1,E_2}|^2$ approaches a constant and so we can do the integral approximately
\begin{equation}
\braket{\psi\psi}_{\text{rhd}(3),0}\sim e^{-S_0}\int dE_1dE_1e^{-\tau E_1}e^{-(\beta-\tau)E_2}\rho_{1/2}(0)^2|\psi_{0,0}|^2\sim\frac{e^{-S_0}}{t^2}\rho_{1/2}(0)^2|\psi_{0,0}|^2
\end{equation}
Thus this answer decays with time. The Pin$^-$ structure just gives a multiplicative factor of $F_{cc}(N)^2$ so the whole thing decays with time.

\pagebreak

\textbf{(4)} A geodesic that goes through no crosscap and divide the geometry into two disconnected pieces with the two crosscaps both in one piece

\begin{figure}[H]
\centering
\includegraphics[width=0.6\textwidth]{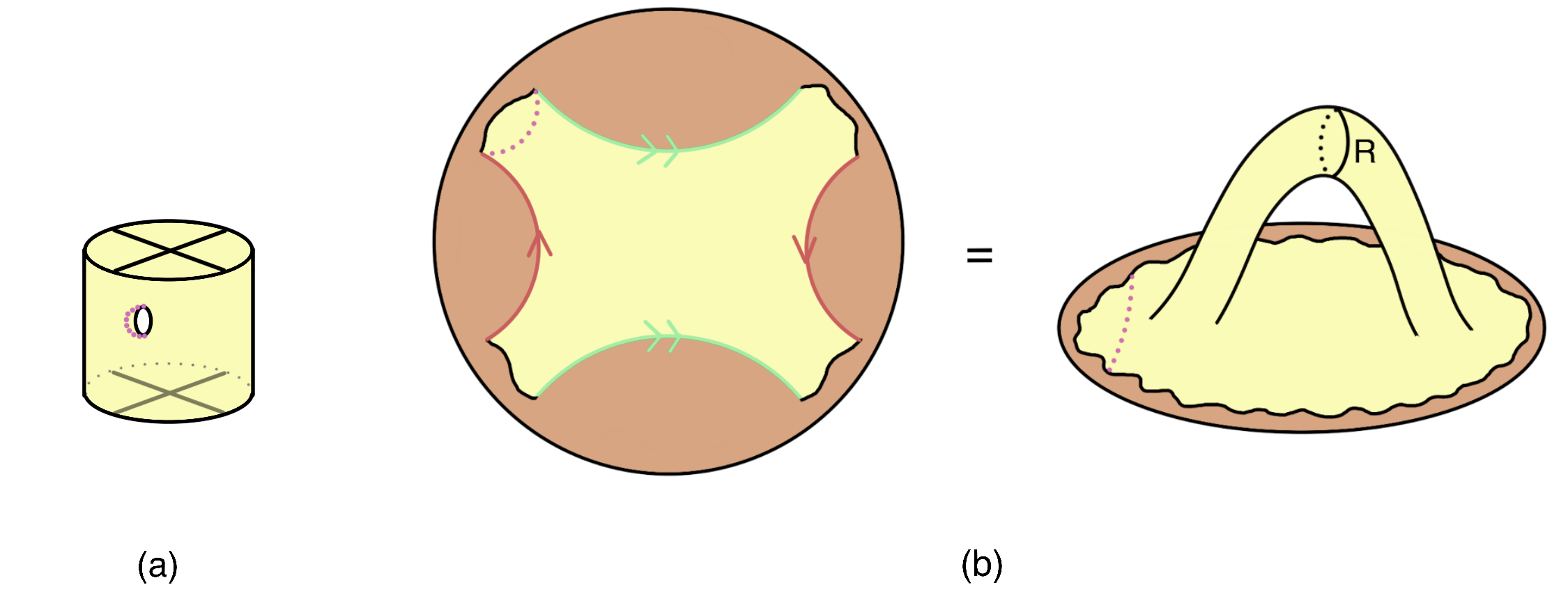}
\caption{(a) On a reflected-handle-disk, type (4) geodesic goes through no crosscap and divide the geometry into two connected pieces with the two crosscaps both in one piece (b) draw a reflected-handle-disk as a quotient of a hyperbolic disk (c) dividing the geometry into pieces we know the function forms of}
\end{figure}

For this configuration, the geodesic doesn't explore the non-trivial geometry, so this contribution to the two-point correlation function should also decay with time.

\noindent Summarizing theses four cases, we know that for even $N$, the ramp contribution to fermionic two-point correlation function is given by (1)
\begin{equation}
\braket{\psi\psi}_{\text{rhd}(1)}=4\sin^2\left(\frac{2\pi N}{8}\right)\braket{\psi\psi}_{\chi=-1,0}\label{kbpin}
\end{equation}
and the plateau contribution is given by one-crosscap 
\begin{equation}
\braket{\psi\psi}_{cc}=2\sin\left(\frac{2\pi N}{8}\right)\braket{\psi \psi }_{\chi=0,0}\label{ccpin}
\end{equation}
combined with the already existing plateau in oriented case.

\subsection{RMT}
In the previous section, we computed the contribution to the two-point function from genus-$1/2$ and genus-$1$ surfaces, with the sum over Pin$^-$ structures weighted by a topological field theory. We would like to compare this to the predictions of a corresponding random matrix ensemble. 

The SYK model with $N$ Majorana fermions is a convenient stepping stone that allows us to relate a particular topological field theory to a particular random matrix symmetry class. This is because, on one hand, the boundary SYK model has the same anomalies as the bulk $e^{-i N\pi\eta/2}$, and on the other hand, the algebra of time reversal and $(-1)^F$ operators in SYK determines a RMT symmetry class \cite{StanfordWitten19}.

The action of SYK model is given by
\begin{equation}
I=\int dt\,\left(\frac{i}{2}\sum_k\psi_k\frac{d\psi_k}{dt}-H\right)
\end{equation}
where H is the SYK Hamiltonian (\ref{syk}). Classically, the fermions $\psi_k$ are treated as Grassmann numbers and the action naturally has two symmetries: the symmetry $(-1)^F$ acting by $\psi_k\mapsto-\psi_k$, and the Lorentzian time-reversal symmetry $T$ acting by $\psi_k\mapsto\psi_k$. Classically, these two symmetries satisfy the relations
\begin{equation}
T^2=1\quad\quad T(-1)^F=(-1)^FT
\end{equation} 
However, these relations are not always satisfied quantum mechanically if we quantize the Hamiltonian and treat the fermions as forming a Clifford algebra. Furthermore, the symmetry $(-1)^F$ cannot be defined for odd $N$. It turns out the anomalies of $T$ and $(-1)^F$ depend on $N\mod8$ and can be summarized in table~\ref{Nmod8}. The last column shows the corresponding random matrix classes as a function of $N\mod8$.

\begin{table}[h]
\centering
\begin{tabular}{ c c c c c c c }
\\
N & $(-1)^F$ & $T^2$ & $T(-1)^F$ & $T\psi_k T^{-1}$ & RMT\\
\hline\\
$0\mod8$& $\checkmark$ & 1 & $(-1)^FT$ & $\psi_k$ & $\begin{pmatrix}\mathrm{GOE}_1&\\&\mathrm{GOE}_2\end{pmatrix}$ \\
\\
$1\mod8$& $\times$ & 1 & N/A & $\psi_k$ & GOE \\
\\
$2\mod8$& $\checkmark$ & 1 & $-(-1)^FT$ & $\psi_k$ & $\begin{pmatrix}\mathrm{GUE}&\\&\mathrm{GUE}\end{pmatrix}$ \\
\\
$3\mod8$& $\times$ & -1 & N/A & $-\psi_k$ & GSE \\
\\
$4\mod8$& $\checkmark$ & -1 & $(-1)^FT$ & $\psi_k$ & $\begin{pmatrix}\mathrm{GSE}_1&\\&\mathrm{GSE}_2\end{pmatrix}$ \\
\\
$5\mod8$& $\times$ & -1 & N/A & $\psi_k$ & GSE \\
\\
$6\mod8$& $\checkmark$ & -1 & $-(-1)^FT$ & $\psi_k$ & $\begin{pmatrix}\mathrm{GUE}&\\&\mathrm{GUE}\end{pmatrix}$ \\
\\
$7\mod8$& $\times$ & 1 & N/A & $-\psi_k$ & GOE \\
\\
\end{tabular}
\caption{Classification of $T$ and $(-1)^F$ anomalies and the corresponding random matrix ensemble (see \cite{StanfordWitten19} for more information).}
\label{Nmod8}
\end{table}

We are able to match all $8$ cases between JT result and RMT calculation, we now show the detailed calculations one by one.

\subsubsection*{N even}

For even $N$, we can choose a basis for the Hilbert space of dimension $2L\times 2L$ such that
\begin{equation}
(-1)^F=\begin{pmatrix}1&0\\0&-1\end{pmatrix}
\end{equation}
and then the constraint that $H$ should commute with $(-1)^F$ tells us that
\begin{equation}
H=\begin{pmatrix}H_1&0\\0&H_2\end{pmatrix}\in\begin{pmatrix}\mathrm{GUE}_1&\\&\mathrm{GUE}_2\end{pmatrix}
\end{equation}
where the two subscripts mean independent random matrix ensembles. Further a Hermitian fermion which anticommute with $(-1)^F$ should look like
\begin{equation}
\psi=\begin{pmatrix}0&\lambda\\\lambda^\dagger&0\end{pmatrix}
\end{equation}
Then depending on $T^2$ and $T$\&$(-1)^F$ commutation relation, we can determine the form of $T$ for each case of $N$mod$8$. With this $T$, the condition that $H$ commutes with $T$ then further constrain the form of the Hamiltonian. 


If $\mathbf{N=0\,mod\,8}$, $T^2=1$ and $T(-1)^F=(-1)^FT$, so we can just take 
\begin{equation}
T=K\begin{pmatrix}1&0\\0&1\end{pmatrix}
\end{equation} 
Then $THT^{-1}=H$ gives
\begin{equation}
H=\begin{pmatrix}H_1&0\\0&H_2\end{pmatrix}\in\begin{pmatrix}\mathrm{GOE}_1&\\&\mathrm{GOE}_2\end{pmatrix}
\end{equation} 
The time evolution operator is then given by
\begin{equation}
U(t)=e^{-iHt}=\begin{pmatrix}e^{-iH_1t}&0\\0&e^{-iH_2t}\end{pmatrix}
\end{equation}
then the two-point function is given by
\begin{equation}
\braket{\psi(t)\psi(0)}=\braket{U(t)^\dagger\lambda(0)U(t)\lambda(0)}=\braket{\begin{pmatrix}\lambda e^{iH_2t}\lambda^\dagger e^{-iH_1t}&0\\0&\lambda^\dagger e^{iH_1t}\lambda e^{-iH_2t}\end{pmatrix}}
\end{equation}
$e^{\pm iH_1t}$ and $e^{\pm iH_2t}$ are independent, so the ensemble average of $\braket{\psi(t)\psi(0)}$ is zero. This is consistent with the fact that on the JT side for $N=0\mod8$ both (\ref{kbpin}) and (\ref{ccpin}) give zero. These all together confirms SYK numerics result that there is no ramp and no plateau (see table~\ref{SYKresult}).

If $\mathbf{N=2\,mod\,8}$, $T^2=1$ and $T(-1)^F=-(-1)^FT$ so we take
\begin{equation}
T=K\begin{pmatrix}0&1\\1&0\end{pmatrix}
\end{equation}
Then $THT^{-1}=H$ gives $H_1=H_2^*$, so
\begin{equation}
H=\begin{pmatrix}H_1&0\\0&H_1^*\end{pmatrix}\in\begin{pmatrix}\mathrm{GUE}&\\&\mathrm{GUE}\end{pmatrix}
\end{equation} 
where we have removed the subscripts saying that the two blocks are the same GUE ensemble. Let $h$ be real and diagonal then we can model time evolution using Haar random unitary matrices $u$ as
\begin{equation}
U(t)=e^{-iHt}=\begin{pmatrix}u^\dagger&0\\0&u^T\end{pmatrix}\begin{pmatrix}e^{-iht}&0\\0&e^{-iht}\end{pmatrix}\begin{pmatrix}u&0\\0&u^*\end{pmatrix}
\end{equation}
Having the form of $T$ we can also act on $\psi$ by it
\begin{equation}
T\psi T^{-1}=\begin{pmatrix}0&\lambda^T\\\lambda^*&0\end{pmatrix}
\end{equation}
Therefore, we can write a two-point function of fermions as
\begin{align}
&\int du\,\braket{\psi(t)\psi(0)}\nonumber\\
=&\int du\,\braket{U(t)^\dagger\lambda(0)U(t)\lambda(0)}\\
=&\int du\,\braket{\begin{pmatrix}u^\dagger &0\\0&u^T\end{pmatrix}\begin{pmatrix}e^{iht}&0\\0&e^{iht}\end{pmatrix}\begin{pmatrix}u&0\\0&u^*\end{pmatrix}\begin{pmatrix}0&\lambda\\\lambda^\dagger&0\end{pmatrix}\begin{pmatrix}u^\dagger&0\\0&u^T\end{pmatrix}\begin{pmatrix}e^{-iht}&0\\0&e^{-iht}\end{pmatrix}\begin{pmatrix}u&0\\0&u^*\end{pmatrix}\begin{pmatrix}0&\lambda\\\lambda^\dagger&0\end{pmatrix}}\\
=&\int du\,\braket{u^\dagger e^{iht}u\lambda u^Te^{-iht}u^*\lambda^\dagger}+h.c.
\end{align}
Using formula (\ref{weingartenu}) we get
\begin{align}
\int du\,\braket{\psi(t)\psi(0)}&=\frac{2L^2}{L^2-1}\braket{\lambda\lambda^\dagger}\braket{e^{iht}}\braket{e^{-iht}}+\frac{2L}{L^2-1}\braket{\lambda\lambda^*}\nonumber\\
&\quad-\frac{1}{L^2-1}\braket{\lambda\lambda^\dagger}(\braket{e^{2iht}}+\braket{e^{-2iht}})-\frac{2L}{L^2-1}\braket{\lambda\lambda^*}\braket{e^{iht}}\braket{e^{-iht}}\\
&=\frac{L^2}{L^2-1}\braket{\psi\psi}\braket{e^{iht}}\braket{e^{-iht}}+\frac{L}{L^2-1}\braket{\psi T\psi T^{-1}}\nonumber\\
&\quad-\frac{1}{2(L^2-1)}\braket{\psi\psi}(\braket{e^{2iht}}+\braket{e^{-2iht}})-\frac{L}{L^2-1}\braket{\psi T\psi T^{-1}}\braket{e^{iht}}\braket{e^{-iht}}
\end{align}
Using the fact that $T\psi T^{-1}=\psi$ and in the limit $L\rightarrow\infty$
\begin{equation}
\int du\,\braket{\psi(t)\psi(0)}\approx \underbrace{\braket{\psi\psi}\braket{e^{iht}}\braket{e^{-iht}}}_{\text{ramp+plateau}}+\underbrace{\frac{1}{L}\braket{\psi \psi }}_{\text{crosscap offset}}\label{2mod8}
\end{equation}
Note that here $H$ is of dimension $2L\times 2L$, so we can identify $2L$ with $\rho_0(E)e^{S_0}$. Also we can identify $\rho_0(E)|\psi_{E,E}|^2=\braket{E|\psi\psi|E}$. Thus using (\ref{ramprelation}) the first term of (\ref{2mod8}) reduces to $\min\{t/L^2,1/L\}\braket{\psi\psi}$. The early time (ramp) part identifies with the reflected-handle-disk contribution, note that when $N=2\mod8$ the Pin$^-$ structure in (\ref{kbpin}) gives a prefactor 4
\begin{equation}
\frac{\braket{\psi\psi}_{\text{rhd}(1)}}{\braket{1}_{\mathrm{disk}}}=\frac{4\braket{\psi\psi}_{\chi=-1,0}}{\int dE\,e^{S_0}\rho_0(E)e^{-\beta E}}\sim \frac{4t}{(\rho_0(E)e^{S_0})^2}\braket{E|\psi\psi|E}\sim \frac{t}{L^2}\braket{E|\psi\psi|E}
\end{equation}
where we have used the results of handle-disk \cite{Saadsingleauthor}. The second term identifies with the contribution from crosscap, note that when $N=2\mod8$ the Pin$^-$ structure in (\ref{ccpin}) gives a prefactor 2.
\begin{equation}
\frac{\braket{\psi\psi}_{cc}}{\braket{1}_{\mathrm{disk}}}=2\frac{\int dE\,e^{-\beta E}\braket{E|\psi \psi |E}}{\int dE\,e^{S_0}\rho_0(E)e^{-\beta E}}\sim \frac{2}{\rho_0(E)e^{S_0}}\braket{E|\psi \psi |E}\sim \frac{1}{L}\braket{E|\psi \psi |E}
\end{equation}
The crosscap contribution doubles the plateau. Thus our computation confirms the SYK numerics result that there is a ramp and a plateau (see table~\ref{SYKresult}).

If $\mathbf{N=4\,mod\,8}$, $T^2=-1$ and $T(-1)^F=(-1)^FT$, so we can just take 
\begin{equation}
T=K\omega\begin{pmatrix}1&0\\0&1\end{pmatrix}
\end{equation} 
Then $THT^{-1}=H$ gives
\begin{equation}
H=\begin{pmatrix}H_1&0\\0&H_2\end{pmatrix}\in\begin{pmatrix}\mathrm{GSE}_1&\\&\mathrm{GSE}_2\end{pmatrix}
\end{equation} 
As for $N=0\mod 8$, $T$ does not mix the two blocks of $H$ so the ensemble average of $\braket{\psi(t)\psi(0)}$ is again zero. This is consistent with the fact that on the JT side for $N=4\mod8$ both (\ref{kbpin}) and (\ref{ccpin}) give zero. These all together confirms SYK numerics result that there is no ramp and no plateau (see table~\ref{SYKresult}).


If $\mathbf{N=6\,mod\,8}$, $T^2=-1$ and $T(-1)^F=-(-1)^FT$ so we take
\begin{equation}
T=K\begin{pmatrix}0&1\\-1&0\end{pmatrix}
\end{equation}
Then $THT^{-1}=H$ gives the same Hamiltonian and time evolution as $N=2\mod 8$. But $T$ acts on $\psi$ differently 
\begin{equation}
T\psi T^{-1}=-\begin{pmatrix}0&\lambda^T\\\lambda^*&0\end{pmatrix}
\end{equation}
Thus the two-point function is given by
\begin{align}
\int du\,\braket{\psi(t)\psi(0)}&=\frac{2L^2}{L^2-1}\braket{\lambda\lambda^\dagger}\braket{e^{iht}}\braket{e^{-iht}}+\frac{2L}{L^2-1}\braket{\lambda\lambda^*}\nonumber\\
&\quad-\frac{1}{L^2-1}\braket{\lambda\lambda^\dagger}(\braket{e^{2iht}}+\braket{e^{-2iht}})-\frac{2L}{L^2-1}\braket{\lambda\lambda^*}\braket{e^{iht}}\braket{e^{-iht}}\\
&=\frac{L^2}{L^2-1}\braket{\psi\psi}\braket{e^{iht}}\braket{e^{-iht}}-\frac{L}{L^2-1}\braket{\psi T\psi T^{-1}}\nonumber\\
&\quad-\frac{1}{2(L^2-1)}\braket{\psi\psi}(\braket{e^{2iht}}+\braket{e^{-2iht}})+\frac{L}{L^2-1}\braket{\psi T\psi T^{-1}}\braket{e^{iht}}\braket{e^{-iht}}
\end{align}
Using the fact that $T\psi T^{-1}=\psi$ and in the limit $L\rightarrow\infty$
\begin{equation}
\int du\,\braket{\psi(t)\psi(0)}\approx \underbrace{\braket{\psi\psi}\braket{e^{iht}}\braket{e^{-iht}}}_{\text{ramp+plateau}}\underbrace{-\frac{1}{L}\braket{\psi \psi }}_{\text{crosscap offset}}\label{6mod8}
\end{equation}
The first term is the same as (\ref{2mod8}) and again identifies with reflected-handle-disk in early time and gives a ramp. But here the second term is negative. Note that it identifies with the contribution from crosscap because the Pin$^-$ structure now gives a prefactor $-2$ which can be seen in (\ref{ccpin}). 
\begin{equation}
\frac{\braket{\psi\psi}_{cc}}{\braket{1}_{\mathrm{disk}}}=-2\frac{\int dE\,e^{-\beta E}\braket{E|\psi \psi |E}}{\int dE\,e^{S_0}\rho_0(E)e^{-\beta E}}\sim -\frac{2}{\rho_0(E)e^{S_0}}\braket{E|\psi \psi |E}\sim -\frac{1}{L}\braket{E|\psi \psi |E}
\end{equation}
The crosscap contribution cancels the plateau. This confirms the SYK numerics result that there is a ramp but no plateau (see table~\ref{SYKresult}).

\subsubsection*{N odd}

For odd $N$, $(-1)^F$ cannot be defined, so naively when we characterize the anomalies we only need to consider $T^2$. But when $(-1)^F$ is defined we can always choose between $T$ and $T(-1)^F$ to ensure that $T\psi_k T^{-1}=\psi_k$ in SYK but this is no longer true when $(-1)^F$ cannot be defined, so there are also anomalies in the commutation relation of $\psi_k$ and $T$. Since we don't constrain our Hamiltonian using $(-1)^F$, $H$ has only one block. Before constraining with $THT^{-1}=H$, our Hamiltonian looks like $H\in\mathrm{GUE}$ of dimension $L\times L$. The calculations become similar to the bosonic cases in section \ref{bosonicRMT}. 


If $\mathbf{N=1\,mod\,8}$, $T^2=1$ so we take $T=K$. $THT^{-1}=H$ gives $H\in\mathrm{GOE}$, so we get the same result as (\ref{VVo}) with $\braket{VV}\mapsto\braket{\psi\psi}$ and $\braket{VV^T}\mapsto\braket{\psi \psi^T}$. Assuming $\psi$ is Hermitian, $T\psi T^{-1}=\psi$ gives $\psi=\psi^T$, so the two-point correlator is
\begin{equation}
\int do\,\braket{\psi(t)\psi(0)}\approx\underbrace{\braket{e^{iht}}\braket{e^{-iht}}\braket{\psi\psi}}_{\text{ramp+plateau}}+\underbrace{\frac{1}{L}\braket{\psi\psi}}_{\text{crosscap offset}}\label{1mod8}
\end{equation}
Recall (\ref{ramprelation}), so the first term gives approximately $\min\{t/L^2,1/L\}\braket{\psi\psi}$. For fermions, \cite{StanfordWitten19} showed that the partition function scales like $\sqrt{2}$ times the dimension of the Hilbert space $L$, so we should identify $\sqrt{2}L$ with $\rho_0(E)e^{S_0}$. The early time (ramp) part identifies with the reflected-handle-disk contribution, note that when $N=1\mod8$ the Pin$^-$ structure in (\ref{kbpin}) gives a prefactor 2 so
\begin{equation}
\frac{\braket{\psi\psi}_{\text{rhd}(1)}}{\braket{1}_{\mathrm{disk}}}=\frac{2\braket{\psi\psi}_{\chi=-1,0}}{\int dE\,e^{S_0}\rho_0(E)e^{-\beta E}}\sim \frac{2t}{(\rho_0(E)e^{S_0})^2}\braket{E|\psi\psi|E}\sim \frac{t}{L^2}\braket{E|\psi\psi|E}
\end{equation}
matching RMT. For disk+crosscap contribution in JT, when $N=1\mod8$ the Pin$^-$ structure in (\ref{ccpin}) gives a prefactor $\sqrt{2}$.
\begin{equation}
\frac{\braket{\psi\psi}_{cc}}{\braket{1}_{\mathrm{disk}}}=\sqrt{2}\frac{\int dE\,e^{-\beta E}\braket{E|\psi \psi |E}}{\int dE\,e^{S_0}\rho_0(E)e^{-\beta E}}\sim \frac{\sqrt{2}}{\rho_0(E)e^{S_0}}\braket{E|\psi \psi |E}\sim \frac{1}{L}\braket{E|\psi \psi |E}
\end{equation}
matching RMT result. 


If $\mathbf{N=3\,mod\,8}$, $T^2=-1$ so we take $T=K\omega$. $THT^{-1}=H$ gives $H\in\mathrm{GSE}$, so we get the same result as (\ref{VVs}) with $\braket{VV}\mapsto\braket{\psi\psi}$ and $\braket{V \omega V^T \omega^{-1}}\mapsto\braket{\psi \omega \psi^T \omega^{-1}}$. Assuming $\psi$ is Hermitian, $T\psi T^{-1}=-\psi$ gives $\omega \psi^T\omega^{-1}=-\psi$, so the two-point correlator is 
\begin{equation}
\int ds\,\braket{\psi(t)\psi(0)}\approx\underbrace{\braket{e^{iht}}\braket{e^{-iht}}\braket{\psi\psi}}_{\text{ramp+plateau}}+\underbrace{\frac{1}{L}\braket{\psi\psi}}_{\text{crosscap offset}}\label{3mod8}
\end{equation}
RMT results and JT results are both the same as $N=1\mod8$, so they match.


If $\mathbf{N=5\,mod\,8}$, $T^2=-1$ so we take $T=K\omega$. $THT^{-1}=H$ gives $H\in\mathrm{GSE}$, so we get the same result as (\ref{VVs}) with $\braket{VV}\mapsto\braket{\psi\psi}$ and $\braket{V \omega V^T \omega^{-1}}\mapsto\braket{\psi \omega \psi^T \omega^{-1}}$. Assuming $\psi$ is Hermitian, $T\psi T^{-1}=\psi$ gives $\omega \psi^T\omega^{-1}=\psi$, so the two-point correlator is 
\begin{equation}
\int ds\,\braket{\psi(t)\psi(0)}\approx\underbrace{\braket{e^{iht}}\braket{e^{-iht}}\braket{\psi\psi}}_{\text{ramp+plateau}}\underbrace{-\frac{1}{L}\braket{\psi\psi}}_{\text{crosscap offset}}\label{5mod8}
\end{equation}
The first term is the same as $N=1\mod8$ and identifies with reflected-handle-disk at early time. For disk+crosscap contribution in JT, when $N=5\mod8$ the Pin$^-$ structure in (\ref{ccpin}) gives a prefactor $-\sqrt{2}$.
\begin{equation}
\frac{\braket{\psi\psi}_{cc}}{\braket{1}_{\mathrm{disk}}}=-\sqrt{2}\frac{\int dE\,e^{-\beta E}\braket{E|\psi \psi |E}}{\int dE\,e^{S_0}\rho_0(E)e^{-\beta E}}\sim -\frac{\sqrt{2}}{\rho_0(E)e^{S_0}}\braket{E|\psi \psi |E}\sim -\frac{1}{L}\braket{E|\psi \psi |E}
\end{equation}
matching RMT result.


If $\mathbf{N=7\,mod\,8}$, $T^2=1$ so we take $T=K$. $THT^{-1}=H$ gives $H\in\mathrm{GOE}$, so we get the same result as (\ref{VVo}) with $\braket{VV}\mapsto\braket{\psi\psi}$ and $\braket{VV^T}\mapsto\braket{\psi \psi^T}$. Assuming $\psi$ is Hermitian, $T\psi T^{-1}=-\psi$ gives $\psi=-\psi^T$, so the two-point correlator is
\begin{equation}
\int do\,\braket{\psi(t)\psi(0)}\approx\underbrace{\braket{e^{iht}}\braket{e^{-iht}}\braket{\psi\psi}}_{\text{ramp+plateau}}\underbrace{-\frac{1}{L}\braket{\psi\psi}}_{\text{crosscap offset}}\label{7mod8}
\end{equation}
RMT results and JT results are both the same as $N=5\mod8$, so they match.

\section{Discussion}

We want to give one possible intuitive way of understanding why the disk+crosscap contributions to two-point correlation functions do not decay over time here. Before doing that we review an intuitive understanding of the handle-disk given by Saad \cite{Saadsingleauthor}. A handle-disk can be viewed as a baby universe being emitted by one Hartle-Hawking state and then being reabsorbed by another Hartle-Hawking state. At late time $t$, the Einstein-Rosen bridge (ERB) of a Hartle-Hawking state becomes very long (proportional to $t$), so its overlap with a second Hartle-Hawking state is small thus causing a decay with time. If a baby universe is emitted, it takes away most of ERB length and leaves behind a short ERB, so the handle-disk gives a non-decaying contribution to two-point correlation functions. The number of ways to match the two baby universes is proportional to $t$. 


A disk with a crosscap is topologically equivalent to a M\"{o}bius band. No time evolution in the traditional sense can happen in this case because if we assume there is time evolution, since a M\"{o}bius band only has one boundary, time cannot flow in one direction along the boundary. Thus the ERB doesn't grow with time and the crosscap gives a non-decaying contribution to the two-point correlation function.

On the other hand, the crosscap contribution to OTOC decays with time. If we take $\beta_1,\beta_3=\frac{\beta}{4}+it$ and $\beta_2,\beta_4=\frac{\beta}{4}-it$ we get 
\begin{align}
\braket{V(0)W(t)V(0)W(t)}_{\chi=-1,0}&=\int e^{\ell}d\ell e^{\ell'}d\ell'\,P_{\text{Disk}}(\beta_1,\beta_3,\ell,\ell')P_{\text{Disk}}(\beta_2,\beta_4,\ell,\ell')e^{-\Delta \ell}e^{-\Delta \ell'}\\
&=\int dE dE'\,\rho_0(E)\rho_0(E')e^{-(\beta_1+\beta_3)E}e^{-(\beta_2+\beta_4)E'}|V_{E,E'}|^2|W_{E,E'}|^2\\
&\sim \frac{1}{t^3}|V_{0,0}|^2|W_{0,0}|^2\quad t\rightarrow\infty
\end{align}

\begin{figure}[H]
\centering
\includegraphics[width=0.5\textwidth]{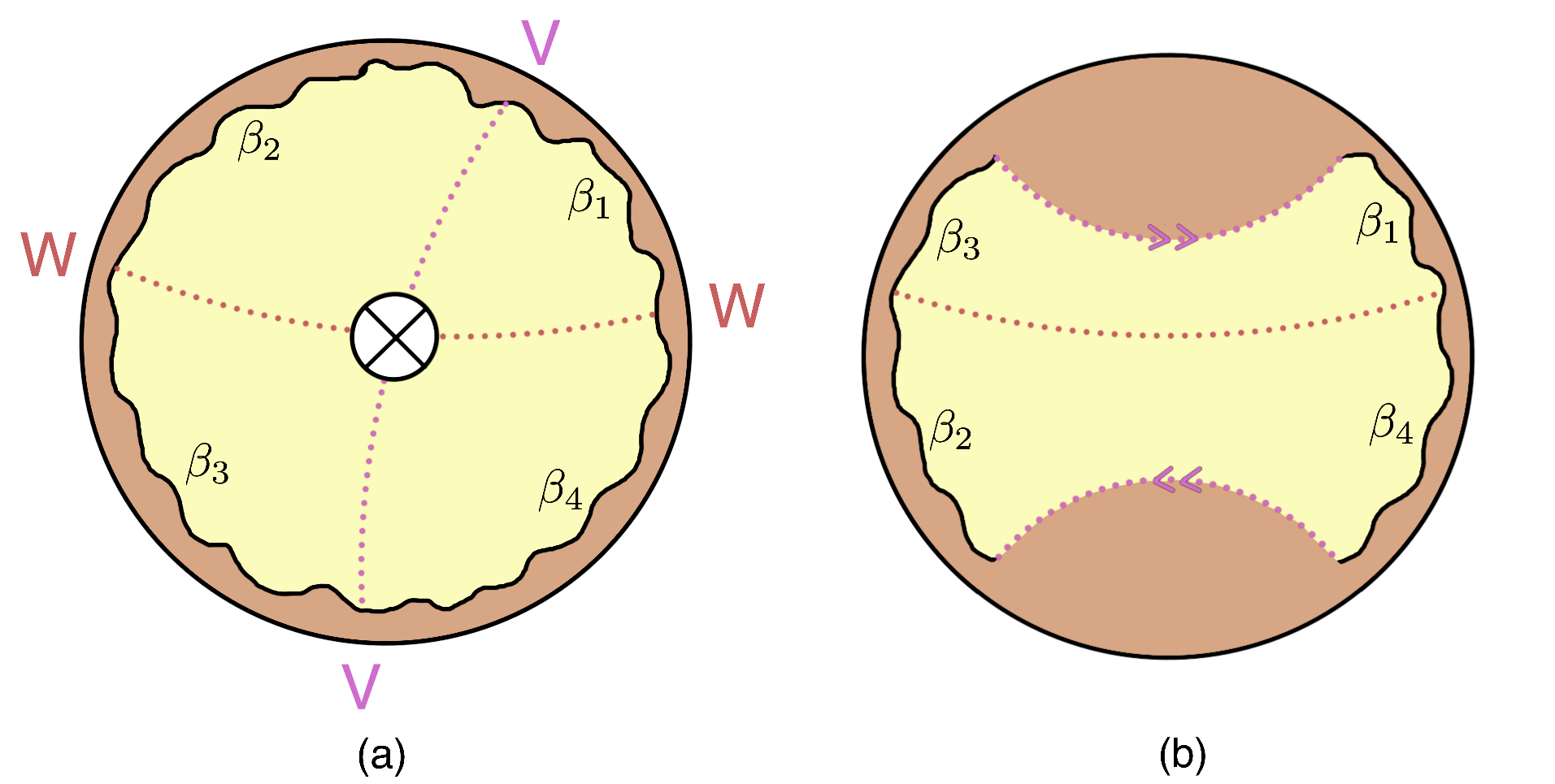}
\caption{disk+crosscap contribution to OTOC (a) the same geometry drawn as a quotient of a hyperbolic disk}
\label{otoc}
\end{figure}

This could be understood intuitively as well. In this case the time evolution has been divided into two parts, so the M\"{o}bius band structure does not hinder the definition of time anymore. Thus the growth of ERB is also restored.

\section*{Acknowledgements} 

I want to give special thanks to Douglas Stanford for patient guidance, extensive discussions, and inspiring comments throughout this project. I am also grateful to Zhenbin Yang and Shunyu Yao for discussions. 
\appendix
\section{Measures}
\label{appendixmeasure}

There are two ways of computing path integrals in JT gravity: the torsion approach in previous calculations \cite{Norbury, Gendulphe, StanfordWitten17} which uses the symplectic measure $\mu_s$ to integrate the boundary wiggles and the wavefunction approach in this paper which uses the ultralocal measure $\mu_u$ to integrate the boundary wiggles. In this Appendix, we want to find the bulk moduli we use in the wavefunction approach of a disk+crosscap by knowing the torsion approach path integral and the relationship between symplectic measure and ultralocal measure for a trumpet. 

Let's start from the very beginning by reviewing metrics on a hyperbolic disk. There are two coordinates systems of hyperbolic disk that will be useful to us (for a more detailed discussion see \cite{Stanfordwormhole}). A hyperbolic disk can be embedded in $1+2$ Minkowski space as $\{X_0,X_1,X_2\}$ such that
\begin{equation}
\mathbf{X}\cdot\mathbf{X}=-X_0^2+X_1^2+X_2^2=-1
\end{equation}
with metric
\begin{equation}
ds^2=-dX_0^2+dX_1^2+dX_2^2
\end{equation}
Note that the origin is at $(1,0,0)$. Geodesic distance $D$ between a pair of points $\mathbf{X}$ and $\mathbf{X}'$ is
\begin{equation}
\cosh D=-\mathbf{X}\cdot\mathbf{X}'
\end{equation}
A common set of coordinates used is the radius and angle such that
\begin{equation}
\mathbf{X}=(X_0,X_1,X_2)=(\cosh\rho,\sinh\rho\sin\theta,\sinh\rho\cos\theta)
\end{equation}
and the metric is given by
\begin{equation}
ds^2=d\rho^2+\sinh^2\rho\,d\theta^2
\end{equation}
Horizontal and vertical translations on the hyperbolic disk are Lorenz boosts in Minkowski space
\begin{equation}
T_1(x)=\begin{pmatrix}\cosh x&\sinh x&0\\\sinh x&\cosh x&0\\0&0&1\end{pmatrix}\quad\quad T_2(y)=\begin{pmatrix}\cosh y&0&\sinh y\\0&1&0\\\sinh y&0&\cosh y\end{pmatrix}
\end{equation}
Thus, moving the origin first to the right by $x$ then up by $y$ we get
\begin{equation}
\mathbf{X}=(X_0,X_1,X_2)=(\cosh x\cosh y,\sinh x,\cosh x\sinh y)
\end{equation}
and the metric in terms of $x$ and $y$ is given by
\begin{equation}
ds^2=dx^2+\cosh^2 x dy^2
\end{equation}
For a trumpet with its hole of geodesic length $b$, we identify $y\sim y+b$. 

\subsection{Trumpet measure}

For a disk+crosscap, the boundary wiggles we need to integrate over is the same as that of a trumpet. To relate symplectic measure with ultralocal measure, we want to find the Pfaffian of the symplectic measure of a trumpet. Again, a trumpet can be drawn as a quotient of a hyperbolic disk whose coordinates in Minkowski space $\mathbf{X}$ can be expressed in two ways
\begin{equation}
(X_1,X_2,X_3)=(\cosh\rho,\sinh\rho\sin\theta,\sinh\rho\cos\theta)=(\cosh x\cosh y,\sinh x,\cosh x\sinh y)
\end{equation}
Then for large $x$, these two sets of coordinates can be related via
\begin{equation}
\cos\theta=\tanh y\quad\Rightarrow\quad \tan\frac{\theta}{2}=e^{-y}
\end{equation}
Then we can rewrite the Schwarzian 
\begin{equation}
\mathrm{Sch}(\tan\frac{\theta}{2},\tau)=\mathrm{Sch}(e^{-y},\tau)
\end{equation}
According to \cite{StanfordWitten17} Schwarzian theory has symplectic form
\begin{equation}
\Omega=\int_0^{2\pi}d\tau\,\left[\left(\frac{dy}{y'}\right)'\wedge\left(\frac{dy}{y'}\right)''-2\mathrm{Sch}(e^{-y},\tau)\left(\frac{dy}{y'}\right)\wedge\left(\frac{dy}{y'}\right)'\right]
\end{equation}
Define
\begin{equation}
\eta=\frac{dy}{y'}\quad\quad T(\tau)=2\mathrm{Sch}(e^{-y},\tau)
\end{equation}
Then noting that there is a residual $U(1)$ symmetry for the trumpet, we can fix $\eta(0)$ and calculate the Pfaffian of the symplectic form of $\mathrm{diff}(S^1)/U(1)$
\begin{align}
\mathrm{Pf}(\Omega)&=\int\mathcal{D}\eta\,\eta(0)\exp\left(\frac{1}{2}\int d\tau\,(\eta'\eta''-T(\tau)\eta\eta')\right)\\
&=\int\mathcal{D}\eta\mathcal{D}\omega\mathcal{D}b\mathcal{D}c\,c(0)\eta(0)\exp\left(\frac{1}{2}\int d\tau\,(\omega(\eta'-b)+bb'-T(\tau)\eta b-cc')\right)\\
&=\int\mathcal{D}\eta\mathcal{D}\omega\mathcal{D}b\mathcal{D}c\,c(0)\eta(0)\exp\left(\frac{1}{2}\int d\tau\,(\omega(\eta'-b)+bb'-T(\tau)\eta b-cc')\right)\\
&=\int\mathcal{D}\eta\mathcal{D}\omega\mathcal{D}b\mathcal{D}c\,c(0)\eta(0)\nonumber\\
&\quad\quad\exp\left(-\frac{1}{2}\int d\tau\,\left(\sum_j\chi_j\chi_j'+i\chi_3(\chi_1-i\chi_2)-iT(\tau)\chi_3(\chi_1+i\chi_2)\right)\right)
\end{align}
where we have used the equation of motion
\begin{equation}
2b'+\omega+T(\tau)\eta=0
\end{equation}
and the reparametrization 
\begin{equation}
\eta=\chi_1+i\chi_2\quad\quad\omega=-\chi_1+i\chi_2\quad\quad b=i\chi_3\quad\quad c=\chi_4
\end{equation}
Using Jordan-Wigner transformation
\begin{equation}
\chi_1=\frac{1}{\sqrt{2}}X\otimes I\quad\quad\chi_2=\frac{1}{\sqrt{2}}Y\otimes I\quad\quad\chi_3=\frac{1}{\sqrt{2}}Z\otimes X\quad\quad\chi_4=\frac{1}{\sqrt{2}}Z\otimes Y
\end{equation}
where
\begin{equation}
X=\begin{pmatrix}0&1\\1&0\end{pmatrix}\quad\quad Y=\begin{pmatrix}0&-i\\i&0\end{pmatrix}\quad\quad Z=\begin{pmatrix}1&0\\0&-1\end{pmatrix}
\end{equation}
so this gives
\begin{multline}
\chi_1=\frac{1}{\sqrt{2}}\begin{pmatrix}0&0&1&0\\0&0&0&1\\1&0&0&0\\0&1&0&0\end{pmatrix}\quad \chi_2=\frac{1}{\sqrt{2}}\begin{pmatrix}0&0&-i&0\\0&0&0&-i\\i&0&0&0\\0&i&0&0\end{pmatrix}\\
\chi_3=\frac{1}{\sqrt{2}}\begin{pmatrix}0&1&0&0\\1&0&0&0\\0&0&0&-1\\0&0&-1&0\end{pmatrix}\quad \chi_4=\frac{1}{\sqrt{2}}\begin{pmatrix}0&-i&0&0\\i&0&0&0\\0&0&0&i\\0&0&-i&0\end{pmatrix}
\end{multline}
Recall that
\begin{equation}
\mathrm{STr}(e^{-\beta H})=\mathrm{Tr}((-1)^Fe^{-\beta H})=\int_Pe^{-\int_0^\beta(\overline{\psi}\dot{\psi}+H(\overline{\psi},\psi))d\tau}\mathcal{D}\psi\mathcal{D}\overline{\psi}
\end{equation}
so if we take $\beta=2\pi$, the Hamiltonian is given by
\begin{equation}
H(\tau)=\frac{1}{2}\left(i\chi_3(\chi_1-i\chi_2)-iT(\tau)\chi_3(\chi_1+i\chi_2)\right)=-\frac{i}{2}\begin{pmatrix}0&0&0&T(\tau)\\0&0&T(\tau)&0\\0&1&0&0\\1&0&0&0\end{pmatrix}
\end{equation}
Recall that the Schwarzian is given by
\begin{equation}
\mathrm{Sch}( y,\tau)=\frac{ y'''}{ y'}-\frac{3 y''^2}{2 y'^2}
\end{equation}
\begin{equation}
\mathrm{Sch}(f\circ g,t)=g'^2\mathrm{Sch}(f,g)+\mathrm{Sch}(g,t)
\end{equation}
Thus we have
\begin{align}
\mathrm{Sch}(e^{- y},\tau)&= y'^2\mathrm{Sch}(e^{- y}, y)+\mathrm{Sch}( y,\tau)\\
&=\mathrm{Sch}( y,\tau)-\frac{1}{2} y'^2\\
&=\frac{ y'''}{ y'}-\frac{3 y''^2}{2 y'^2}-\frac{1}{2} y'^2
\end{align}
We want to compute the time evolution $U(2\pi)=Pe^{-\int_0^{2\pi}H(\tau)d\tau}$ by solving the Schrodinger equation $\Psi'=-H\Psi$. The specific form of $H$ enables us to simplify the Schrodinger equation to a pair of equations  
\begin{equation}
\frac{d}{d\tau}\begin{pmatrix}\psi_1\\\psi_2\end{pmatrix}=\frac{i}{2}\begin{pmatrix}0&T(\tau)\\1&0\end{pmatrix}\begin{pmatrix}\psi_1\\\psi_2\end{pmatrix}\label{schrogingerblcok}
\end{equation}
Eliminating the top component, we get a second order equation
\begin{equation}
\psi_2''+\frac{1}{2}\mathrm{Sch}(e^{- y},\tau)\psi_2=0
\end{equation}
and $\psi_1=-2i\psi_2$. Expanding out the Schwarzian in the differential equation we get
\begin{equation}
(\sqrt{ y'}\psi_2)''-\frac{ y''}{ y'}(\sqrt{ y'}\psi_2)'-\frac{ y'^2}{4}(\sqrt{ y'}\psi_2)=0
\end{equation}
Then (\ref{schrogingerblcok}) has solutions
\begin{equation}
\begin{pmatrix}\psi_1\\\psi_2\end{pmatrix}=\frac{e^{\pm y/2}}{\sqrt{ y'}}\begin{pmatrix}-i(\pm y'-y''/y')\\1\end{pmatrix}\label{blocksol}
\end{equation}
$U(2\pi)$ should satisfy the relation
\begin{equation}
\Psi(\tau+2\pi)=U(2\pi)\Psi(\tau)
\end{equation}
In particular, we should note that $t\rightarrow\tau+2\pi$ corresponds to $y\rightarrow y+b$. Thus the solutions (\ref{blocksol}) are two eigenvectors of $U(2\pi)$ with eigenvalues $e^{\pm b/2}$. We can thus solve for $U(2\pi)$ and get
\begin{equation}
U(2\pi)=\begin{pmatrix}\cosh\frac{b}{2}-\sinh\frac{b}{2}y''/y'^2&0&0&-i\sinh\frac{b}{2}(y'^4-y''^2)/y'^3\\0&\cosh\frac{b}{2}-\sinh\frac{b}{2}y''/y'^2&-i\sinh\frac{b}{2}(y'^4-y''^2)/y'^3&0\\0&i\sinh\frac{b}{2}/y'&\cosh\frac{b}{2}+\sinh\frac{b}{2}y''/y'^2&0\\i\sinh\frac{b}{2}/y'&0&0&\cosh\frac{b}{2}+\sinh\frac{b}{2}y''/y'^2\end{pmatrix}
\end{equation}
Note that these are two eigenvectors of the 
Since we have 
\begin{equation}
(-1)^F=Z\otimes Z
\end{equation}
and that $\chi_i^2=1/2$ and $XY=iZ$
\begin{align}
c\eta &=\chi_4(\chi_1+i\chi_2)\\
&=\chi_4\chi_1+i\chi_4\chi_2\\
&=\frac{i}{2}Y\otimes Y+\frac{1}{2}X\otimes Y\\
&=\frac{1}{2}(X+iY)\otimes Y
\end{align}
Plugging this into the Pfaffian 
\begin{equation}
\mathrm{Pf}(\Omega)=\mathrm{Tr}\left((-1)^FU(2\pi)c\eta(0)\right)=\frac{2\sinh\frac{b}{2}}{y'}
\end{equation}
Therefore, the symplectic measure and the ultralocal measure are related by 
\begin{equation}
\frac{d\mu_s}{d\mu_u}=\frac{2\sinh\frac{b}{2}}{dy}
\end{equation}

\subsection{Crosscap measure}

\begin{figure}[H]
\centering
\includegraphics[width=0.6\textwidth]{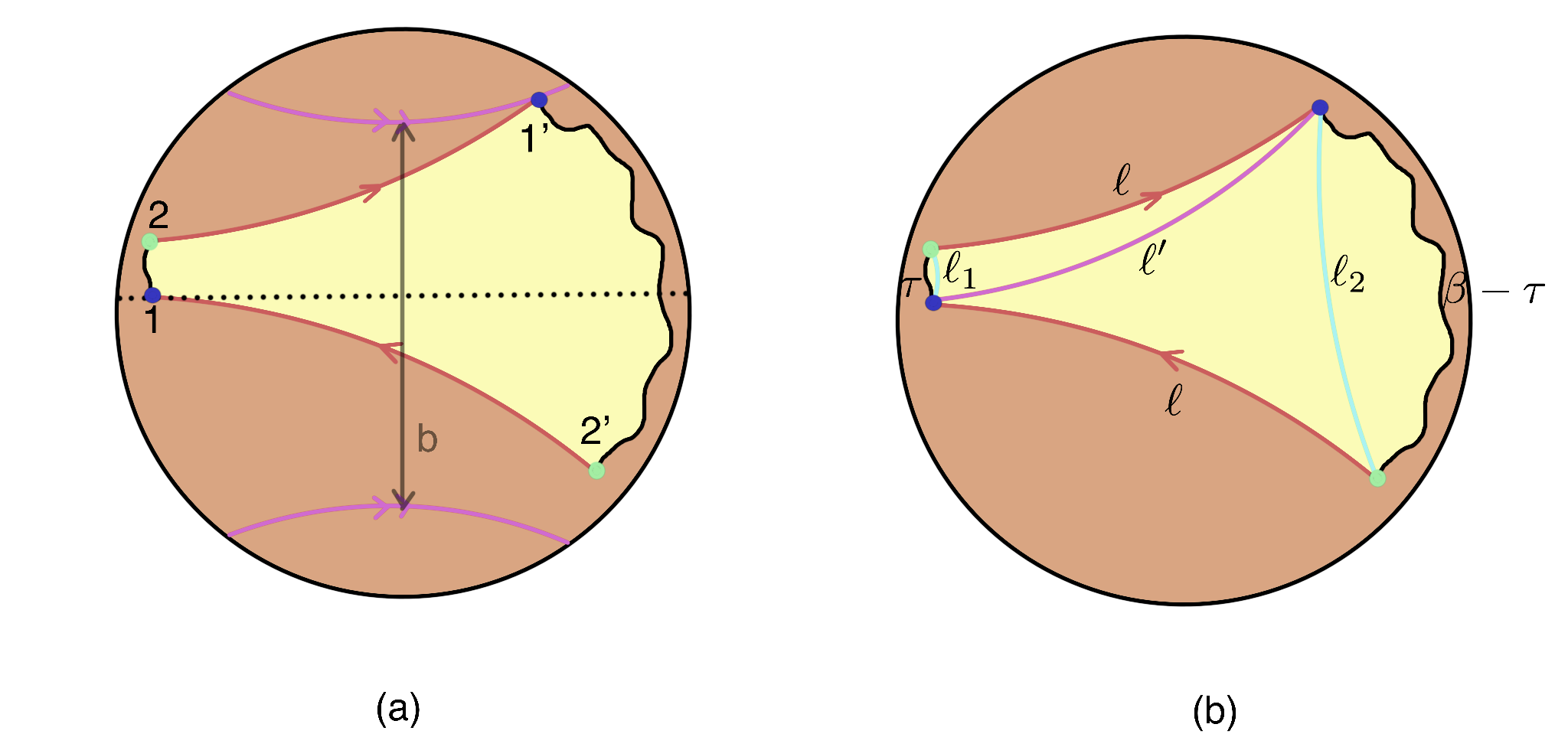}
\caption{(a) Disk+crosscap drawn as a quotient of a hyperbolic disk (b) with labels}
\label{ccmetric}
\end{figure}

As in figure~\ref{ccdrawing}(b), disk+crosscap is the quotient of part of a double trumpet. If we draw both the disk+crosscap and the double trumpet as quotient of a hyperbolic disk, we get figure~\ref{ccmetric}(a). Note that $b$ is the length of a two-sided geodesic, with antipodal points identified so that the length of one-sided geodesic is $b/2$. 

Now consider two points on the boundary of the disk+crosscap and their images $1,2,1',2'$. Let the coordinates of $1,2,1',2'$ be $(x_1,y_1)=(r,0)$, $(x_2,y_2)=(r',\tau)$, $(-r,b/2)$, $(-r',\tau-b/2)$. We can then relate the coordinates we use in Hartle-Hawking wavefunction calculations, i.e. the $\ell$'s, to the coordinates of these points.

\begin{align}
\cosh\ell_1&=\cosh r\cosh r'\cosh\tau-\sinh r\sinh r'\\
\cosh\ell_2&=\cosh r\cosh r'\cosh(b-\tau)-\sinh r\sinh r'\\
\cosh \ell&=\cosh r\cosh r'\cosh(\tau-\frac{b}{2})+\sinh r\sinh r'\\
\cosh \ell'&=\cosh^2r\cosh\frac{b}{2}+\sinh^2 r
\end{align}
Since we are looking at points on the boundary $r,r'\rightarrow\infty$ and the above equations become
\begin{align}
e^{\ell_1}&\approx e^{|r|+|r'|}\sinh^2\frac{\tau}{2}\\
e^{\ell_2}&\approx e^{|r|+|r'|}\sinh^2\frac{b-\tau}{2}\\
e^\ell&\approx e^{|r|+|r'|}\cosh^2\frac{\tau-b/2}{2}\\
e^{\ell'}&\approx e^{2|r|}\cosh^2\frac{b}{4}
\end{align}
This gives a Jacobian matrix of $\{\ell_1,\ell_2,\ell,\ell'\}$ with respect to $\{r,r',b,\tau\}$ with determinant
\begin{equation}
\det\begin{pmatrix}1&1&0&\mathrm{coth}\frac{\tau}{2}\\1&1&\mathrm{coth}\frac{b-\tau}{2}&-\mathrm{coth}\frac{b-\tau}{2}\\1&1&-\frac{1}{2}\tanh\frac{\tau-b/2}{2}&\tanh\frac{\tau-b/2}{2}\\2&0&\frac{1}{2}\tanh\frac{b}{4}&0\end{pmatrix}=\frac{2\cosh^2\frac{b}{4}}{\sinh\frac{\tau}{2}\sinh\frac{b-\tau}{2}}
\end{equation}
Putting these together, we then have
\begin{align}
e^{\ell_1/2+\ell_2/2}d\ell_1d\ell_2d\ell d\ell'&=e^{|r|+|r'|}\sinh\frac{\tau}{2}\sinh\frac{b-\tau}{2}\frac{2\cosh^2\frac{b}{4}}{\sinh\frac{\tau}{2}\sinh\frac{b-\tau}{2}}drdr'dbd\tau\\
&=e^{|r|+|r'|}2\cosh^2\frac{b}{4}drdr'dbd\tau\\
&=e^{|r|+|r'|}\frac{\sinh\frac{b}{2}}{\tanh\frac{b}{4}}drdr'dbd\tau\label{trumpetmeasure}
\end{align}
Note that $x_1=r$, $x_2=r'$, $y_2=\tau$ and we fixed $y_1=0$ but we should note that $y_1$ should still be integrated as part of the wiggle in the ultralocal measure. Combining (\ref{trumpetmeasure}) with $\frac{d\mu_s}{d\mu_u}=\frac{2\sinh\frac{b}{2}}{dy_1}$ from last section, we get
\begin{align}
e^{\ell_1/2+\ell_2/2}d\ell_1d\ell_2d\ell d\ell'd\mu_u&=e^{|x_1|+|x_2|}\frac{\sinh\frac{b}{2}}{\tanh\frac{b}{4}}dx_1dx_2dbdy_2 \frac{dy_1}{2\sinh\frac{b}{2}} d\mu_s\\
&=e^{|x_1|+|x_2|}dx_1dx_2dy_1dy_2\frac{db}{2\tanh\frac{b}{4}}d\mu_s
\end{align}
This is then the measure we should use for disk+crosscap to calculate path integrals. Comparing with the trumpet measure of path integrals, we see an additional factor $\frac{db}{2\tanh\frac{b}{4}}$.


\section{Fermions on a Hyperbolic Disk}
\label{appendixfermion}
To understand the behavior of the bulk field, let us first examine the Dirac equation on a hyperbolic disk. See Appendix F of \cite{AppendixF} and \cite{Foit:2019nsr} for more information.

Recall that the metric on a hyperbolic disk is given by
\begin{equation}
ds^2=dx^2+\cosh^2x\,dy^2
\end{equation}
then Euclidean frame fields $e^a=e^{\phantom{\mu}a}_\mu dx^\mu$ satisfy
\begin{equation}
g_{\mu\nu}=e^{\phantom{\mu}a}_\mu e^{\phantom{\nu}b}_\nu\delta_{ab}=\begin{pmatrix}1&0\\0&\cosh^2x\end{pmatrix}
\end{equation}
so we can choose
\begin{equation}
e^1=dx\quad\quad e^2=\cosh x\,dy
\end{equation}
Remember that a torsion-free spin connection one-form satisfies
\begin{equation}
de^a=-\omega^a_{\phantom{a}b}\wedge e^b
\end{equation}
Evaluating $de^a$ explicitly we then get
\begin{align}
de^1&=-\omega^1_{\phantom{a}b}\wedge e^b=0\\
de^2&=-\omega^2_{\phantom{a}b}\wedge e^b=\sinh x \,dx\wedge dy=\tanh x\,e^1\wedge e^2
\end{align}
Using the above two equations we can solve for $\omega$ and get
\begin{equation}
\omega^1_{\phantom{1}2}=-\tanh x\,e^2=-\sinh x\,dy
\end{equation}
i.e.
\begin{equation}
\omega^{12}_y=-\sinh x
\end{equation}
Recall that in Euclidean signature, the gamma matrices are given by
\begin{equation}
\gamma_1=\begin{pmatrix}0&-1\\1&0\end{pmatrix}\quad\quad\gamma_2=\begin{pmatrix}0&i\\i&0\end{pmatrix}
\end{equation}
Then gamma matrices on a hyperbolic disk then look like
\begin{equation}
\gamma_x=e^a_x\gamma_a=\gamma_1=\begin{pmatrix}0&-1\\1&0\end{pmatrix}\quad\quad\gamma_y=e^a_y\gamma_a=\cosh x\,\gamma_2=\begin{pmatrix}0&i\cosh x\\i\cosh x&0\end{pmatrix}
\end{equation}
For Pin$^-$ structure, the spin matrices are given by
\begin{equation}
\Sigma_{12}=-\frac{1}{4}[\gamma_1,\gamma_2]=\frac{i}{2}\begin{pmatrix}1&0\\0&-1\end{pmatrix}=\Lambda_0
\end{equation}
According to \cite{Witten16} section 5, for a 2-manifold endowed with a Pin$^-$ structure the Dirac equation is given by $(\overline{\gamma}\slashed{D}+M)\Psi=0$ with $\overline{\gamma}$ given in equation (\ref{gammabar}). The Dirac operator is then
\begin{align}
\overline{\gamma}\slashed{D}+M&=\overline{\gamma}\gamma^\mu D_\mu+M=\overline{\gamma}\gamma^\mu(\partial_\mu+\frac{1}{2}\omega^{ab}_\mu\Sigma_{ab})+M\\
&=\begin{pmatrix}0&-1\\-1&0\end{pmatrix}\frac{\partial}{\partial x}+\begin{pmatrix}0&i/\cosh x\\-i/\cosh x&0\end{pmatrix}(\frac{\partial}{\partial y}-\frac{1}{2}2\sinh x\begin{pmatrix}i/2&0\\0&-i/2\end{pmatrix})+M\\
&=\begin{pmatrix}M&-\frac{\partial}{\partial x}+\frac{i}{\cosh x}\frac{\partial}{\partial y}-\frac{1}{2}\tanh x\\-\frac{\partial}{\partial x}-\frac{i}{\cosh x}\frac{\partial}{\partial y}-\frac{1}{2}\tanh x&M\end{pmatrix}
\end{align}
For $x\rightarrow\infty$, the Dirac equation is approximately
\begin{equation}
\begin{pmatrix}M&-\frac{\partial}{\partial x}-\frac{1}{2}\\-\frac{\partial}{\partial x}-\frac{1}{2}&M\end{pmatrix}\Psi=0
\end{equation}
For plane-wave solution $e^{\alpha x}$ this gives
\begin{equation}
\begin{pmatrix}M&-\alpha-\frac{1}{2}\\-\alpha-\frac{1}{2}&M\end{pmatrix}\Psi=0
\end{equation}
This has solution if $\alpha=-\frac{1}{2}\pm M$ with null space $\eta_+=\begin{pmatrix}1\\1\end{pmatrix}$ and $\eta_-=\begin{pmatrix}1\\-1\end{pmatrix}$ respectively. Given these two solutions we can then expand the bulk fermion $\Psi$ in terms of the boundary fermions $\psi_\pm$. 
\begin{equation}
\Psi(y,x\rightarrow\infty)=e^{(-\frac{1}{2}+M)x}\psi_+(y)\eta_++e^{(-\frac{1}{2}-M)x}\psi_-(y)\eta_-
\end{equation}
For $x\rightarrow-\infty$, the Dirac equation is approximately
\begin{equation}
\begin{pmatrix}M&-\frac{\partial}{\partial x}+\frac{1}{2}\\-\frac{\partial}{\partial x}+\frac{1}{2}&M\end{pmatrix}\Psi=0
\end{equation}
For plane-wave solution $e^{\alpha x}$ this gives
\begin{equation}
\begin{pmatrix}M&-\alpha+\frac{1}{2}\\-\alpha+\frac{1}{2}&M\end{pmatrix}\Psi=0
\end{equation}
This has solution if $\alpha=\frac{1}{2}\pm M$ with null space $\eta_+=\begin{pmatrix}1\\1\end{pmatrix}$ and $\eta_-=\begin{pmatrix}1\\-1\end{pmatrix}$ respectively. Given these two solutions we can then expand the bulk fermion $\Psi$ in terms of the boundary fermions $\psi_\pm$. 
\begin{equation}
\Psi(y,x\rightarrow-\infty)=e^{(\frac{1}{2}+M)x}\psi_+(y)\eta_++e^{(\frac{1}{2}-M)x}\psi_-(y)\eta_-
\end{equation}
We pick a boundary condition by setting either $\psi_+$ or $\psi_-$ to zero and the other one becomes the boundary fermion operator.

Now let us calculate the two-point correlation functions of the bulk fermions near the boundary, which allows us to compute the two-point correlation functions of the boundary fermions. This was computed in \cite{AppendixF} which solved the bulk fermion near the boundary exactly with the metric
\begin{equation}
ds^2=\frac{4dzd\overline{z}}{(1-z\overline{z})^2}\quad\quad z=x^1+ix^2
\end{equation}
Expressed using variables $u$ and $\varphi$ defined by
\begin{equation}
z=\sqrt{u} e^{i\varphi}\quad\quad\overline{z}=\sqrt{u}e^{-i\varphi}
\end{equation}
the bulk fermion in the $(x^1,x^2)$ local frames with Neumann boundary condition on the asymptotic boundary is given by
\begin{equation}
\Psi\approx\psi(\varphi)(1-u)^\Delta\begin{pmatrix}e^{i\varphi/2}\\-e^{-i\varphi/2}\end{pmatrix}\quad\quad u\rightarrow1
\end{equation}
where the phases $e^{\pm i\varphi/2}=e^{-i\nu\varphi}$ come from spinors with spin $\nu=\mp\frac{1}{2}$ respectively. We can relate $(u,\varphi)$ to $(x,y)$ on the boundary by
\begin{equation}
1-u\approx \frac{e^{-|x|}}{\cosh y}\quad\quad \sin\varphi=\tanh y
\end{equation}
From \cite{Kitaev:2017hnr}, we know that we can rotate $\Psi$ in $(x^1,x^2)$ frame to that in $(u,\varphi)$ frame by multiplying the result by $e^{i\nu\varphi}$. And then we can further get $\Psi$ on the boundary in $(x,y)$ frame which is the same as the $(u,\varphi)$ frame on the right boundary and related to $(u,\varphi)$ frame by a $\pi$-rotation on the left boundary, i.e. multiplying $\Psi$ by $e^{i\nu\pi}$. Thus
\begin{equation}
\Psi\approx\begin{cases}\frac{\psi(y)}{(\cosh y)^\Delta}e^{-\Delta x}\begin{pmatrix}1\\-1\end{pmatrix}&x\rightarrow\infty\\ -\frac{i\psi(y)}{(\cosh y)^\Delta}e^{\Delta x}\begin{pmatrix}1\\1\end{pmatrix}&x\rightarrow-\infty\end{cases}
\end{equation}
But we should note that \cite{AppendixF} worked in Pin$^+$ instead of Pin$-$ with Dirac equations and $\gamma$ matrices different. By comparing the different conventions, we find that in our case
\begin{equation}
\Psi\approx\begin{cases}\frac{\psi(y)}{(\cosh y)^\Delta}e^{-\Delta x}\begin{pmatrix}1\\1\end{pmatrix}&x\rightarrow\infty\\ \frac{i\psi(y)}{(\cosh y)^\Delta}e^{\Delta x}\begin{pmatrix}1\\-1\end{pmatrix}&x\rightarrow-\infty\end{cases}
\end{equation}
In the large mass approximation, this reduces to our previous result. 

Let $(u_0\rightarrow1,\varphi_0)$ and $(u_1\rightarrow1,\varphi_1)$ be two points on the asymptotic boundary one on the left side and one on the right side, respectively. From \cite{AppendixF} (F.50), we know that if we insert two bulk fermions at these two points, the Green's function is given by
\begin{align}
\braket{\Psi(u_0,\varphi_0)\Psi^T(u_1,\varphi_1)}&=\braket{\Psi(u_0,\varphi_0)\overline{\Psi}(u_1,\varphi_1)}\gamma^0\\
&\approx\frac{i\Gamma(\Delta+\frac{1}{2})^2}{4\pi\Gamma(2\Delta)}\left(2\sin\frac{\varphi_1-\varphi_0}{2}\right)^{-2\Delta}(1-u_0)^{\Delta}(1-u_1)^{\Delta}\begin{pmatrix}1\\-1\end{pmatrix}\begin{pmatrix}1&1\end{pmatrix}\\
&\approx\frac{i\Gamma(\Delta+\frac{1}{2})^24^\Delta }{4\pi\Gamma(2\Delta)}e^{-\Delta\ell}\begin{pmatrix}1\\-1\end{pmatrix}\begin{pmatrix}1&1\end{pmatrix}
\end{align}
where we have used the fact that
\begin{equation}
\frac{e^\ell}{4}\approx  \frac{4\sin^2\left((\varphi_1-\varphi_0)/2\right)}{(1-u_1)(1-u_0)}
\end{equation}
The free two-point correlation function of boundary fermions is then given by
\begin{equation}
\braket{\psi_-\psi_+}\propto\eta_-^T\braket{\Psi\Psi^T}\eta_+\propto ie^{-\Delta\ell}
\end{equation}
as we expected. 


\section{Crosscap $\eta$}
\label{appendixeta}

In order to calculate $\eta$ for a crosscap, we study $\mathbb{RP}^2$, which is a hemisphere with boundary pair of antipodal points identified. Thus we can start from solving fermions on a hemisphere. For the hemisphere, use metric
\begin{equation}
ds^2=d\theta^2+\sin^2\theta\,d\phi^2
\end{equation}
We want to solve the eigenvalue problem
\begin{equation}
-i\slashed{D}\Phi=\lambda\Phi
\end{equation}
This is solved by(for more detail see Appendix F of \cite{hemisphere})
\begin{equation}
-i\slashed{D}\Phi_{\pm n,m}=\pm n\Phi_{\pm n,m}
\end{equation}
where $m=\frac{1}{2}-n,\ldots,n-\frac{1}{2}$. The boundary condition is given by
\begin{equation}
\Phi_{\pm n,m}(\pi-\theta,\phi+\pi)=\mp(-1)^ni\sigma_y\Phi_{\pm n,m}(\theta,\phi)
\end{equation}
If reflection $R$ acts as left multiplication by $-i\sigma_y$, the modes we choose are $\Phi_{n,m}(\theta,\phi)$ for odd $n>0$ and $\Phi_{-n,m}(\theta,\phi)$ for even $n>0$. Then the topological invariant $\eta$ is given by
\begin{align}
\eta&=\sum_{n>0,\text{odd}}\frac{1}{n^s}-\sum_{n>0,\text{even}}\frac{1}{n^s}\\
&=\sum_{k=0}^\infty\frac{1}{(2k+1)^s}-\sum_{k=1}^\infty\frac{1}{(2k)^s}\\
&=\lim_{\epsilon\rightarrow0}e^{-\epsilon}\sum_{k=0}^\infty e^{-2k\epsilon}-\sum_{k=1}^\infty e^{-2k\epsilon}\\
&=\lim_{\epsilon\rightarrow0}\frac{e^{-\epsilon}-e^{-2\epsilon}}{1-e^{-2\epsilon}}\\
&=\lim_{\epsilon\rightarrow0}\frac{e^{-\epsilon}}{1+e^{-\epsilon}}\\
&=\frac{1}{2}
\end{align}
Similarly, if reflection $R$ acts as left multiplication by $i\sigma_y$, the topological invariant $\eta=-\frac{1}{2}$.


\bibliography{references}

\bibliographystyle{utphys}

\end{document}